\documentclass[floatfix,twocolumn,showpacs,preprintnumbers,amsmath,amssymb,pre,superscriptaddress]{revtex4-1}

\usepackage{color}    
\usepackage{graphicx}
\usepackage{braket} 
\usepackage{dcolumn}
\usepackage{bm}
\usepackage{subfigure}
\usepackage{amssymb}
\usepackage{multirow}
\usepackage{natbib}
\usepackage{adjustbox}
\usepackage{tikz}
\usepackage{upgreek}
\usetikzlibrary{shapes,backgrounds,calc,patterns}
\tikzset{subfiglabel/.style={left, minimum size=4ex}}%
\usepackage{mathtools}
\graphicspath{{plots/}}
\usepackage[english]{babel}

\usepackage{xcolor}
\usepackage{placeins}

\usepackage{newcommands}

\newcommand{\euler}{\mathrm{e}}
\newcommand{\im}{\mathrm{i}}

\renewcommand{\pi}{\uppi}

\begin{document}


\title{Restricted configuration path integral Monte Carlo\footnote{Dedicated to Jim Dufty on the occasion of his 80th birthday}}

\author{A.~Yilmaz}
\affiliation{Institut f\"ur Theoretische Physik und Astrophysik, Christian-Albrechts-Universit\"at zu Kiel,
 Leibnizstra{\ss}e 15, 24098 Kiel, Germany}
 
 \author{K.~Hunger}
\affiliation{Institut f\"ur Theoretische Physik und Astrophysik, Christian-Albrechts-Universit\"at zu Kiel,
 Leibnizstra{\ss}e 15, 24098 Kiel, Germany}

\author{T.~Dornheim}
\affiliation{Center for Advanced Systems Understanding (CASUS), D-02826 G\"orlitz, Germany}

 \author{S.~Groth}
\affiliation{Institut f\"ur Theoretische Physik und Astrophysik, Christian-Albrechts-Universit\"at zu Kiel,
 Leibnizstra{\ss}e 15, 24098 Kiel, Germany}

  \author{M.~Bonitz}
\affiliation{Institut f\"ur Theoretische Physik und Astrophysik, Christian-Albrechts-Universit\"at zu Kiel,
 Leibnizstra{\ss}e 15, 24098 Kiel, Germany}

\begin{abstract}
Quantum Monte Carlo belongs to the most accurate simulation techniques for quantum many-particle systems. However, for fermions, these simulations are hampered by the sign problem that prohibits simulations in the regime of strong degeneracy. The situation changed with the development of configuration path integral Monte Carlo (CPIMC)
by Schoof \textit{et al.} [T. Schoof \textit{et al.}, Contrib. Plasma Phys. \textbf{51}, 687 (2011)] that allowed for the first \textit{ab initio} simulations for dense quantum plasmas [T. Schoof \textit{et al.}, Phys. Rev. Lett. \textbf{115}, 130402 (2015)]. CPIMC also has a sign problem that occurs when the density is lowered, i.e. in a parameter range that is complementary to traditional QMC formulated in coordinate space. Thus, CPIMC simulations for the warm dense electron gas are limited to small values of the Brueckner parameter -- the ratio of the interparticle distance to the Bohr radius -- $r_s=\overline{r}/a_B \lesssim 1$. In order to reach the regime of stronger coupling (lower density) with CPIMC, here we investigate additional restrictions on the Monte Carlo procedure. In particular, we introduce two different versions of ``restricted CPIMC'' -- called RCPIMC and RCPIMC+ -- where certain sign changing Monte Carlo updates are being omitted. Interestingly, one of the methods (RCPIMC) has no sign problem at all, but it is less accurate than RCPIMC+ which neglects only a smaller class of the Monte Carlo steps. Here we report extensive simulations for the ferromagnetic uniform electron gas with which we investigate the properties and accuracy of RCPIMC and RCPIMC+. Further, we establish the parameter range in the density-temperature plane where these simulations are both feasible and accurate. The conclusion is that RCPIMC and RCPIMC+ work best at temperatures in the range of $\Theta \sim 0.1\dots 0.5$ allowing to reach density parameters up to $r_s \sim 3\dots 5$, 
thereby partially filling a gap left open by existing \textit{ab initio} QMC methods.
\end{abstract}

\maketitle

 \section{Introduction}\label{s:intro}
Warm dense matter has recently become one of the most active research fields in plasma physics, being situated on the border of plasma physics and condensed matter physics, e.g.~Refs.~\onlinecite{graziani-book,Fortov2016, moldabekov_pre_18, dornheim_physrep_18}. 
Examples include dense degenerate matter in brown and white dwarf stars \cite{saumon_the_role_1992, chabrier_quantum_1993,chabrier_cooling_2000}, giant planets, e.g. \cite{schlanges_cpp_95,bezkrovny_pre_4, vorberger_hydrogen-helium_2007, militzer_massive_2008, redmer_icarus_11,nettelmann_saturn_2013}  and the outer crust of neutron stars \cite{Haensel,daligault_electronion_2009}. 
In the laboratory, WDM is being routinely produced via laser or ion beam compression or with Z-pinches, see Ref.~\onlinecite{falk_2018} for a recent review. The important experimental facilities include  the National Ignition facility at Lawrence Livermore National Laboratory \cite{moses_national_2009,hurricane_inertially_2016}, the Z-machine at Sandia National Laboratory \cite{matzen_pulsed-power-driven_2005,knudson_direct_2015}, the Omega laser at the University of Rochester \cite{nora_gigabar_2015},  the Linac Coherent Light Source (LCLS) at Stanford\cite{sperling_free-electron_2015,glenzer_matter_2016}, the European free electron  laser facilities FLASH and X-FEL \cite{zastrau_resolving_2014,tschentscher_photon_2017},  and the upcoming FAIR facility at GSI Darmstadt  \cite{hoffmann_cpp_18,tahir_cpp19}.  
Among the important applications is inertial confinement fusion \cite{moses_national_2009,matzen_pulsed-power-driven_2005,hurricane_inertially_2016} where electronic quantum effects are important during the initial phase. 
Aside from dense plasmas, also many condensed matter systems exhibit WDM behavior -- if they are subject to strong excitation, e.g. by lasers or free electron lasers~\cite{Ernstorfer1033,PhysRevX.6.021003}. 

Characteristic of all these  diverse systems is the governing role of electronic quantum effects, moderate to strong Coulomb correlations and finite temperature effects. 
Quantum effects of electrons are of relevance at low temperature and/or if matter is very highly compressed, such that the temperature is of the order of (or lower than) the Fermi temperature, for a recent overview, see Ref.~\onlinecite{bonitz_pop_20}. 

Due to the simultaneous relevance of these effects, a theoretical description of WDM is difficult. Among the actively used approaches are quantum kinetic theory
\cite{kadanoff-baym,keldysh65,green-book, bonitz_aip_12, balzer_pra_10,balzer_pra_10_2,schluenzen_19_prl,schluenzen_cpp_18}, quantum hydrodynamics \cite{bonitz_pre_13,zhandos_pop18,zhandos_cpp_19} and density functional theory simulations because they, for the first time, enabled the selfconsistent simulation of realistic warm dense matter, that includes both, plasma and condensed matter phases, e.g. Refs.~\onlinecite{collins_pre_95,plagemann_njp_2012,witte_prl_17}. The most accurate results for warm dense matter, so far, were obtained via first principle computer simulations such as quantum Monte Carlo (QMC) \cite{sign_cite,militzer_path_2000,filinov_ppcf_01,filinov_pss_00,schoof_cpp11,dornheim_physrep_18,filinov_pre15,schoof_prl15,dornheim_njp15,pierleoni_cpp19}.
The first QMC results for dense quantum plasmas and WDM were obtained by Ceperley and Militzer \textit{et al.}, e.g. Refs.~\onlinecite{Ceperley1991,militzer_path_2000} and Filinov \textit{et al.} \cite{filinov_ppcf_01,filinov_jetpl_01}. However, the simulations of the electronic part of WDM were strongly hampered by the fermion sign problem (see Ref.~\onlinecite{dornheim_pre_2019} for a recent overview) that made simulations at strong electronic degeneracy impossible. One approach to relieve this problem is the fixed node approximation \cite{Ceperley1991}, which works well in the ground state. However, the associated systematic error at finite temperatures is largely unclear. On the other hand, in the direct fermionic QMC simulations of Filinov \textit{et al.} the sign problem was reduced by optimized Monte Carlo updates \cite{bonitz_prl_5} but the simulations remained restricted to moderate quantum degeneracy.

Recently, a number of breakthroughs in QMC simulations occurred. The first was the idea to disentangle the complex warm dense matter system in the simulations and first develop improved simulations for the most challenging component -- the partially degenerate electrons. Brown \textit{et al.} \cite{Brown_2014,Brown_PRB_2013} presented the first restricted QMC (RPIMC) simulations for the jellium model at finite temperature, $0.0625 \le\Theta \le 8$, 
where $\Theta=k_BT/E_F$ ($E_F$ denotes the Fermi energy), for which ground state QMC approaches are not applicable. Even though these simulations have no sign problem they could not be extended into the degenerate regime, $r_s\lesssim 1$, where $r_s$ is the ratio of the mean interparticle distance to the Bohr radius,  $r_s=\bar r/a_B$, and even for moderate $r_s$, in the range between $1$ and $5$, the potential energy showed an unexpected non-monotonic behavior. 
This behavior was clarified by novel configuration PIMC (CPIMC) simulations -- i.e. PIMC formulated in Fock space \cite{schoof_cpp11} -- and has no sign problem at strong degeneracy (small $r_s$). In combination with an improved configuration space approach -- permutation blocking PIMC (PB-PIMC) \cite{dornheim_njp15,dornheim_jcp15,dornheim_cpp_19} -- and confirmed by independent density matrix QMC (DM-QMC) \cite{Malone_PRL_2016}, it became clear that RPIMC in this parameter range exhibits large systematic errors. On the other hand, the combination of CPIMC and PB-PIMC made it possible to obtain exact thermodynamic results for the warm dense uniform electron gas (UEG) over the entire density range, for $\Theta \gtrsim 0.5$. These data were extended to the thermodynamic limit by Dornheim \textit{et al}. in Ref.~\onlinecite{dornheim_prl16} and connected to existing ground state results~\cite{Spink_PRB_2013} by Groth \textit{et al.} \cite{groth_prl17} which led to the first accurate analytical parametrization of the free energy of the UEG \cite{dornheim_physrep_18}, for a discussion see Ref. \onlinecite{karasiev_PRB_2019}. Finally, we mention recent breakthroughs in the \textit{ab initio} computation of static and dynamic response functions using path integral MC approaches by Dornheim \textit{et al.} \cite{dornheim_prl_18,dornheim_jcp_19-nn,groth_prb_19,dornheim_HEDP_2020,dornheim_prl_20}.

\begin{figure}
    \centering
    \includegraphics[width=0.899\columnwidth]{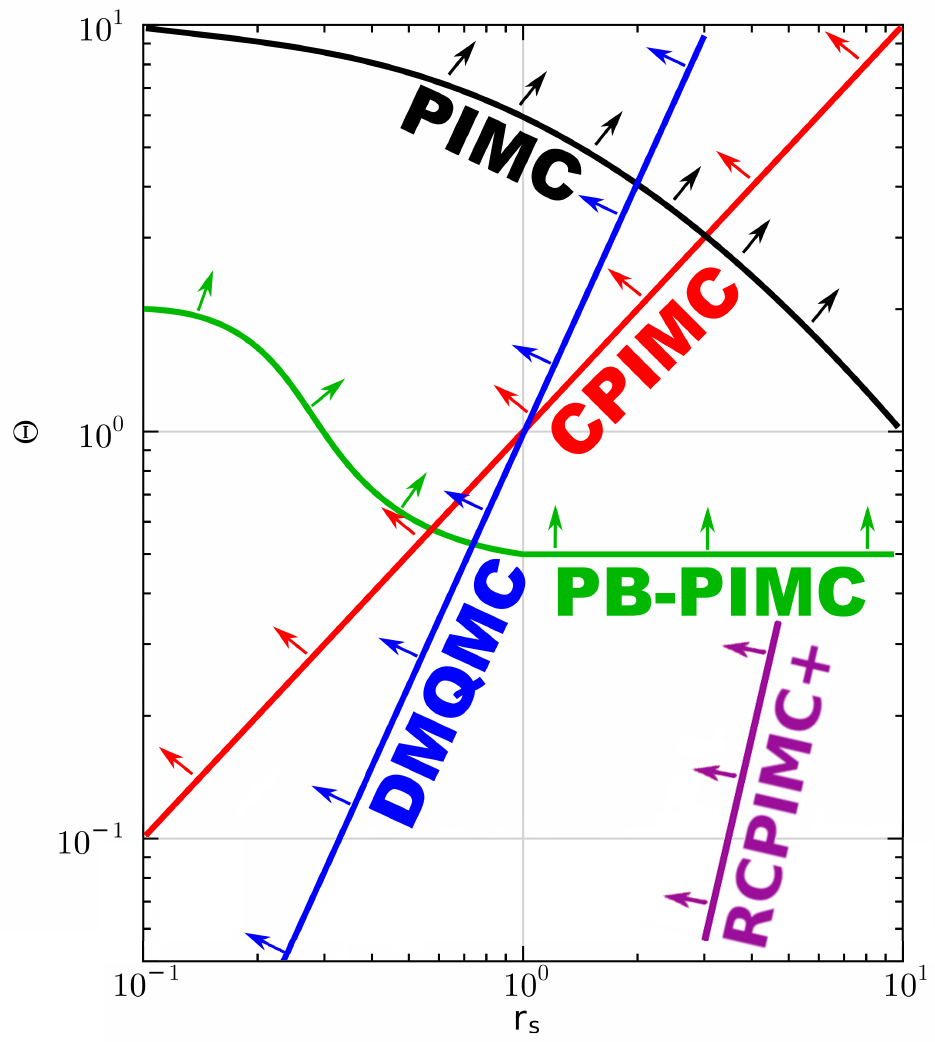}
    \caption{Qualitative overview on the status of \textit{ab initio} QMC simulations for the warm dense electron gas. Arrows indicate where the different methods (PIMC: fermionic PIMC, CPIMC: configuration PIMC, PB-PIMC: permutation blocking PIMC and DMQMC: density matrix QMC) are currently applicable. The presented restricted CPIMC (RCPIMC) approaches, for the first time, fill part of the two lower  quadrants. In contrast to the other methods RCPIMC has a systematic error, and the drawn line corresponds roughly to a $3\%$ error in the potential energy. Figure modified from Ref.~\onlinecite{dornheim_pop17}.
    }
    \label{fig:rs-theta-overview}
\end{figure}

Aside from these achievements, the fermion sign problem still puts severe limitations on the range of WDM parameters that are accessible by first principle QMC methods. Let us summarize these limitations, for an illustration, see Fig.~\ref{fig:rs-theta-overview}.
\begin{enumerate}
    \item Fermionic PIMC (``PIMC'' in Fig.~\ref{fig:rs-theta-overview}) in configuration space cannot access strong degeneracy, i.e. low $r_s$ and low $\Theta$
    \item PB-PIMC, as an optimized coordinate space method is able to reach low $r_s$ values but only above a minimum temperature, $\Theta\ge 0.5$.
    \item Fock space approaches (CPIMC and DMQMC) 
    cannot access strong coupling, i.e. they are restricted to $r_s\lesssim 1$, except for high temperature, $\Theta\gtrsim 1$.
    \item The above estimates refer to finite simulation sizes of the order of $N=33$ for the polarized and $N=66$ for the paramagnetic UEG. For smaller $N$ the parameter range increases. These data are sufficient for a reliable extrapolation to the thermodynamic limit using the accurate finite size correction of Ref.~\onlinecite{dornheim_prl16}.
\end{enumerate}
From this summary it is clear that there remains a ``white area'' in the lower half of the temperature  temperature-density plane where no ab initio data exist (even though the analytical parametrization of Ref.~\onlinecite{groth_prl17} does not leave a gap).
This yields the motivation for the present paper: 
The goal is to investigate whether CPIMC can be extended towards stronger coupling by omitting some or all of the sign changing Monte Carlo updates. Even though with this we give up the \textit{ab initio} character of the method this would be acceptable if the error introduced by these  approximations is sufficiently small and the trends are predictable. 
After carefully analyzing all of the Monte Carlo steps we identify two new ``restricted'' CPIMC methods that will be called RCPIMC and RCIPMC+. While RCPIMC omits all sign changing kinks (Monte Carlo updates), RCPIMC+ neglects only a part of them and is, therefore, more accurate than the former. At the same time, RCPIMC has no sign problem at all, whereas the sign problem of RCPIMC+ is considerably reduced, compared to full CPIMC.
With this the two methods overcome, in part, the limitations of CPIMC allowing for approximate, but still accurate simulations in a broader parameter range than the latter. We analyze in detail the accuracy and the applicability limits of the two new approximations by performing extensive simulations for the uniform ferromagnetic electron gas. Our main result is that, indeed RCPIMC and RCPIMC+ allow one to enter parts of the parameter range that was previously unaccessible to existing \textit{ab initio} approaches such as CPIMC, PB-PIMC, and Density Matrix QMC. In particular, simulations are possible in the range of $0.05 \lesssim \Theta \lesssim 0.5$ and $r_s$ up to  $3\dots 5$.\\

 This paper is organized as follows: In Sec.~\ref{s:cpimc} we give an overview on CPIMC and the involved Monte Carlo updates. There we also introduce RCPIMC and RCPIMC+. After this, in Sec.~\ref{s:results} we present extensive new thermodynamic results for the uniform electron gas for particle numbers ranging from $N=4$ to $N=33$. Furthermore, we investigate how well these methods can reproduce the thermodynamic properties of the macroscopic UEG. The paper concludes with a discussion and outlook in Sec.~\ref{s:conclusion}.

\section{Configuration PIMC}\label{s:cpimc}
Here we give a brief overview on the CPIMC method and present  the main formulas.
Throughout this paper we use atomic units where \(\hbar=m_e=e=1\). 
We start by considering a generic quantum many-body system at finite temperature with the hamiltonian 
\begin{align}
    \hat H = \hat T + \hat W \,,
    \label{eq:h}
\end{align}
where $\hat T$ denotes the single-particle energy and $\hat W$ the interaction energy. We will keep the following discussion of CPIMC as general as possible before concentrating on the UEG hamiltonian, 
in Sec.~\ref{ss:cpimc}.
The thermodynamic properties of the system (\ref{eq:h}), including the expectation value of an arbitrary operator $\hat A$, are determined by the density operator and its normalization -- the partition function,
\begin{align}
    \hat \rho &= e^{-\beta \hat H}\,,
    \label{eq:rho-def}\\
    Z(\beta) &= \mbox{Tr}\, \hat \rho\,,
    \label{eq:z-def}\\
    \langle \hat A\rangle(\beta) &=\frac{1}{Z}
    \mbox{Tr}\,  \hat A \hat \rho\,,
    \label{eq:mean-a}
\end{align}
where $\beta = 1/k_BT$. Below we will work in the canonical ensemble, i.e. system volume and particle number $N$ are considered fixed.

The main problem for the computation of expressions $(\ref{eq:z-def})$ and $(\ref{eq:mean-a})$ is that the density operator is not known because kinetic and potential energy operators do not commute. This problem is solved in quantum Monte Carlo by transforming the density operator to a form where perturbation methods can be applied. Standard PIMC in coordinate space 
uses a Trotter decomposition in high-temperature factors allowing for perturbation theory around the strongly coupled semiclassical limit, e.g. Ref.~\onlinecite{filinov_ppcf_01}. Due to the fermion sign problem (FSP) this approach is limited to the weakly and moderately degenerate regime, for the UEG this corresponds to $r_s \gtrsim 1$ \cite{Brown_2014,filinov_pre15}, the precise value depending on temperature, cf. Fig.~\ref{fig:rs-theta-overview}. The origin of the FSP, in this approach, is the anti-symmetrization of the density operator, $\hat \rho \to \hat \rho^A$ which gives rise to $N!$ sign alternating terms causing an exponential loss of accuracy with the increase of $\beta$ and $N$.

Configuration PIMC (CPIMC) was introduced with the goal to access the high-density regime of strong degeneracy. There, the properly anti-symmetrized expectation value (\ref{eq:mean-a}) is computed without anti-symmetrization of the density operator. Instead, the trace is computed using a complete set of anti-symmetrized $N$-particle states \cite{schoof_cpp11,schoof_cpp15, cpimc_springer_14}, as will be explained below. Subsequently  a perturbation expansion of the density operator with respect to the interaction energy is performed. This approach is complementary to the one in standard PIMC and, therefore, allows to access complementary parameter regions, e.g., for the UEG, the region where $r_s \lesssim 1$.

In the following we introduce the perturbation expansion with respect to the interaction strength by using the interaction representation of quantum mechanics. 

\subsection{Perturbation Expansion with respect to the interaction}\label{ss:perturbation-expansion}
The perturbation expansion of the partition function [Sec.~\ref{ss:perturbation-z}] is analogous to time-dependent perturbation theory in standard quantum mechanics which we, therefore, consider first.

Recall the interaction (Dirac) picture of quantum mechanics for a system with the hamiltonian (\ref{eq:h}). The idea is to rewrite the hamiltonian as  $\Hmt = \Do + \Yo$, such that the time evolution is split into an ``easy'' part, driven by the operator $\hat D$, and a ``difficult'' part, driven by the operator $\hat Y$. Using the stationary eigenfunctions of $\hat D$ as a basis, the dynamics of $\hat D$ are given by a simple exponential operator, $e^{-i\hat D t}$, which is diagonal in that basis. The remaining part is non-diagonal with the dynamics given by 
\begin{equation}
    Y(t) = \euler^{\im\Do t} \hat{Y} \euler^{-\im\Do t}\,.
    \label{eq:y-interaction-picture}
\end{equation}
Consequently, the full time evolution operator has the form 
\begin{align}
    U(t,t') = e^{-i\hat D (t-t')} \hat T e^{-i\int_{t'}^t d\bar t\, \hat Y(\bar t)}\,,
    \label{eq:full-U}
\end{align}
where $\hat T$ is the time ordering operator. 
This equation can be rewritten as a Taylor series. Alternatively, we can start with the integral representation
\begin{equation}
    U(t,t') = \one - \im \int\limits_{t'}^t d\bar t \;\hat{Y}(\bar t)\,U(\bar t, t')\,,
\end{equation}
that can be rewritten as an iteration series
    \begin{align}
    \hat{U}(t,t') = \one
    &-\im \int\limits_{t'}^t \dx t_1\;\hat{Y}(t_1)
\label{eq:u-expansion}    \\
    &
    + \left(-\im\right)^2
    \int\limits_{t'}^t \dx t_1 \int\limits_{t'}^{t_1} \dx t_2
    \;\hat{Y}(t_1) \hat{Y}(t_2)
    \nonumber\\
    &
    +
    \left(-\im\right)^3
    \int\limits_{t'}^t \dx t_1 \int\limits_{t'}^{t_1} \dx t_2 \int\limits_{t'}^{t_2} \dx t_3
    \;\hat{Y}(t_1) \hat{Y}(t_2) \hat{Y}(t_3)
    \nonumber\\\nonumber
    &
    +
    \ldots
    .
\end{align}
So far, this is a completely general result, in particular, concerning the subdivision of the hamiltonian into the two operators $\hat D$ and $\hat Y$. Below, in Sec.~\ref{ss:cpimc}, we will introduce the choice that is appropriate for CPIMC and the uniform electron gas and show how the two operators are related to the kinetic and interaction energy operators in Eq.~(\ref{eq:h}). But first, we need to apply the interaction picture  to the computation of the canonical density operator. 

\subsection{Perturbation theory for the partition function}\label{ss:perturbation-z}
It is well known that the density operator (\ref{eq:rho-def}) can be written as a time evolution operator of a stationary system in imaginary ``time'' (Wick rotation) 
\begin{align}
\hat{\rho} = 
\euler^{-\beta\Hmt} \equiv U(-i\beta,0) \,.   
\end{align}
With the replacements $t'\to 0$ and $t\to -i\beta$ we can directly use the solution (\ref{eq:full-U}) or (\ref{eq:u-expansion}) and obtain for the sum of the iteration series
\begin{widetext}
\begin{align}\label{eq:do_identity}
 e^{-\beta\hat{H}}=
  e^{-\beta\hat{D}}\sum_{K=0}^\infty (-1)^K \int\limits_{0}^{\beta} d t_1 \int\limits_{t_1}^{\beta} d t_2 \ldots \int\limits_{t_{K-1}}^\beta d t_K
\hat{Y}(t_K)\hat{Y}(t_{K-1})\cdot\ldots\cdot\hat{Y}(t_1)\,.
\end{align}
\end{widetext}

Statistical expectation values for a system of $N$ fermions can be calculated from the above (canonical) statistical operator via the trace, Eq.~(\ref{eq:mean-a}),
for which a key quantity is the partition function, Eq.~(\ref{eq:z-def}).
For the evaluation of the trace in Eqs.~(\ref{eq:mean-a}) and (\ref{eq:z-def}), we consider a complete orthonormal set of anti-symmetric $N$-particle states, i.e. Slater determinants,
which we write in \emph{occupation number representation} for a fixed particle number $N$,
\begin{equation}
    \ket{\onvv} = \ket{n_1, n_2, n_3, \ldots }
    ,\quad
    n_i \in \set{0,1}\,.
\end{equation}
In this representation the hamiltonian of an ideal Fermi gas will be diagonal, i.e., this representation is well adopted to the limit of high degeneracy (such as the high density electron gas, $r_s\to 0$), and the ideal Fermi gas can be simulated exactly without any sign problem \cite{schoof_cpp11}. 
For the general case of an interacting system, we obtain from (\ref{eq:do_identity}) for the partition function
\begin{widetext}
\begin{equation}
\label{eq:partition_perturbation_identity_inserted}
    \begin{aligned}
        Z(\beta) 
        &   =
        \sum\limits_{K=0}^{\infty} 
        (-1)^K
        \sum\limits_{\onvv}
        \int\limits_0^{\beta} \dx t_1 \int\limits_0^{t_1} \dx t_2  \ldots  \int\limits_0^{t_{K-1}} \dx t_K
        \;
        \braket{ \onvv | 
        \euler^{-\beta\Do }
        \hat{Y}(t_1) \hat{Y}(t_2) \ldots \hat{Y}(t_K)
        | \onvv }
        \\
        &
        =
        \sum\limits_{K=0}^{\infty} (-1)^K
        \sum\limits_{\onvv} \ldots \sum\limits_{\onv{K-1}}
        \int\limits_0^{\beta} \dx t_1 \int\limits_{t_1}^{\beta} \dx t_2  \ldots  \int\limits_{t_{K-1}}^{\beta} \dx t_K
        \;
        \braket{ \onvv | \euler^{-\Do \beta} \hat{Y}(t_1) | \onv{1} }
        \ldots
        \braket{ \onv{K-1} | \hat{Y}(t_K)  | \onvv }\,,
    \end{aligned}
\end{equation}
\end{widetext}
where the sum over $\{n\}$ arises from the trace, and
the second line is obtained by inserting \((K-1)\) identities \(\one = \sum_{\onvv} \ket{\onvv} \bra{\onvv}\) of the appropriate Fock space, and the indexing and the integration boundaries
are chosen as such to have \(t_1 \leq t_2 \leq \ldots \leq t_K\). Now consider the matrix elements, using the  interaction picture representation  (\ref{eq:y-interaction-picture}),
\begin{equation}
\begin{aligned}
    \braket{ \onv{\alpha} | \hat{Y}(t) | \onv{\gamma} }
    = &  
    \braket{ \onv{\alpha} | \euler^{\Do t} \hat{Y} \euler^{-\Do t}| \onv{\gamma} }
    \\
     \qquad\qquad\qquad = & \, \euler^{\Di{\alpha}t}
    \underbrace{\braket{ \onv{\alpha} | \hat{Y} | \onv{\gamma} }}_{\eqqcolon\Yij{\alpha}{\gamma}}
    \euler^{-\Di{\gamma}t}\,.
\end{aligned}
\nonumber
\end{equation}
Thus, the matrix elements arising in the partition function can be straightforwardly calculated. The  matrix elements of the diagonal part of the hamiltonian are the sums of the single-particle energies of the occupied orbitals
\begin{equation}
    \Di{\alpha} = \sum\limits_{i=0}^{\infty} b_{ii} n_i^{(\alpha)}\,,
\end{equation}
whereas the matrix elements of the off-diagonal part can be computed with the Slater-Condon rules \cite{schoof_cpp11,schoof_cpp15}. Explicit results for the uniform electron gas and the choice of a plane wave basis will be given in Sec.~\ref{ss:cpimc}. 
In order to shorten the notation and taking explicitly into account the periodicity of the trace we 
rename the time-arguments and the indexes, according to  \(t_0 = 0, t_{K+1} = \beta\) and \(\onv{0} = \onv{K} = \onvv\) (respectively \(\Di{K} = \Di{0}\) and \(\Yij{1}{0} = \Yij{1}{K}\)). As a result, the partition function (\ref{eq:partition_perturbation_identity_inserted}) becomes
\begin{widetext}
\begin{equation}
    \label{eq:partition_perturbation_compact}
    Z(\beta)
    =
    \sum\limits_{K=0}^{\infty} 
    \sum\limits_{\onvv} \ldots \sum\limits_{\onv{K-1}}
    \int\limits_0^{\beta} \dx t_1 
    \int\limits_{t_1}^{\beta} \dx t_2 \ldots 
    \int\limits_{t_{K-1}}^{\beta} \dx t_K
    \;
    (-1)^K
    \left(
    \prod\limits_{i=0}^{K} 
    \euler^{
    -
    \Di{i}(t_{i+1} - t_i)
    }
    \right)
    \times
    \left(
    \prod\limits_{i=1}^{K} 
    \Yij{i}{i-1}
    \right)
\end{equation}
\end{widetext}

\subsection{Configurations and paths in CPIMC}\label{ss:cpimc-paths}
The partition function (\ref{eq:partition_perturbation_compact}) can be understood as a sum over micro-configurations $C$, 
\begin{equation}
    Z = \intsum_C W(C)\,,
\end{equation}
which have the weight
\begin{equation}
    W(C) 
    =
    (-1)^K
    \left(
    \prod\limits_{i=0}^{K} 
    \euler^{
    -
    \Di{i}(t_{i+1} - t_i)
    }
    \right)
    \left(
    \prod\limits_{i=1}^{K} 
    \Yij{i}{i-1}
    \right)\,,
\label{eq:weight-c}
\end{equation}
which allows one to rewrite thermodynamic expectation values (\ref{eq:mean-a}) as 
\begin{equation}
    \braket{A} = \intsum_C W(C) A(C)\,  .
\end{equation}
What is left is to specify the configurations \(C\) that appear in the integrand in Eq.~(\ref{eq:partition_perturbation_compact}).
Evidently, the sum contains contributions (``paths'') in Slater determinant space, $|\{n\}\rangle\to |\{n^{(1)}\}\rangle\to |\{n^{(2)}\}\rangle\to \dots$, running through a total of $K$  steps. Due to the trace all paths are periodic, i.e. start and end with the same state $|\{n\}\rangle$. This means a path is specified by the full information about all states involved and their respective imaginary times,
\begin{equation}
    C \coloneqq \Set{K, \onvv, \onv{1},\dots, \onv{K-1}, t_1,\ldots,t_K}\,.
    \label{eq:path-1}
\end{equation}

\begin{figure}[ht]
    \centering

\begin{tikzpicture}[xscale=0.7, yscale=0.35]
\newcommand{\xrange}{10}
\newcommand{\yrange}{1}

\newcommand{\taunull}{0}
\newcommand{\taunulllb}{$0$}
\newcommand{\taueins}{\xrange*0.08}
\newcommand{\taueinslb}{$t_1$}
\newcommand{\tauzwei}{\xrange*0.2}
\newcommand{\tauzweilb}{$t_2$}
\newcommand{\taudrei}{\xrange*0.69}
\newcommand{\taudreilb}{$t_3$}
\newcommand{\tauvier}{\xrange*0.83}
\newcommand{\tauvierlb}{$t_4$}
\newcommand{\taufuenf}{\xrange*0.9}
\newcommand{\taufuenflb}{$t_5$}
\newcommand{\taulast}{\xrange}
\newcommand{\taulastlb}{$\beta$}
\newcommand{\zero}{\yrange*1}
\newcommand{\eins}{\yrange*2}
\newcommand{\zwei}{\yrange*3}
\newcommand{\drei}{\yrange*4}
\newcommand{\vier}{\yrange*5}
\newcommand{\fuenf}{\yrange*6}

\draw (0,0) -- +(\xrange,0) coordinate (xlabel);
\draw[->] (0,0) -- +(0,\yrange*7) coordinate (ylabel);
\foreach \i in {0,...,5} {
	\draw (-0.1,\i*\yrange+\yrange) node[left] {$\i$} -- (0.1,\i*\yrange+\yrange);
}
\foreach \i/\l in {\taunull/\taunulllb,\taueins/\taueinslb,\tauzwei/\tauzweilb,\taudrei/\taudreilb,\tauvier/\tauvierlb,\taufuenf/\taufuenflb,\taulast/\taulastlb} {
\draw (\i,0.1) -- (\i,-0.1) node[below] {\l};
}
\draw (0,0.1) -- (0,-0.1) node[below] {$0$};
\node at (0.5*\xrange,-1.7) {imaginary time};
\node[rotate=90] at (-1,3.5*\yrange) {orbital $i$};

\foreach \i in {1,...,6} {
\draw[semithick,dotted] (0,\i*\yrange) -- (\xrange,\i*\yrange);
}

\begin{scope}[thick]
\draw (0,\zero) -| (\tauzwei,\eins) -| (\taudrei,\zero) -- (\xrange,\zero);
\draw (0,\eins) -| (\taueins,\drei) -| (\taudrei,\zwei) -| (\tauvier,\vier) -- (\xrange,\vier);
\draw (0,\vier) -| (\taueins,\fuenf) -| (\taufuenf,\eins) -- (\xrange,\eins);
\draw (\taueins, \drei) -- (\taueins, \vier);
\draw (\taudrei, \eins) -- (\taudrei, \zwei);
 \end{scope}

\node at (\taueins*1.4,\yrange*7) {$q_{1,0}(s_1)$};


\draw[dashed] (\tauzwei+0.2,0.5) rectangle (\taudrei-0.2,\yrange*6.5);
\node at (\tauzwei*2.2,\fuenf+\yrange+0.25) {$\ket{\onv{3}}$};

\end{tikzpicture}
    
    \caption{Illustration of a path $C$, Eq.~(\ref{eq:path2}), with five kinks. The kink 1 at time $t_1$ involves four orbitals, $s_1=(1,4;3,5)$, whereas the kink at $t_2$ involves two orbitals, $s_2=(0;1)$. The third Slater determinant $|n^{(3)}\rangle$ exists between the imaginary ``times'' $t_2$ and $t_3$ and contains three occupied orbitals (1, 3, 5) and zeroes otherwise.}
    \label{fig:Kink_path}
\end{figure}
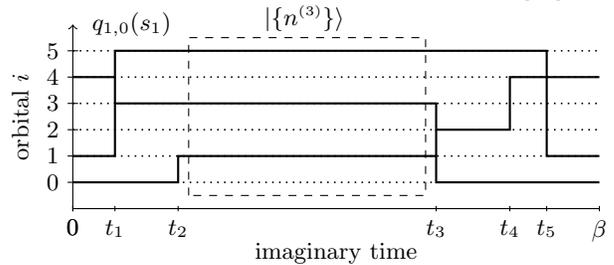

However, this information can be drastically reduced by taking into account the properties of the matrix elements of $\hat Y$.
Indeed, two adjacent Slater determinants are not independent of each other but are linked by the properties of the matrix \(\Yij{\alpha}{\alpha-1}\). As we will see below, in the case of the UEG, the off-diagonal operator $\hat Y$ is determined by the pair interaction operator $\hat W$, the matrix elements of which, in a basis of Slater determinants, are known and given by the Slater-Condon rules \cite{schoof_cpp11}.
The matrix elements \(\Yij{\alpha}{\gamma}\) are only nonzero in three cases: if the basis states \(\ket{\onv{\alpha}}\) and \(\ket{\onv{\gamma}}\) differ in zero, two or four orbitals and, thus, no error is introduced if the sum is restricted to only such paths
Thus, it is clear that, instead of considering each of the \(K\) states 
forming the path, it is equally possible to consider just the first state $|\{n\}\rangle$, together with the \(K\) \emph{changes} that yield nonzero matrix elements. 
These changes are called \emph{kinks} and are specified by the changed orbitals between the states forming  matrix element in Eq.~(\ref{eq:partition_perturbation_identity_inserted}). These can be either two indices, \(s=(p;q)\), in case of a one-particle excitation, or four indices, \(s=(p,q;r,s)\), in case of a two-particle excitation. Accordingly, a configuration is now specified via the initial state, \(|\onvv\rangle\), and \(K-1\) kinks, \(s_1,\ldots,s_K\), together with their respective times, \(t_1,\ldots,t_K\)
\begin{equation}
    C \coloneqq \Set{K,\onvv, t_1,\ldots,t_K, s_1,\ldots,s_K}\,,
    \label{eq:path2}
\end{equation}
and the partition function~(\ref{eq:partition_perturbation_identity_inserted}) may be written as
\begin{widetext}
\begin{equation}
    \label{eq:partition_perturbation_compact_kinks}
    Z(\beta)
    =
    \sum_
{\substack{K=0 \\ K \neq 1}}^{\infty}
    \sum\limits_{\onvv}
    \sum\limits_{s_1} \ldots \sum\limits_{s_K}
    \int\limits_0^{\beta} \dx t_1 
    \int\limits_{t_1}^{\beta} \dx t_2 \ldots 
    \int\limits_{t_{K-1}}^{\beta} \dx t_K
    \;
    (-1)^K
    \left(
    \prod\limits_{i=0}^{K} 
    \euler^{
    -
    \Di{i}(t_{i+1} - t_i)
    }
    \right)
    \times
    \left(
    \prod\limits_{i=1}^{K} 
    \qi{i}(s_i)
    \right)\,,
\end{equation}
\end{widetext}
where paths with $K=1$ violate the periodicity and have to be excluded, and 
the \emph{kink matrix elements} \(\qi{i}(s_i)\) represent the off-diagonal matrix elements with respect to the possible choices of 2- or 4-tuples \(s_i\).

Figure~\ref{fig:Kink_path} shows an example of a single path $C$ of $N=3$ particles that initially ($t=0$) occupy the three orbitals $0, 1, 4$, whereas all other orbitals are empty, corresponding to the state $|\{n\}\rangle = |\{n^{(0)}\}\rangle$. Due to the periodicity this initial state coincides with the final state at $t=\beta$, and the path is interrupted by five kinks at times $t_1 \dots t_5$. These include one-particle kinks (involving two orbitals), at times $t_2, t_4, t_5$ and two two-particle kinks (involving four orbitals), at $t_1$ and $t_3$. The orbitals involved in each of the kinks are visualized in the figure. The probability of this particular path is given by Eq.~(\ref{eq:weight-c}) and sensitively depends on the interaction matrix elements that involve the orbitals associated with each of the kinks.

Using this kink-based path integral representation of the partition function  macroscopic observables follow from standard thermodynamic relations. For example, for the mean total energy we obtain \cite{schoof_cpp11}
\begin{widetext}
\begin{eqnarray}
\langle\hat{H}\rangle= - \frac{\partial}{\partial \beta} \ln Z 
=
\sum_{\substack{K=0 \\ K \neq 1}}^{\infty} \sum_{\{n\}}
\sum_{s_1\ldots s_{K-1}}\,
\int\limits_{0}^{\beta} d t_1 \int\limits_{t_1}^{\beta} d t_2 \ldots \int\limits_{t_{K-1}}^\beta d t_K 
\biggl( \sum_{i=0}^K D_{\{n^{(i)}\}}\frac{t_{i+1}-t_i}{\beta} -\frac{K}{\beta}\biggr) W(\mathbf{C})\ . \label{eq:CPIMC_energy_estimator}
\end{eqnarray}
\end{widetext}
It is interesting to note that, for a noninteracting system where $\hat Y=0$ (see below), the internal energy is determined completely by the diagonal operator $\hat D$. In contrast, for an interacting many-body system, there appears an additional term: evidently, the interaction energy is directly related to the mean number of kinks $\langle K\rangle$. Similarly, other thermodynamic observables can be computed, for details see Refs.~\cite{cpimc_springer_14,dornheim_physrep_18}.

The present path integral representation  
in Slater determinant space, together with the kink formulation of the paths, is well suited for efficient quantum Monte Carlo simulations (CPIMC) and has been applied to a variety of interacting fermion systems, including homogeneous and inhomogeneous systems, e.g. Refs.~\onlinecite{schoof_cpp15,schoof_prl15,cpimc_springer_14,groth_jcp17}. Particularly extensive applications were developed for the uniform electron gas at finite temperature which is the subject of the remainder of this article. An important advantage of homogenous systems is that, due to momentum conservation, no single-particle excitations (kinks involving only two orbitals) are allowed but only those that involve four orbitals. The application of CPIMC to the warm dense UEG is explained in the next section. 

\subsection{CPIMC approach to the warm dense uniform electron gas. plane wave basis}\label{ss:cpimc}
In case of the UEG, second quantization is naturally performed with respect to plane wave spin orbitals \cite{dornheim_physrep_18}, 
$|i\rangle \to |\mathbf{k}_i\sigma_i\rangle$, with the momentum and spin eigenvalues $\mathbf{k}_i$ and $\sigma_i$, respectively.
In coordinate representation they are written as $\langle \mathbf{r} \sigma \;|\mathbf{k}_i\sigma_i\rangle = \frac{1}{L^{3/2}} e^{i\mathbf{k}_i \cdot \mathbf{r}}\delta_{\sigma,\sigma_i}$ with $\mathbf{k}_i=\frac{2\uppi}{L}\mathbf{m}_i$, $\mathbf{m}_i\in \mathbb{Z}^3$ and $\sigma_i\in\{\uparrow,\downarrow\}$,   so that the UEG Hamiltonian, Eq.~(\ref{eq:h}), becomes 
\begin{eqnarray}\label{eq:UEG_Ham_second}
\hat{H}=
\frac{1}{2}\sum_{i}\mathbf{k}_i^2 \hat a^\dagger_{i}\hat a_{i} + \smashoperator{\sum_{\substack{i<j,k<l \\ i\neq k,j\neq l}}} 
w^-_{ijkl}\hat a^\dagger_{i}\hat a^\dagger_{j} \hat a_{l} \hat a_{k} + N\frac{\xi_\text{M}}{2}\ ,
\end{eqnarray}
where $\xi_{\text{M}}$ is the Madelung constant that is due to the neutralizing background. 
In the hamiltonian (\ref{eq:UEG_Ham_second}), the creation (annihilation) operator $\hat a^\dagger_i$ ($\hat a_i$) creates (annihilates) an electron in the $i$-th spin orbital, and for electrons (fermions) the operators obey the standard anti-commutation relations. Further,  $w^-_{ijkl} =w_{ijkl}-w_{ijlk}$ denotes the anti-symmetrized two-electron (Coulomb) integral with
\begin{eqnarray}\label{eq:two_ints}
w_{ijkl}=\frac{4\pi e^2}{L^3 (\mathbf{k}_{i} - \mathbf{k}_{k})^2}\delta_{\mathbf{k}_i+\mathbf{k}_j, \mathbf{k}_k + \mathbf{k}_l}\delta_{\sigma_i,\sigma_k}\delta_{\sigma_j,\sigma_l}\ ,
\end{eqnarray}
that involves the Fourier transform of the Coulomb potential.

Applying the Slater-Condon rules to the UEG Hamiltonian (\ref{eq:UEG_Ham_second}) we readily compute its matrix elements according to
\begin{align}\label{eq:matrix_elements}
&\braket{\{n\}|\hat{H}|\{\bar{n}\}} =
\\
&\begin{cases}
 \displaystyle D_{\{n\}}=\frac{1}{2}\sum_l \mathbf{k}_l^2 n_{l} + \frac{1}{2}\sum_{l<k}w^-_{lklk}n_{l}n_{k}, &\{n\}=\{\bar{n}\}\ ,\\[0.2cm]
 Y_{\{n\},\{\bar{n}\}} =w_{pqrs}^- (-1)^{\alpha_{\{n\},pq}+\alpha_{\{\bar{n}\},rs}}, & \{n\}=\{\bar{n}\}_{r<s}^{p<q}\ ,
\end{cases}
\nonumber
\end{align}  
where the first line contains the diagonal component, with respect to the Slater determinants, and the second the off-diagonal component. Note that the Coulomb interaction matrix contains a diagonal part (Hartree term) contributing to $D$ and a two-particle excitation (second line) contributing to $Y$ where 
$\ket{\{\bar{n}\}_{r<s}^{p<q}}$ refers to the Slater determinant that is obtained by exciting two electrons from the orbitals $r$ and $s$ to $p$ and $q$ in  $\ket{\{\bar{n}\}}$. Thus, in the path integral picture, this matrix element is related to kinks involving four orbitals. Note that, due to spatial homogeneity, $Y$ does not contain contributions from single-particle excitations (that would involve two orbitals) as was illustrated in Fig.~\ref{fig:Kink_path}.
With the matrix elements $D_{\{n\}}$ and $Y_{\{n\},\{n'\}}$, the kink-based CPIMC algorithm of the previous section can be directly applied to the warm dense UEG.

For completeness, we also provide the definition of the phase factor in $Y$,
\begin{eqnarray}\label{eq:phase_factor}
\alpha_{\{n\},pq} &= \displaystyle\sum_{l=\min(p,q)+1}^{\max(p,q)-1}n_l\ ,
\end{eqnarray}
which causes sign changes in the matrix elements of $Y$ and, thus, is important for evaluating the average sign in the CPIMC simulations. There are two other sources of sign changes in the partition function (\ref{eq:partition_perturbation_compact_kinks}): the first is the factor $(-1)^K$ and the second is the sign of the matrix elements $w^-$. All three factors determine the fermion sign problem in CPIMC and will be investigated below.

\subsection{CPIMC algorithm for the warm dense uniform electron gas. Monte Carlo updates}\label{ss:cpimc-updates}
In the following we outline a Monte Carlo algorithm for the computation of expectation values (\ref{eq:mean-a}) via sampling of paths of type (\ref{eq:path2}).
We use five different types of 
 Monte Carlo updates that will be labeled $A \dots E$ below. These updates were found to be ergodic (within numerical accuracy) and were the basis of extensive numerical results for the UEG, for an overview, see Ref.~\cite{dornheim_physrep_18}.

\subsubsection{Update A: Change of Orbital Occupations} 
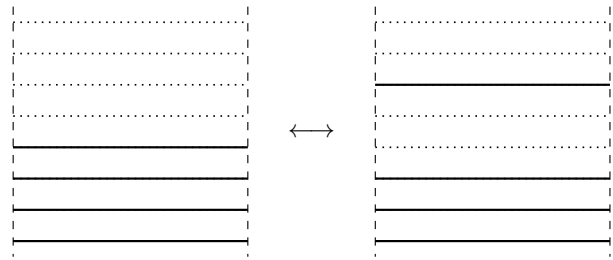
\begin{figure}[h]
\centering
\begin{adjustbox}{width=0.49\textwidth}
\begin{tikzpicture}[baseline=-0.5ex, yscale=0.4]
	\node at (-0.3,0) {}; \node at (3.3,0) {};
	\newcommand{\pp}{3.5}
	\newcommand{\qq}{2.5}
	\newcommand{\ii}{1.5}
	\newcommand{\jj}{0.5}
	\newcommand{\kk}{-0.5}
	\renewcommand{\ll}{-1.5}
	\newcommand{\mm}{-2.5}
	\newcommand{\nn}{-3.5}
	\draw[dashed] (0,4) -- (0,-4) ;
	\draw[dashed] (3,4) -- (3,-4) ;
	\draw[semithick, dotted] (0,\jj) -- (3,\jj);
	\draw[semithick, dotted] (0,\ii) -- (3,\ii);
	\draw[semithick, dotted] (0,\kk) -- (3,\kk);
	\draw[semithick, dotted] (0,\ll) -- (3,\ll);
	\draw[semithick, dotted] (0,\mm) -- (3,\mm);
	\draw[semithick, dotted] (0,\nn) -- (3,\nn);
	\draw[semithick, dotted] (0,\pp) -- (3,\pp);
	\draw[semithick, dotted] (0,\qq) -- (3,\qq);
	\draw[thick] (0,\ll) -- (3,\ll);
	\draw[thick] (0,\kk) -- (3,\kk);
	\draw[thick] (0,\mm) -- (3,\mm);
	\draw[thick] (0,\nn) -- (3,\nn);
\end{tikzpicture}
$\longleftrightarrow$
\begin{tikzpicture}[baseline=-0.5ex, yscale=0.4]
	\node at (-0.3,0) {}; \node at (3.3,0) {};
	\newcommand{\pp}{3.5}
	\newcommand{\qq}{2.5}
	\newcommand{\ii}{1.5}
	\newcommand{\jj}{0.5}
	\newcommand{\kk}{-0.5}
	\renewcommand{\ll}{-1.5}
	\newcommand{\mm}{-2.5}
	\newcommand{\nn}{-3.5}
	\draw[dashed] (0,4) -- (0,-4) ;
	\draw[dashed] (3,4) -- (3,-4) ;
	\draw[semithick, dotted] (0,\jj) -- (3,\jj);
	\draw[semithick, dotted] (0,\ii) -- (3,\ii);
	\draw[semithick, dotted] (0,\kk) -- (3,\kk);
	\draw[semithick, dotted] (0,\ll) -- (3,\ll);
	\draw[semithick, dotted] (0,\mm) -- (3,\mm);
	\draw[semithick, dotted] (0,\nn) -- (3,\nn);
	\draw[semithick, dotted] (0,\pp) -- (3,\pp);
	\draw[semithick, dotted] (0,\qq) -- (3,\qq);
	\draw[thick] (0,\ll) -- (3,\ll);
	\draw[thick] (0,\ii) -- (3,\ii);
	\draw[thick] (0,\mm) -- (3,\mm);
	\draw[thick] (0,\nn) -- (3,\nn);
\end{tikzpicture}
\end{adjustbox}
\caption{Illustration of Update A for four electrons: the electron occupying the highest orbital is excited to an unoccupied orbital above, and vice versa.}
\label{fig:type-a}
\end{figure}

In this step we change the occupation of two orbitals one of which was occupied before the step over the entire imaginary time interval, whereas the other one was unoccupied. Note that, for 
an ideal Fermi gas, the hamiltonian is completely diagonal in the present basis which means that only paths with no kinks are realized. In that case, the update A would be sufficient to achieve ergodicity and exact results. Therefore, CPIMC is able to simulate the ideal Fermi gas without any sign problem including averages and fluctuations of observables. An illustration of update A is shown in Fig.~\ref{fig:type-a}. There a system containing four occupied orbitals is shown, and the uppermost orbital is excited (de-excited). 

\subsubsection{Update B: Add or remove two type 4 kinks (two-particle kinks involving four orbitals)}
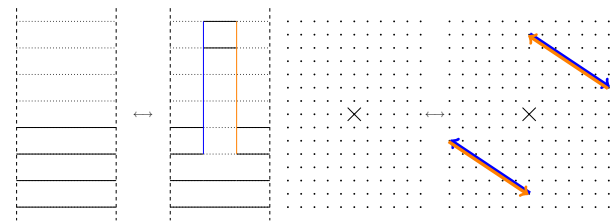
\begin{figure}[!htbp]
\begin{adjustbox}{width=0.45\textwidth}
\begin{tikzpicture}[baseline=-0.5ex, yscale=0.8]
	\node at (-0.3,0) {}; \node at (3.3,0) {};
	\newcommand{\pp}{3.5}
	\newcommand{\qq}{2.5}
	\newcommand{\ii}{1.5}
	\newcommand{\jj}{0.5}
	\newcommand{\kk}{-0.5}
	\renewcommand{\ll}{-1.5}
	\newcommand{\mm}{-2.5}
	\newcommand{\nn}{-3.5}
	\draw[dashed] (0,4) -- (0,-4) ;
	\draw[dashed] (3,4) -- (3,-4) ;
	\draw[semithick, dotted] (0,\jj) -- (3,\jj);
	\draw[semithick, dotted] (0,\ii) -- (3,\ii);
	\draw[semithick, dotted] (0,\kk) -- (3,\kk);
	\draw[semithick, dotted] (0,\ll) -- (3,\ll);
	\draw[semithick, dotted] (0,\mm) -- (3,\mm);
	\draw[semithick, dotted] (0,\nn) -- (3,\nn);
	\draw[semithick, dotted] (0,\pp) -- (3,\pp);
	\draw[semithick, dotted] (0,\qq) -- (3,\qq);
	\draw[thick] (0,\ll) -- (3,\ll);
	\draw[thick] (0,\kk) -- (3,\kk);
	\draw[thick] (0,\mm) -- (3,\mm);
	\draw[thick] (0,\nn) -- (3,\nn);

\end{tikzpicture}
$\longleftrightarrow$
\begin{tikzpicture}[baseline=-0.5ex, yscale=0.8]
	\node at (-0.3,0) {}; \node at (3.3,0) {};
	\newcommand{\pp}{3.5}
	\newcommand{\qq}{2.5}
	\newcommand{\ii}{1.5}
	\newcommand{\jj}{0.5}
	\newcommand{\kk}{-0.5}
	\renewcommand{\ll}{-1.5}
	\newcommand{\mm}{-2.5}
	\newcommand{\nn}{-3.5}
	\draw[dashed] (0,4) -- (0,-4) ;
	\draw[dashed] (3,4) -- (3,-4) ;
	\draw[semithick, dotted] (0,\jj) -- (3,\jj);
	\draw[semithick, dotted] (0,\ii) -- (3,\ii);
	\draw[semithick, dotted] (0,\kk) -- (3,\kk);
	\draw[semithick, dotted] (0,\ll) -- (3,\ll);
	\draw[semithick, dotted] (0,\mm) -- (3,\mm);
	\draw[semithick, dotted] (0,\nn) -- (3,\nn);
	\draw[semithick, dotted] (0,\pp) -- (3,\pp);
	\draw[semithick, dotted] (0,\qq) -- (3,\qq);
	\draw[thick] (0,\ll) -- (1,\ll);
	\draw[thick] (0,\kk) -- (1,\kk);
	\draw[thick] (0,\mm) -- (1,\mm);
	\draw[thick] (0,\nn) -- (1,\nn);
	\draw[thick] (1,\pp) -- (2,\pp);
	\draw[thick] (1,\qq) -- (2,\qq);
	\draw[thick] (1,\mm) -- (2,\mm);
	\draw[thick] (1,\nn) -- (2,\nn);
	\draw[thick] (2,\kk) -- (3,\kk);
	\draw[thick] (2,\ll) -- (3,\ll);
	\draw[thick] (2,\nn) -- (3,\nn);
	\draw[thick] (2,\mm) -- (3,\mm);
	\draw[thick,blue] (1,\ll) -- (1,\pp);
	\draw[thick,orange] (2,\pp) -- (2,\ll);
\end{tikzpicture}

\begin{tikzpicture} [baseline=-0.5ex, scale = 0.8]
    \foreach \x in {-2.5,...,3}
    \foreach \y in {-3.5,...,4}
    {
    \fill (\x,\y) circle (1pt);
    }
    \foreach \x in {-2.5,...,3}
    \foreach \y in {-3,...,3.5}
    {
    \fill (\x,\y) circle (1pt);
    }
    \foreach \x in {-2,...,2.5}
    \foreach \y in {-3.5,...,4}
    {
    \fill (\x,\y) circle (1pt);
    }
    \foreach \x in {-2,...,2.5}
    \foreach \y in {-3,...,3.5}
    {
    \fill (\x,\y) circle (1pt);
    }

    \coordinate (startorb) at (0cm,3cm);
    \coordinate (endorb) at (2cm,2.5cm);
    \coordinate (secondstart) at (-0cm,-3cm);
    \coordinate (secondend) at (-2cm,-2.5cm);

        \draw (-0.25cm,0.25cm) -- (0.25cm,-0.25cm);
        \draw (-0.25cm,-0.25cm) -- (0.25cm,0.25cm);
\end{tikzpicture}
$\longleftrightarrow$
\begin{tikzpicture} [baseline=-0.5ex, scale = 0.8]
    \foreach \x in {-3,...,3.5}
    \foreach \y in {-3.5,...,4}
    {
    \fill (\x,\y) circle (1pt);
    }
    \foreach \x in {-3,...,3.5}
    \foreach \y in {-3,...,3.5}
    {
    \fill (\x,\y) circle (1pt);
    }
    \foreach \x in {-2.5,...,3}
    \foreach \y in {-3.5,...,4}
    {
    \fill (\x,\y) circle (1pt);
    }
    \foreach \x in {-2.5,...,3}
    \foreach \y in {-3,...,3.5}
    {
    \fill (\x,\y) circle (1pt);
    }

    \coordinate (startorb) at (0cm,3cm);
    \coordinate (midorb) at (3cm,1);
    \coordinate (endorb) at (2cm,2.5cm);

    \coordinate (secondstart) at (-0cm,-3cm);
    \coordinate (secondmid) at (-3cm,-1);
    \coordinate (secondend) at (-2cm,-2.5cm);

        \draw (-0.25cm,0.25cm) -- (0.25cm,-0.25cm);
        \draw (-0.25cm,-0.25cm) -- (0.25cm,0.25cm);

        \draw[->,color = blue, line width=1mm, transform canvas={xshift = 0.25mm, yshift=0.25mm}](startorb) -- ($(startorb) !1! (midorb)$);
        \draw[->,color = orange, line width=1mm, , transform canvas={xshift = -0.25mm, yshift=-0.25mm}](midorb) -- ($(midorb) !1! (startorb)$);
        
        \draw[->,color = blue, line width=1mm, transform canvas={xshift = 0.25mm, yshift = 0.25mm}](secondstart) -- ($(secondstart) !1! (secondmid)$);
        \draw[->,color = orange, line width=1mm, transform canvas={xshift = -0.25mm, yshift = -0.25mm}](secondmid) -- ($(secondmid) !1! (secondstart)$);
\end{tikzpicture}
\end{adjustbox}
\caption{Illustration of Update B for four electrons. Two electrons are simultaneously excited to two previously unoccupied (de-excited from two occupied) orbitals corresponding to two type-4 kinks. \textbf{Right:} Arrows show the associated momentum changes of the two electrons in momentum space. The first (second) kink corresponds to the two blue (yellow) arrows. Momentum conservation is obvious from adding the two arrows.}
\label{fig:type-b}

\end{figure}

Here we add or remove a pair of two symmetric kinks that involve four orbitals (two-particle excitation).  As is illustrated in  Fig.~\ref{fig:type-b} the same orbitals that where exited by the first kink are de-excited by the second kink. Due to the symmetry of the interaction operator this update does not lead to a sign change in  the weight function.

\subsubsection{Updates C, D, E: Addition or removal of a kink}
\begin{figure}[!htbp]
\begin{adjustbox}{width=0.45\textwidth}
\begin{tikzpicture}[baseline=-0.5ex, yscale=0.8]
	\node at (-0.3,0) {}; \node at (3.3,0) {};
	\newcommand{\pp}{3.5}
	\newcommand{\qq}{2.5}
	\newcommand{\ii}{1.5}
	\newcommand{\jj}{0.5}
	\newcommand{\kk}{-0.5}
	\renewcommand{\ll}{-1.5}
	\newcommand{\mm}{-2.5}
	\newcommand{\nn}{-3.5}
	\draw[dashed] (0,4) -- (0,-4) ;
	\draw[dashed] (3,4) -- (3,-4) ;
	\draw[semithick, dotted] (0,\jj) -- (3,\jj);
	\draw[semithick, dotted] (0,\ii) -- (3,\ii);
	\draw[semithick, dotted] (0,\kk) -- (3,\kk);
	\draw[semithick, dotted] (0,\ll) -- (3,\ll);
	\draw[semithick, dotted] (0,\mm) -- (3,\mm);
	\draw[semithick, dotted] (0,\nn) -- (3,\nn);
	\draw[semithick, dotted] (0,\pp) -- (3,\pp);
	\draw[semithick, dotted] (0,\qq) -- (3,\qq);
	\draw[thick] (0,\ll) -- (1,\ll);
	\draw[thick] (0,\kk) -- (1,\kk);
	\draw[thick] (0,\mm) -- (1,\mm);
	\draw[thick] (0,\nn) -- (1,\nn);
	\draw[thick] (1,\kk) -- (2,\kk);
	\draw[thick] (1,\ll) -- (2,\ll);
	\draw[thick] (1,\mm) -- (2,\mm);
	\draw[thick] (1,\nn) -- (2,\nn);
	\draw[thick] (2,\ii) -- (3,\ii);
	\draw[thick] (2,\jj) -- (3,\jj);
	\draw[thick] (2,\nn) -- (3,\nn);
	\draw[thick] (2,\mm) -- (3,\mm);
	\draw[thick,red] (2,\ll) -- (2,\ii);

\end{tikzpicture}
$\longleftrightarrow$
\begin{tikzpicture}[baseline=-0.5ex, yscale=0.8]
	\node at (-0.3,0) {}; \node at (3.3,0) {};
	\newcommand{\pp}{3.5}
	\newcommand{\qq}{2.5}
	\newcommand{\ii}{1.5}
	\newcommand{\jj}{0.5}
	\newcommand{\kk}{-0.5}
	\renewcommand{\ll}{-1.5}
	\newcommand{\mm}{-2.5}
	\newcommand{\nn}{-3.5}
	\draw[dashed] (0,4) -- (0,-4) ;
	\draw[dashed] (3,4) -- (3,-4) ;
	\draw[semithick, dotted] (0,\jj) -- (3,\jj);
	\draw[semithick, dotted] (0,\ii) -- (3,\ii);
	\draw[semithick, dotted] (0,\kk) -- (3,\kk);
	\draw[semithick, dotted] (0,\ll) -- (3,\ll);
	\draw[semithick, dotted] (0,\mm) -- (3,\mm);
	\draw[semithick, dotted] (0,\nn) -- (3,\nn);
	\draw[line width=0.6mm, dotted] (0,\pp) -- (3,\pp);
	\draw[line width=0.6mm, dotted] (0,\qq) -- (3,\qq);
	\draw[thick] (0,\ll) -- (1,\ll);
	\draw[thick] (0,\kk) -- (1,\kk);
	\draw[thick] (0,\mm) -- (1,\mm);
	\draw[thick] (0,\nn) -- (1,\nn);
	\draw[line width=0.75mm] (1,\pp) -- (2,\pp);
	\draw[line width=0.75mm] (1,\qq) -- (2,\qq);
	\draw[thick] (1,\mm) -- (2,\mm);
	\draw[thick] (1,\nn) -- (2,\nn);
	\draw[thick] (2,\ii) -- (3,\ii);
	\draw[thick] (2,\jj) -- (3,\jj);
	\draw[thick] (2,\nn) -- (3,\nn);
	\draw[thick] (2,\mm) -- (3,\mm);
	\draw[thick,blue] (1,\ll) -- (1,\pp);
	\draw[thick,orange] (2,\pp) -- (2,\jj);
\end{tikzpicture}

\begin{tikzpicture} [baseline=-0.5ex, scale = 0.8]
    \foreach \x in {-2.5,...,3}
    \foreach \y in {-3.5,...,4}
    {
    \fill (\x,\y) circle (1pt);
    }
    \foreach \x in {-2.5,...,3}
    \foreach \y in {-3,...,3.5}
    {
    \fill (\x,\y) circle (1pt);
    }
    \foreach \x in {-2,...,2.5}
    \foreach \y in {-3.5,...,4}
    {
    \fill (\x,\y) circle (1pt);
    }
    \foreach \x in {-2,...,2.5}
    \foreach \y in {-3,...,3.5}
    {
    \fill (\x,\y) circle (1pt);
    }

    \coordinate (startorb) at (0cm,3cm);
    \coordinate (endorb) at (2cm,2.5cm);
    \coordinate (secondstart) at (-0cm,-3cm);
    \coordinate (secondend) at (-2cm,-2.5cm);

        \draw (-0.25cm,0.25cm) -- (0.25cm,-0.25cm);
        \draw (-0.25cm,-0.25cm) -- (0.25cm,0.25cm);
        \draw[->,color = red, line width=1mm](startorb) -- ($(startorb) !1! (endorb)$);
        \draw[->,color = red, line width=1mm](secondstart) -- ($(secondstart) !1! (secondend)$);
\end{tikzpicture}
$\longleftrightarrow$
\begin{tikzpicture} [baseline=-0.5ex, scale = 0.8]
    \foreach \x in {-3,...,3.5}
    \foreach \y in {-3.5,...,4}
    {
    \fill (\x,\y) circle (1pt);
    }
    \foreach \x in {-3,...,3.5}
    \foreach \y in {-3,...,3.5}
    {
    \fill (\x,\y) circle (1pt);
    }
    \foreach \x in {-2.5,...,3}
    \foreach \y in {-3.5,...,4}
    {
    \fill (\x,\y) circle (1pt);
    }
    \foreach \x in {-2.5,...,3}
    \foreach \y in {-3,...,3.5}
    {
    \fill (\x,\y) circle (1pt);
    }

    \coordinate (startorb) at (0cm,3cm);
    \coordinate (midorb) at (3cm,1);
    \coordinate (endorb) at (2cm,2.5cm);

    \coordinate (secondstart) at (-0cm,-3cm);
    \coordinate (secondmid) at (-3cm,-1);
    \coordinate (secondend) at (-2cm,-2.5cm);

        \draw (-0.25cm,0.25cm) -- (0.25cm,-0.25cm);
        \draw (-0.25cm,-0.25cm) -- (0.25cm,0.25cm);

        \draw[->,color = blue, line width=1mm](startorb) -- ($(startorb) !1! (midorb)$);
        \draw[->,color = orange, line width=1mm](midorb) -- ($(midorb) !1! (endorb)$);
        
        \draw[->,color = blue, line width=1mm](secondstart) -- ($(secondstart) !1! (secondmid)$);
        \draw[->,color = orange, line width=1mm](secondmid) -- ($(secondmid) !1! (secondend)$);
\end{tikzpicture}
\end{adjustbox}
\caption{Illustration of Update C for four electrons. The original kink is shown by the vertical red line and the two red arrows in the right part. Update C adds another kink (blue lines) and, in addition, replaces the previous (red) kink by the yellow one. After Update C the two kinks share two previously unoccupied orbitals (drawn bold). The momentum change associated with the kink (excitation) after update C (blue plus yellow arrow) equals the momentum change before the kink (red arrow). Note that the states at the left and right boundary remain unchanged by Update C.}
\label{fig:update-c}
\end{figure}
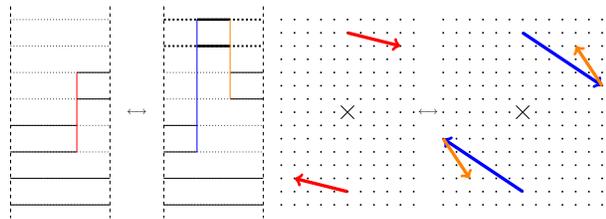

\begin{figure}[!htbp]
\begin{adjustbox}{width=0.45\textwidth}
\begin{tikzpicture}[baseline=-0.5ex, yscale=0.8]
	\node at (-0.3,0) {}; \node at (3.3,0) {};
	\newcommand{\pp}{-2.5}
	\newcommand{\qq}{-3.5}
	\newcommand{\ii}{1.5}
	\newcommand{\jj}{0.5}
	\newcommand{\kk}{-0.5}
	\renewcommand{\ll}{-1.5}
	\newcommand{\mm}{3.5}
	\newcommand{\nn}{2.5}
	\draw[dashed] (0,4) -- (0,-4) ;
	\draw[dashed] (3,4) -- (3,-4) ;
	\draw[semithick, dotted] (0,\jj) -- (3,\jj);
	\draw[semithick, dotted] (0,\ii) -- (3,\ii);
	\draw[semithick, dotted] (0,\kk) -- (3,\kk);
	\draw[semithick, dotted] (0,\ll) -- (3,\ll);
	\draw[semithick, dotted] (0,\mm) -- (3,\mm);
	\draw[semithick, dotted] (0,\nn) -- (3,\nn);
	\draw[semithick, dotted] (0,\pp) -- (3,\pp);
	\draw[semithick, dotted] (0,\qq) -- (3,\qq);
	\draw[thick] (0,\ll) -- (1,\ll);
	\draw[thick] (0,\kk) -- (1,\kk);
	\draw[thick] (0,\mm) -- (1,\mm);
	\draw[thick] (0,\nn) -- (1,\nn);
	\draw[thick] (1,\kk) -- (2,\kk);
	\draw[thick] (1,\ll) -- (2,\ll);
	\draw[thick] (1,\mm) -- (2,\mm);
	\draw[thick] (1,\nn) -- (2,\nn);
	\draw[thick] (2,\ii) -- (3,\ii);
	\draw[thick] (2,\jj) -- (3,\jj);
	\draw[thick] (2,\nn) -- (3,\nn);
	\draw[thick] (2,\mm) -- (3,\mm);
	\draw[thick,red] (2,\ll) -- (2,\ii);

\end{tikzpicture}
$\longleftrightarrow$
\begin{tikzpicture}[baseline=-0.5ex, yscale=0.8]
	\node at (-0.3,0) {}; \node at (3.3,0) {};
	\newcommand{\pp}{-2.5}
	\newcommand{\qq}{-3.5}
	\newcommand{\ii}{1.5}
	\newcommand{\jj}{0.5}
	\newcommand{\kk}{-0.5}
	\renewcommand{\ll}{-1.5}
	\newcommand{\mm}{3.5}
	\newcommand{\nn}{2.5}
	\draw[dashed] (0,4) -- (0,-4) ;
	\draw[dashed] (3,4) -- (3,-4) ;
	\draw[semithick, dotted] (0,\jj) -- (3,\jj);
	\draw[semithick, dotted] (0,\ii) -- (3,\ii);
	\draw[semithick, dotted] (0,\kk) -- (3,\kk);
	\draw[semithick, dotted] (0,\ll) -- (3,\ll);
	\draw[line width=0.6mm, dotted] (0,\mm) -- (3,\mm);
	\draw[line width=0.6mm, dotted] (0,\nn) -- (3,\nn);
	\draw[semithick, dotted] (0,\pp) -- (3,\pp);
	\draw[semithick, dotted] (0,\qq) -- (3,\qq);
	\draw[thick] (0,\ll) -- (1,\ll);
	\draw[thick] (0,\kk) -- (1,\kk);
	\draw[line width=0.75mm] (0,\mm) -- (1,\mm);
	\draw[line width=0.75mm] (0,\nn) -- (1,\nn);
	\draw[thick] (1,\kk) -- (2,\kk);
	\draw[thick] (1,\ll) -- (2,\ll);
	\draw[thick] (1,\ii) -- (2,\ii);
	\draw[thick] (1,\jj) -- (2,\jj);
	\draw[thick] (2,\ii) -- (3,\ii);
	\draw[thick] (2,\jj) -- (3,\jj);
	\draw[line width=0.75mm] (2,\nn) -- (3,\nn);
	\draw[line width=0.75mm] (2,\mm) -- (3,\mm);
	\draw[thick,blue] (1,\mm) -- (1,\jj);
	\draw[thick,orange] (2,\ll) -- (2,\mm);
\end{tikzpicture}

\begin{tikzpicture} [baseline=-0.5ex, scale = 0.8]
    \foreach \x in {-2.5,...,3}
    \foreach \y in {-3.5,...,4}
    {
    \fill (\x,\y) circle (1pt);
    }
    \foreach \x in {-2.5,...,3}
    \foreach \y in {-3,...,3.5}
    {
    \fill (\x,\y) circle (1pt);
    }
    \foreach \x in {-2,...,2.5}
    \foreach \y in {-3.5,...,4}
    {
    \fill (\x,\y) circle (1pt);
    }
    \foreach \x in {-2,...,2.5}
    \foreach \y in {-3,...,3.5}
    {
    \fill (\x,\y) circle (1pt);
    }

    \coordinate (startorb) at (0cm,3cm);
    \coordinate (endorb) at (2cm,2.5cm);
    \coordinate (secondstart) at (-0cm,-3cm);
    \coordinate (secondend) at (-2cm,-2.5cm);

        \draw (-0.25cm,0.25cm) -- (0.25cm,-0.25cm);
        \draw (-0.25cm,-0.25cm) -- (0.25cm,0.25cm);
        \draw[->,color = red, line width=1mm](startorb) -- ($(startorb) !1! (endorb)$);
        \draw[->,color = red, line width=1mm](secondstart) -- ($(secondstart) !1! (secondend)$);
\end{tikzpicture}
$\longleftrightarrow$
\begin{tikzpicture} [baseline=-0.5ex, scale = 0.8]
    \foreach \x in {-3,...,3.5}
    \foreach \y in {-3.5,...,4}
    {
    \fill (\x,\y) circle (1pt);
    }
    \foreach \x in {-3,...,3.5}
    \foreach \y in {-3,...,3.5}
    {
    \fill (\x,\y) circle (1pt);
    }
    \foreach \x in {-2.5,...,3}
    \foreach \y in {-3.5,...,4}
    {
    \fill (\x,\y) circle (1pt);
    }
    \foreach \x in {-2.5,...,3}
    \foreach \y in {-3,...,3.5}
    {
    \fill (\x,\y) circle (1pt);
    }

    \coordinate (startorb) at (0cm,3cm);
    \coordinate (midorb) at (3cm,1);
    \coordinate (endorb) at (2cm,2.5cm);

    \coordinate (secondstart) at (-0cm,-3cm);
    \coordinate (secondmid) at (-3cm,-1);
    \coordinate (secondend) at (-2cm,-2.5cm);

        \draw (-0.25cm,0.25cm) -- (0.25cm,-0.25cm);
        \draw (-0.25cm,-0.25cm) -- (0.25cm,0.25cm);

        \draw[->,color = orange, line width=1mm](startorb) -- ($(startorb) !1! (midorb)$);
        \draw[->,color = blue, line width=1mm](midorb) -- ($(midorb) !1! (endorb)$);
        
        \draw[->,color = orange, line width=1mm](secondstart) -- ($(secondstart) !1! (secondmid)$);
        \draw[->,color = blue, line width=1mm](secondmid) -- ($(secondmid) !1! (secondend)$);
\end{tikzpicture}
\end{adjustbox}
\caption{Illustration of Update D for four electrons. The original kink is shown by the vertical red line and the two red arrows in the right part. Update D adds another kink (blue) and, in addition, replaces the previous (red) kink by the yellow one. In Update D the new (blue) and changed (yellow) kink share two previously occupied orbitals (bold). The momentum change associated with the kink (excitation) after update D (blue plus yellow arrow) equals the momentum change before the kink (red arrow). Note that the states at the left and right boundary remain unchanged by  Update D.}
\label{fig:update-d}
\end{figure}
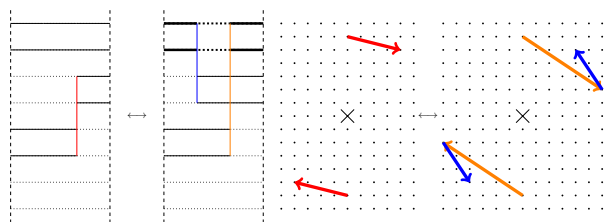

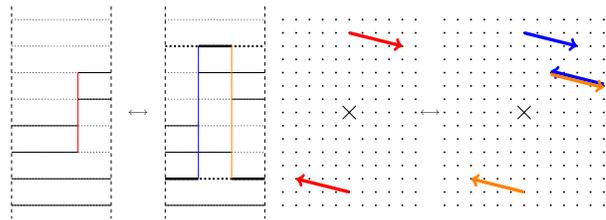
\begin{figure}[!htbp]
\begin{adjustbox}{width=0.45\textwidth}
\begin{tikzpicture}[baseline=-0.5ex, yscale=0.8]
	\node at (-0.3,0) {}; \node at (3.3,0) {};
	\newcommand{\pp}{3.5}
	\newcommand{\qq}{2.5}
	\newcommand{\ii}{1.5}
	\newcommand{\jj}{0.5}
	\newcommand{\kk}{-0.5}
	\renewcommand{\ll}{-1.5}
	\newcommand{\mm}{-2.5}
	\newcommand{\nn}{-3.5}
	\draw[dashed] (0,4) -- (0,-4) ;
	\draw[dashed] (3,4) -- (3,-4) ;
	\draw[semithick, dotted] (0,\jj) -- (3,\jj);
	\draw[semithick, dotted] (0,\ii) -- (3,\ii);
	\draw[semithick, dotted] (0,\kk) -- (3,\kk);
	\draw[semithick, dotted] (0,\ll) -- (3,\ll);
	\draw[semithick, dotted] (0,\mm) -- (3,\mm);
	\draw[semithick, dotted] (0,\nn) -- (3,\nn);
	\draw[semithick, dotted] (0,\pp) -- (3,\pp);
	\draw[semithick, dotted] (0,\qq) -- (3,\qq);
	\draw[thick] (0,\ll) -- (1,\ll);
	\draw[thick] (0,\kk) -- (1,\kk);
	\draw[thick] (0,\mm) -- (1,\mm);
	\draw[thick] (0,\nn) -- (1,\nn);
	\draw[thick] (1,\kk) -- (2,\kk);
	\draw[thick] (1,\ll) -- (2,\ll);
	\draw[thick] (1,\mm) -- (2,\mm);
	\draw[thick] (1,\nn) -- (2,\nn);
	\draw[thick] (2,\ii) -- (3,\ii);
	\draw[thick] (2,\jj) -- (3,\jj);
	\draw[thick] (2,\nn) -- (3,\nn);
	\draw[thick] (2,\mm) -- (3,\mm);
	\draw[thick,red] (2,\ll) -- (2,\ii);

\end{tikzpicture}
$\longleftrightarrow$
\begin{tikzpicture}[baseline=-0.5ex, yscale=0.8]
	\node at (-0.3,0) {}; \node at (3.3,0) {};
	\newcommand{\pp}{3.5}
	\newcommand{\qq}{2.5}
	\newcommand{\ii}{1.5}
	\newcommand{\jj}{0.5}
	\newcommand{\kk}{-0.5}
	\renewcommand{\ll}{-1.5}
	\newcommand{\mm}{-2.5}
	\newcommand{\nn}{-3.5}
	\draw[dashed] (0,4) -- (0,-4) ;
	\draw[dashed] (3,4) -- (3,-4) ;
	\draw[semithick, dotted] (0,\jj) -- (3,\jj);
	\draw[semithick, dotted] (0,\ii) -- (3,\ii);
	\draw[semithick, dotted] (0,\kk) -- (3,\kk);
	\draw[semithick, dotted] (0,\ll) -- (3,\ll);
	\draw[line width=0.6mm, dotted] (0,\mm) -- (3,\mm);
	\draw[semithick, dotted] (0,\nn) -- (3,\nn);
	\draw[semithick, dotted] (0,\pp) -- (3,\pp);
	\draw[line width=0.6mm, dotted] (0,\qq) -- (3,\qq);
	\draw[thick] (0,\ll) -- (1,\ll);
	\draw[thick] (0,\kk) -- (1,\kk);
	\draw[line width=0.75mm] (0,\mm) -- (1,\mm);
	\draw[thick] (0,\nn) -- (1,\nn);
	\draw[thick] (1,\ii) -- (2,\ii);
	\draw[thick] (1,\ll) -- (2,\ll);
	\draw[line width=0.75mm] (1,\qq) -- (2,\qq);
	\draw[thick] (1,\nn) -- (2,\nn);
	\draw[thick] (2,\ii) -- (3,\ii);
	\draw[thick] (2,\jj) -- (3,\jj);
	\draw[thick] (2,\nn) -- (3,\nn);
	\draw[line width=0.75mm] (2,\mm) -- (3,\mm);
	\draw[thick,blue] (1,\mm) -- (1,\qq);
	\draw[thick,orange] (2,\mm) -- (2,\qq);
\end{tikzpicture}

\begin{tikzpicture} [baseline=-0.5ex, scale = 0.8]
    \foreach \x in {-2.5,...,3}
    \foreach \y in {-3.5,...,4}
    {
    \fill (\x,\y) circle (1pt);
    }
    \foreach \x in {-2.5,...,3}
    \foreach \y in {-3,...,3.5}
    {
    \fill (\x,\y) circle (1pt);
    }
    \foreach \x in {-2,...,2.5}
    \foreach \y in {-3.5,...,4}
    {
    \fill (\x,\y) circle (1pt);
    }
    \foreach \x in {-2,...,2.5}
    \foreach \y in {-3,...,3.5}
    {
    \fill (\x,\y) circle (1pt);
    }

    \coordinate (startorb) at (0cm,3cm);
    \coordinate (endorb) at (2cm,2.5cm);
    \coordinate (secondstart) at (-0cm,-3cm);
    \coordinate (secondend) at (-2cm,-2.5cm);

        \draw (-0.25cm,0.25cm) -- (0.25cm,-0.25cm);
        \draw (-0.25cm,-0.25cm) -- (0.25cm,0.25cm);
        \draw[->,color = red, line width=1mm](startorb) -- ($(startorb) !1! (endorb)$);
        \draw[->,color = red, line width=1mm](secondstart) -- ($(secondstart) !1! (secondend)$);
\end{tikzpicture}
$\longleftrightarrow$
\begin{tikzpicture} [baseline=-0.5ex, scale = 0.8]
    \foreach \x in {-3,...,3.5}
    \foreach \y in {-3.5,...,4}
    {
    \fill (\x,\y) circle (1pt);
    }
    \foreach \x in {-3,...,3.5}
    \foreach \y in {-3,...,3.5}
    {
    \fill (\x,\y) circle (1pt);
    }
    \foreach \x in {-2.5,...,3}
    \foreach \y in {-3.5,...,4}
    {
    \fill (\x,\y) circle (1pt);
    }
    \foreach \x in {-2.5,...,3}
    \foreach \y in {-3,...,3.5}
    {
    \fill (\x,\y) circle (1pt);
    }

    \coordinate (startorb) at (0cm,3cm);
    \coordinate (endorb) at (2cm,2.5cm);

    \coordinate (secondstart) at (-0cm,-3cm);
    \coordinate (secondend) at (-2cm,-2.5cm);
    
    \coordinate (firste) at (3cm,1cm);
    \coordinate (seconde) at (1cm,1.5cm);

        \draw (-0.25cm,0.25cm) -- (0.25cm,-0.25cm);
        \draw (-0.25cm,-0.25cm) -- (0.25cm,0.25cm);

        \draw[->,color = blue, line width=1mm](startorb) -- ($(startorb) !1! (endorb)$);
        \draw[->,color = orange, line width=1mm](secondstart) -- ($(secondstart) !1! (secondend)$);
        
        \draw[->,color = blue, line width=1mm, transform canvas={yshift=0.5mm}](firste) -- ($(firste) !1! (seconde)$);
        \draw[->,color = orange, line width=1mm, transform canvas={yshift=-0.5mm}](seconde) -- ($(seconde) !1! (firste)$);
\end{tikzpicture}
\end{adjustbox}
\caption{Illustration of Update E for four electrons. The original kink is shown by the vertical red line and the two red arrows in the right part. Update E adds another kink (blue lines) and, in addition, replaces the previous (red) kink by the yellow one. In Update E the new (blue) and changed (yellow) kink share a previously occupied and a previously unoccupied orbital (bold).
The momentum change associated with the kink (excitation) after update E (blue plus yellow arrow) equals the momentum change before the kink (red arrow). Note that the states at the left and right boundary remain unchanged by  Update E.}
\label{fig:update-e}
\end{figure}

For the CPIMC procedure it is also required to be able to change a path by adding or removing a single kink. As we will see, there exist three different ways to achieve this that will be called Updates C, D and E. They are illustrated in Figs.~\ref{fig:update-c}--\ref{fig:update-e}, respectively.

By executing an update of one of these types we either add or remove a single kink (shown in blue in the figures). Since we are sampling only closed paths, this update cannot alter all orbital occupations to the left and to the right of the imaginary time interval. This is only possible if, in addition to adding (removing) a kink, we also properly change one neighboring kink (indicated by the replacement of the red lines by the yellow lines). 

Note that, to reverse the change of the occupations of the added (removed blue) kink, by only changing one other (yellow) kink, the latter needs to have exactly two orbitals in common with the former kink for the update to be allowed. It turns out that there exist three topologically distinct ways how the created/removed kink and the changed kink are related. Correspondingly, we call such an update to be either of type C, D or E. 

\textbf{Type C Update}: the two orbitals common to both kinks were both unoccupied, before the update. In the imaginary time interval between the two kinks these common orbitals are occupied. This is illustrated in Fig.~\ref{fig:update-c} where the common orbitals are the two uppermost orbitals.

\textbf{Type D Update}: the two orbitals common to both kinks were both occupied before the update. In the imaginary time interval between the two kinks, these orbitals are un-occupied. This is illustrated in Fig.~\ref{fig:update-d}.

\textbf{Type E Update}: of the two orbitals common to both kinks one was occupied and one was unoccupied, before the update. This is illustrated in Fig.~\ref{fig:update-e}, where the lower common (bold) orbital was occupied and the upper common orbital was previously un-occupied.

\subsubsection{Fermion sign Problem associated to \\ updates B, C, D, and E}\label{ss:fsp-updates}
As we noted previously, while removing the sign problem at high degeneracy, CPIMC suffers a sign problem at low density, i.e. with increasing interaction energy (coupling). As we have seen in formula (\ref{eq:CPIMC_energy_estimator}) the interaction energy is directly related to the mean number of kinks $\langle K\rangle$. Indeed, it was confirmed in previous investigations \cite{groth_prb16} that an increase of $\langle K\rangle$ causes  a reduction of the average sign. This is also illustrated in Fig.~\ref{fig:n33TWKS} below. An interesting question is whether this effect applies to each of the four updates (B--E) where kinks are introduced (or removed). The answer is ``no''. The update B involves kinks but does not involve sign changes. This is due to the specific structure of the matrix elements of $\hat Y$. In fact, 
using exclusively updates of type B will create only configurations where every kink is accompanied by a mirror-symmetric second kink.
Since the interaction operator $\hat W^\dagger = \hat W$, its matrix elements are real in a momentum basis, and the phase factors $\alpha$ are invariant under exchange of the start and end orbitals, the contributions of the two kinks are identical.
Any possible sign change will, therefore, be compensated.

In contrast, the three updates, C -- E, involve kinks and sign changes and are, therefore, critical for the fermion sign problem in CPIMC. It is, therefore, of high interest to investigate the relative importance of the updates C, D, and E. This analysis is carried out in the remainder of this article.

\subsection{Restricted CPIMC}\label{ss:rcpimc}
The properties of the different Monte Carlo updates studied in Sec.~\ref{ss:cpimc-updates}, in particular, their effect on the fermion sign problems of CPIMC suggests to consider various modifications of the algorithm. In the following we consider two approximations where we artificially ``turn off'' some of the Monte Carlo updates, thereby restricting the space of configurations and paths. The hope is that this allows to eliminate some of the kinks that strongly affect the sign problem and, we thereby will be able to extend the simulations to parameters not accessible to CPIMC, i.e., in particular, to stronger coupling ($r_s \gtrsim 1$). 
Of course, such restrictions  will introduce a systematic error that is unknown beforehand and has to be tested against exact CPIMC simulations. We will consider two approximations that are explained in the following.

\begin{description}
\item[RCPIMC] We introduce ``Restricted CPIMC'' (RCPIMC)
as an approximation to CPIMC where only Updates A and B (and the opposite updates) are performed. On the other hand, Updates C--E are excluded from the algorithm.
As was explained in Sec.~\ref{ss:fsp-updates}, Update B [cf. Fig.~\ref{fig:type-b}] introduces only two symmetric kinks and, therefore, does not lead to sign changes. The same is true for Update A which does not involve kinks at all, cf. Fig.~\ref{fig:type-a}. Therefore, RCPIMC is not afflicted by a sign problem at all.
\item[RCPIMC+] We will also consider a modified version of ``Restricted CPIMC'' that will be called RCPIMC+ because, in addition to the updates of RCPIMC, two other updates will be allowed: Updates C and D. Thus, RCPIMC+ only neglects Update E.

With this it is clear that 
 RCPIMC+ does involve sign changes. At the same time, the complexity of the occuring configurations is reduced significantly and it is much easier to leave certain configurations which improves the ergodicity. Also, the correlations of subsequent configurations are substantially reduced.
 As we will see, for some range of parameters of the warm dense uniform electron gas the mean number of kinks is significantly reduced, compared to full CPIMC, which reduces the fermion sign problem.
\end{description}

In the next section we perform extensive tests of RCPIMC and RCPIMC+ for the spin-polarized uniform electron gas at finite temperature. We will consider, both, finite systems as well as the thermodynamic limit, over a broad range of temperatures, $\Theta$ and densities, $r_s$.

\section{Numerical results for the ferromagnetic uniform electron gas}\label{s:results}

\subsection{Test for $N=4$ particles}
\label{ss:n=4}
Let us start with a small system containing just 4 electrons. This has the advantage that extensive benchmark data are available to test RCPIMC and RCPIMC+, for a broad range of temperatures and densities.
For $N=4$,
the hamiltonian can be diagonalized and the thermodynamic properties can be computed exactly via thermodynamic configuration interaction (CI) methods. On the other hand, we can perform independent CPIMC simulations that are also exact and agree with CI within their range of availability. For 
CI, the limitations arise from the dimension $N_b$ of the single-particle basis that because the computational effort scales exponentially with $N_b$. On the other hand, CPIMC uses the same single-particle basis but can handle much larger values of  $N_b$~e.g.~Ref.~\onlinecite{cpimc_springer_14}, but here the limit is set by the fermion sign problem which restricts the simulations to parameters corresponding to sufficiently small $r_s$. For a reasonable comparison we set the basis dimension to $N_b=19$ -- a value that is handable in CI, and use the same $N_b$ in CPIMC as well as in the RCPIMC and RCPIMC+ simulations. In addition we perform CPIMC, RCPIMC and RCPIMC+ simulations for a converged (much larger) basis.
\begin{figure}[t]
\includegraphics[width=0.45\textwidth]{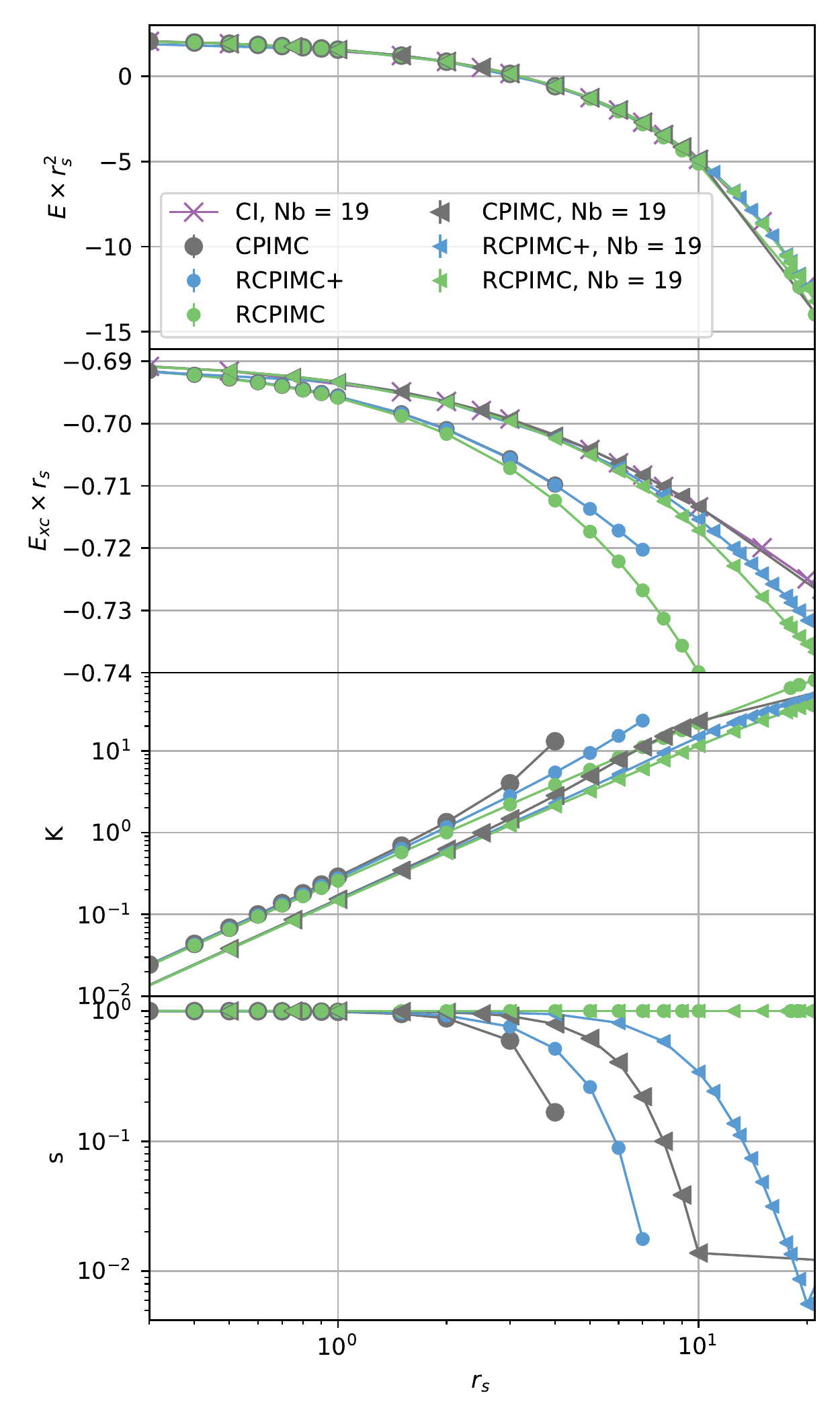}
\caption{Thermodynamic properties for $N = 4$ particles and $\Theta = 0.0625$.  Comparison of CI, CPIMC, RCPIMC and RCPIMC+ simulations. (a): total energy per particle times $r_s^2$, (b) Exchange-correlation energy per particle times $r_s$, (c) average kink number and (d) average sign. Simulations using a single-particle basis of a fixed size, $N_b=19$ are compared to simulations using a converged (much larger) basis.}
\label{fig:n4Nb19}
\end{figure}

The simulation results for a low temperature of \(\Theta=0.0625\) are shown in Fig.~\ref{fig:n4Nb19}. There we plot the total energy (top) and the exchange-correlation energy. There is very good agreement of all simulations for the total energy, up to about $r_s=10$ because this quantity is dominated by the kinetic energy that is common to all methods. Much larger (relative) deviations are observed for the exchange-correlation energy. There, CI and CPIMC are in perfect agreement, but RCPIMC and RCPIMC+ exhibit systematic deviations with increasing $r_s$. Consider first the results for the fixed basis, $N_b=19$. There RCPIMC and RCPIMC+ are accurate up to $r_s \approx 10$ where RCPIMC+ turns out to be more accurate. In this range, the errors in $E_{xc}$  are below $0.5\%$. Even at $r_s=20$ the relative deviations of RCPIMC+  (RCPIMC) are below $1\%$ ($2\%$).

In the two bottom panels we plot the key parameters that characterize the efficiency of the kink-based CPIMC simulations: the average kink number and the average sign. Consider first again the case of a fixed basis size, $N_b=19$ (triangles).
The average sign of the CPIMC simulations (grey trianges) starts to drop exponentially around $r_s=4$. Interestingly the sign for RCPIMC+ also decreases exponentially but at significantly larger values of $r_s$,  and it is about $0.004$, at $r_s=20$. A value of $s=10^{-3}$ is typically the limit for efficient simulations. 
An unexpected observation is that, for larger coupling, $r_s > 20$, the average sign of CPIMC does not show a further decrease (see the last point), and a similar behavior is observed for RCPIMC+ (not shown in the figure). Consequently the agreement of both RCPIMC+ and RCPIMC with the benchmark data improves again. 
The explanation of this behavior is a basis size effect. With increasing $r_s$, particles are pushed into higher orbitals, and these higher excitations are the main cause of the increasing kink number. Cutting the basis at a constant dimension, $N_b=19$, artificially reduces the error of the simulations. The same behavior is observed in the CI simulations and for full CPIMC which are missing essential correlation effects. This explanation is confirmed by the second set of QMC simulations that use a converged basis size that increases with $r_s$. In that case, for CPIMC, the average kink number (grey circles) is significantly higher than for the smaller basis, and the sign decreases already for a smaller value of $r_s\approx 2$. Similar trends are observed for RCPIMC+ where the sign drops around $r_s\approx 3$ and the kink number increases much slower than for CPIMC. On the other hand, as expected, the average sign of RCPIMC is always equal to one even though the kink number (corresponding to Updates B) is increasing monotonically with $r_s$.
\begin{figure}[h]
\includegraphics[width=0.5\textwidth]{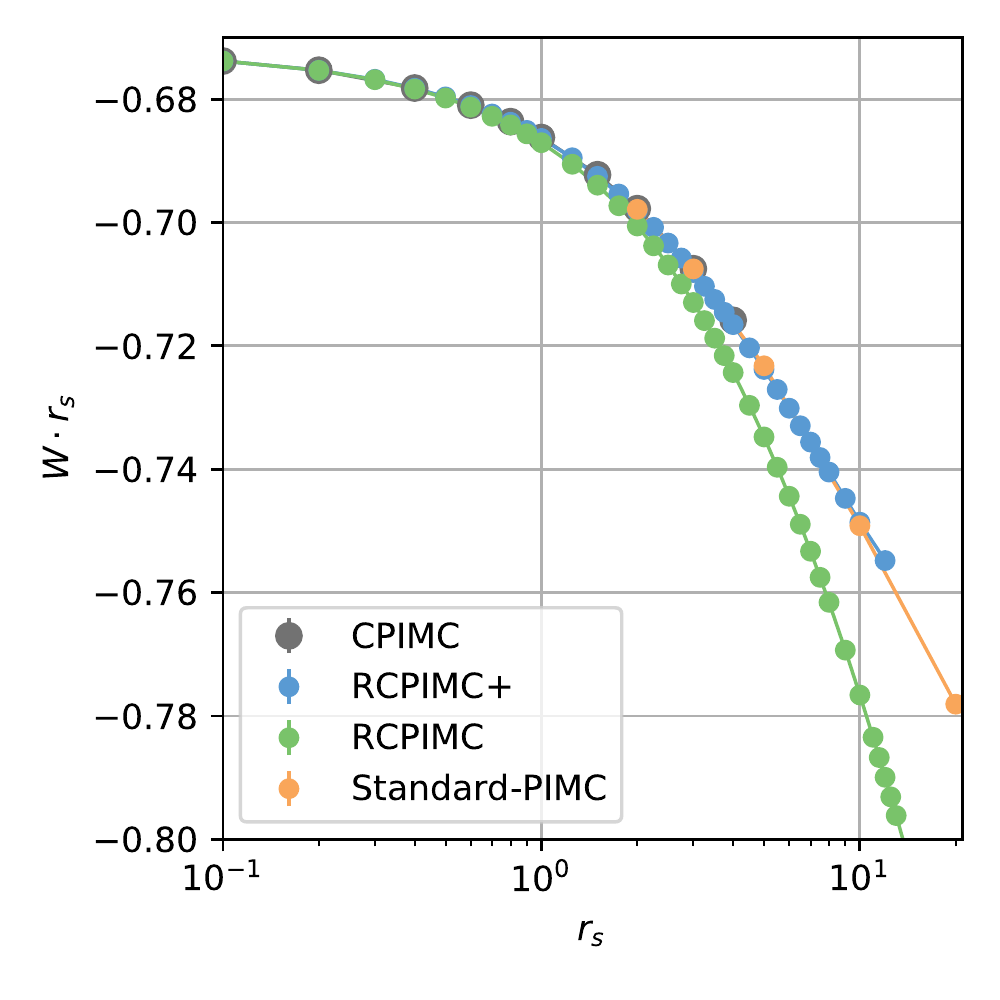}
\caption{Interaction energy per particle times $r_s$ for $N=4$ and $\Theta=0.5$. RCPIMC and RCPIMC+ data are compared to first principle CPIMC (available for $r_s \le 4$) and PB-PIMC benchmarks. }
\label{fig:n4,theta0.5W}
\end{figure}

\begin{figure}[h]
\includegraphics[width=0.5\textwidth]{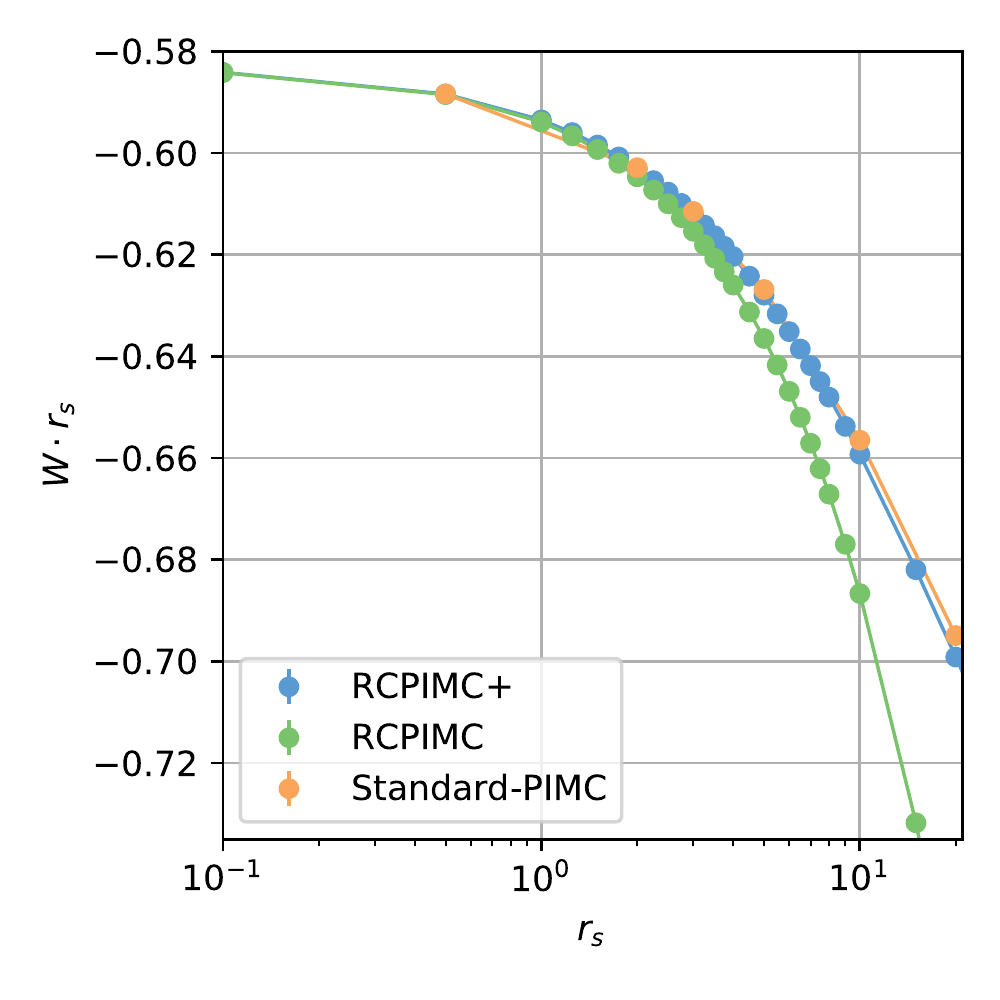}
\caption{Same as Fig.~\ref{fig:n4,theta0.5W}, but for $\Theta=5$. }
\label{fig:n4,theta5W}
\end{figure}

Let us now turn to higher temperatures, $0.5\le \Theta \le 5$. Here, CI simulations are difficult due to the rapid increase of the required basis size. Instead, for benchmark purposes we have generated new CPIMC and PB-PIMC data and always use a converged single particle basis.
In Figs.~\ref{fig:n4,theta0.5W} 
and \ref{fig:n4,theta5W} we show two sets of isotherms for the interaction energy at $N=4$. The comparison with CPIMC and PB-PIMC results reveals very good agreement. Even for $\Theta = 5$ the relative deviations remain below $2.5\%$ up to $r_s = 10$, for RCPIMC. On the other hand, the RCPIMC+ data are almost indistinguishable from the benchmarks in the whole parameter range. While \textit{ab initio} CPIMC simulations are feasible only up to $r_s\approx 4$, in this temperature interval, the reduced sign problem of RCPIMC+ allows us to extend the simulations to about $r_s=15$ ($r_s=20$) for $\Theta = 0.5$ ($\Theta=5$), corresponding to a density reduction by approximately two orders of magnitude.

\subsection{Results for $N=33$}
\begin{figure}[t]
\includegraphics[width=0.5\textwidth]{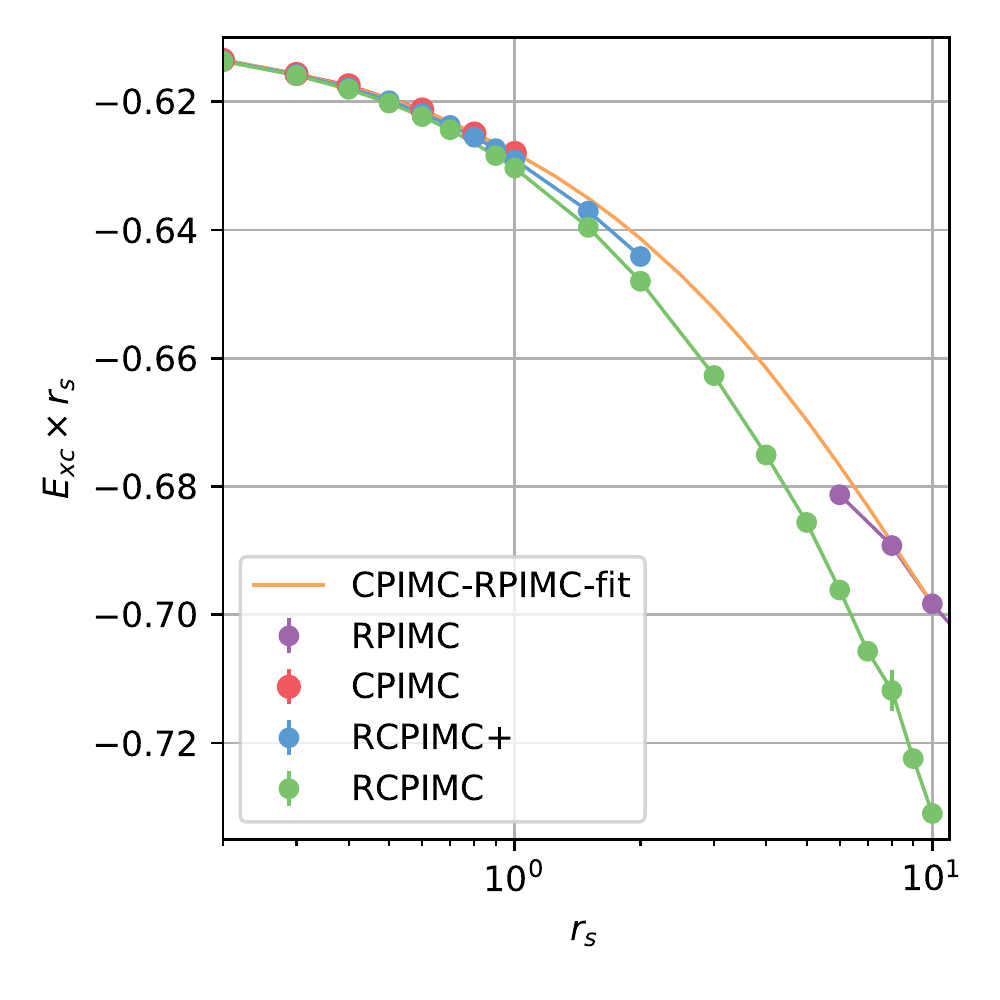}
\caption{Exchange-correlation energy per particle times $r_s$ for $N=33$ and $\Theta=0.0625$. CPIMC results of Ref.~\onlinecite{schoof_prl15} are compared to RPIMC data of Ref.~\onlinecite{Brown_2014} and the present restricted CPIMC approximation. The fit is produced according to 
Eq.~(\ref{eq:groundstate_exc_fit_form}).
\label{fig:n33,theta0.0625}
}
\end{figure}
\begin{figure}[t]
\includegraphics[width=0.5\textwidth]{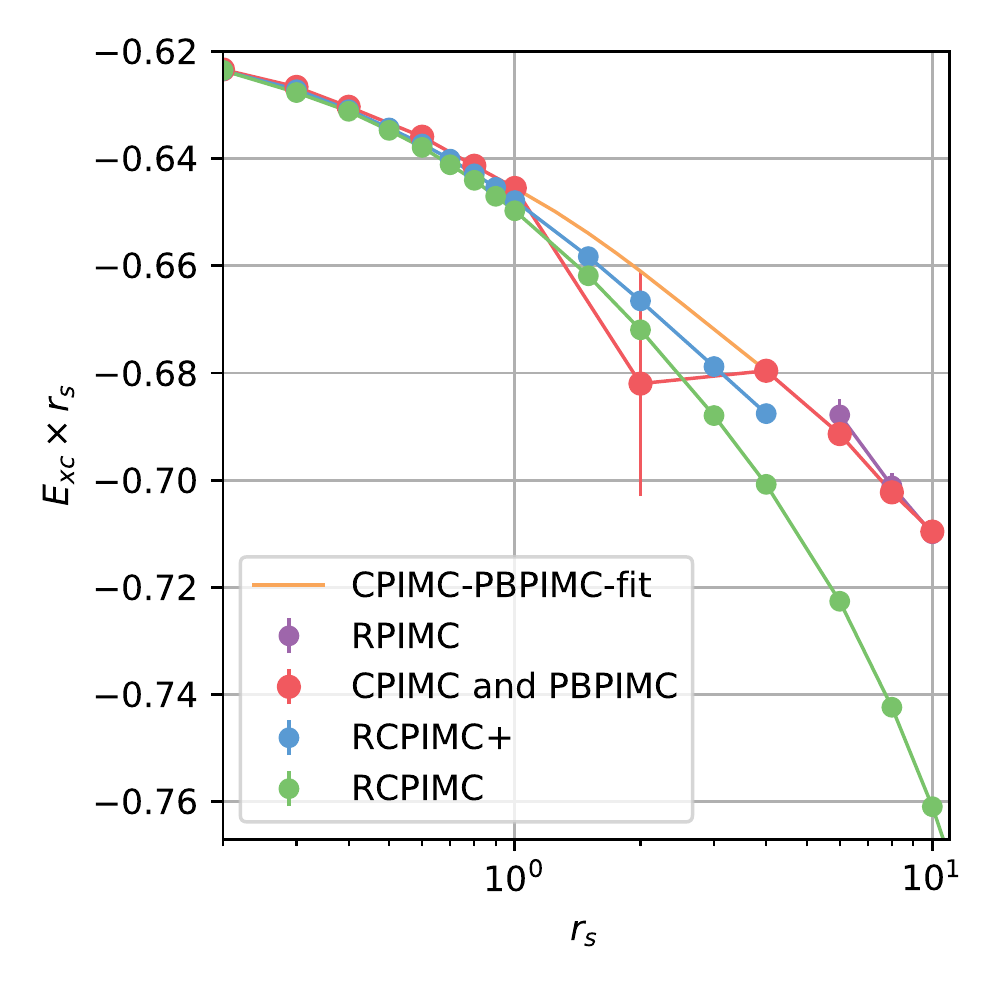}
\caption{Same as Fig.~\ref{fig:n33,theta0.0625}, but for $\Theta=0.5$. For $r_s>1$ the red points are from PB-PIMC simulations, as CPIMC is not feasible, the orange line is an interpolation \cite{groth_prb16}. }
\label{fig:n33,theta0.5}
\end{figure}

We now turn to the second finite system containing $N=33$ electrons. This example of a ``magic number'' (closed shell) cluster was studied in the first simulations of the warm dense UEG that were performed with restricted PIMC (RPIMC) by Brown \textit{et al.} \cite{Brown_2014}, because it is expected that its properties are close to the UEG in the thermodynamic limit. This system was subsequently analyzed by many groups using different methods, including fermionic PIMC \cite{filinov_pre15}, CPIMC \cite{schoof_prl15}, PB-PIMC \cite{dornheim_prb15}, density matrix QMC \cite{Malone_PRL_2016}, and Green functions \cite{Kas_PRL_2017}. Even though RPIMC has no sign problem, the simulations were only possible for $r_s \ge 1$. The first \textit{ab initio} results were obtained by CPIMC \cite{schoof_prl15} and later confirmed by DM-QMC \cite{Malone_PRL_2016} and revealed that the RPIMC data in the range $1\le r_s \le 5$ are surprisingly inaccurate, pointing to a significant nodal error. Improved simulation results for $r_s \gtrsim 1$ became available with the development of PB-PIMC by Dornheim \textit{et al.} \cite{dornheim_njp15}. The combination of CPIMC and PB-PIMC, finally allowed to close the gap in densities that can be simulated \cite{dornheim_prb16,groth_prb16}, however, only for temperatures exceeding $\Theta=0.5$. The missing simulation data for temperatures below this value is one of the main motivations for the development of RCPIMC and RCPIMC+.

%

In Figs.~\ref{fig:n33,theta0.0625} and \ref{fig:n33,theta0.5} we present our new RCPIMC and RCPIMC+ data for the exchange-correlation energy of the polarized UEG with $N=33$ particles, for temperatures in the range of $\Theta = 0.0625 \dots 0.5$ and compare to the available reference results.
Let us start with the lowest temperature, $\Theta=0.0625$, Fig.~\ref{fig:n33,theta0.0625}. Here, CPIMC data are available from simulations with a kink potential up to $r_s=1$ \cite{schoof_prl15}. On the low density side, at these temperatures, no \textit{ab initio} data are available. There exist RPIMC data that are, however, increasingly inaccurate when $r_s$ is lowered, as discussed above. A reliable interpolation is possible by combining all CPIMC data with the RPIMC points for $r_s \ge 10$, see the orange line in Fig.~\ref{fig:n33,theta0.0625}. Here we have used a three-parameter fit that is guided by the known form of the ground state exchange-correlation energy,  
\begin{equation}
\label{eq:groundstate_exc_fit_form}
    f(x) \coloneqq A + B\log(x) x - C x
    .
\end{equation}
Consider first the RCPIMC+ data (blue line). These simulations are very close to the CPIMC results and can be extended up to $r_s=2$ before the sign problem becomes too severe. This is a factor four improvement over CPIMC without an additional kink potential, see Ref.~\onlinecite{dornheim_prb16}. At this point the RCPIMC+ results are still very accurate, with an error on the order of $0.5\%$. The RCPIMC results, on the other hand, are possible for all $r_s$-values. They are systematically too low with an error approximately twice the one of RCPIMC+, so up to $r_s=2$ the error remains within $1\%$.

Next, consider the highest temperature, $\Theta=0.5$. Here, the CPIMC data can be smoothly connected to PB-PIMC simulations \cite{groth_prb16}. So for these and higher temperatures \textit{ab initio} data are available, for the present finite system. At this temperature the behavior of the RCPIMC+ algorithm improves improves further: simulations are possible up to $r_s=4$ where the deviation from the benchmark is again negative and of the order of $1\%$. The behavior of RCPIMC, on the other hand, is similar to the $\Theta=0.0625$ isotherm: the deviations are negative and about twice as large as for RCPIMC+. A $1\%$ deviation is observed around $r_s=2$. For larger $r_s$ values the deviations of RCPIMC continue to increase monotonically. 

\begin{figure}[h]
\includegraphics[width=0.5\textwidth]{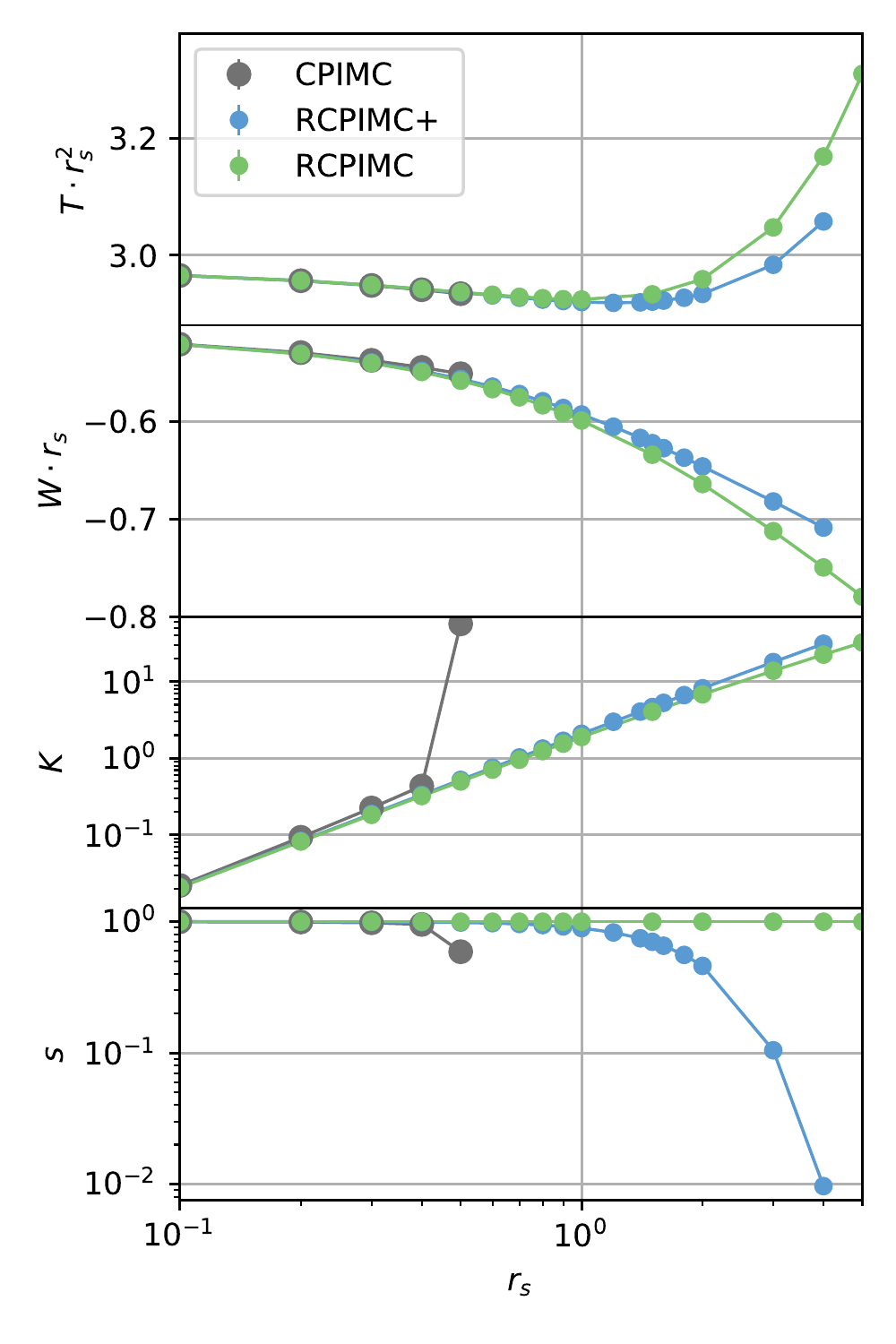}
\caption{Thermodynamic properties of $N = 33$ spin-polarized particles and $\Theta = 0.5$. Comparison of CPIMC, RCPIMC and RCPIMC+ simulations. (a): kinetic energy per particle times $r^2_s$, (b) potential energy per particle times $r_s$, (c) average kink number and (d) average sign. 
}
\label{fig:n33TWKS}
\end{figure}

Finally, we analyze the simulations for the high temperature case in more detail.
In Fig.~\ref{fig:n33TWKS} we present results for the kinetic energy, the potential energy, the average number of kinks and the average sign (top to bottom panels), for $N=33$ and $\Theta=0.5$. First, we observe that both, kinetic and potential energy of CPIMC, are accurately reproduced by RCPIMC+ as well as by RCPIMC, as far as CPIMC data are available. It is interesting to note that the deviations of kinetic energy from the benchmark are of similar magnitude but opposite sign than the interaction energy. This means that the total energy produced by RCPIMC and RCPIMC+ is significantly more accurate than the individual contributions. 

The two bottom panels show that CPIMC experiences a drastic increase of the mean kink number around $r_s=0.4$ and, correspondingly, a sudden decrease of the average sign (here we do not include data with a kink potential that allow to extend CPIMC to $r_s=1$, cf. Ref.~\cite{schoof_prl15}). Let us now consider the behavior of the RCPIMC+ simulations. Here there is no drastic increase of the average kink number but rather a continuous increase of $\langle K\rangle$ with $r_s$. As a consequence, the average sign remains managable up to $r_s\sim 2$.

\subsection{Predictions of RCPIMC and RCPIMC+ for the macroscopic UEG}
\label{ss:tdl}
Our tests against available benchmark data for finite systems with $N=4$ and $N=33$ particles
indicate that RCPIMC+ but also RCPIMC allow to reliably extend simulations to larger couplings ($r_s$) than was possible so far with existing simulations. Unfortunately, no data for larger finite systems or lower temperatures are available that would allow to better quantify the predictions.

On the other hand, there exist accurate thermodynamic data for the thermodynamic limit \cite{groth_prl17,dornheim_physrep_18} that were obtained from a combination of CPIMC and PB-PIMC together with an accurate finite size correction \cite{dornheim_prl16}. Therefore, in the following, we explore the quality of the predictions of RCPIMC and RCPIMC+ for the thermodynamic limit.

\subsubsection{Applying the finite size corrections to the RCPIMC+ data}
\begin{figure}[t]
\includegraphics[width=0.5\textwidth]{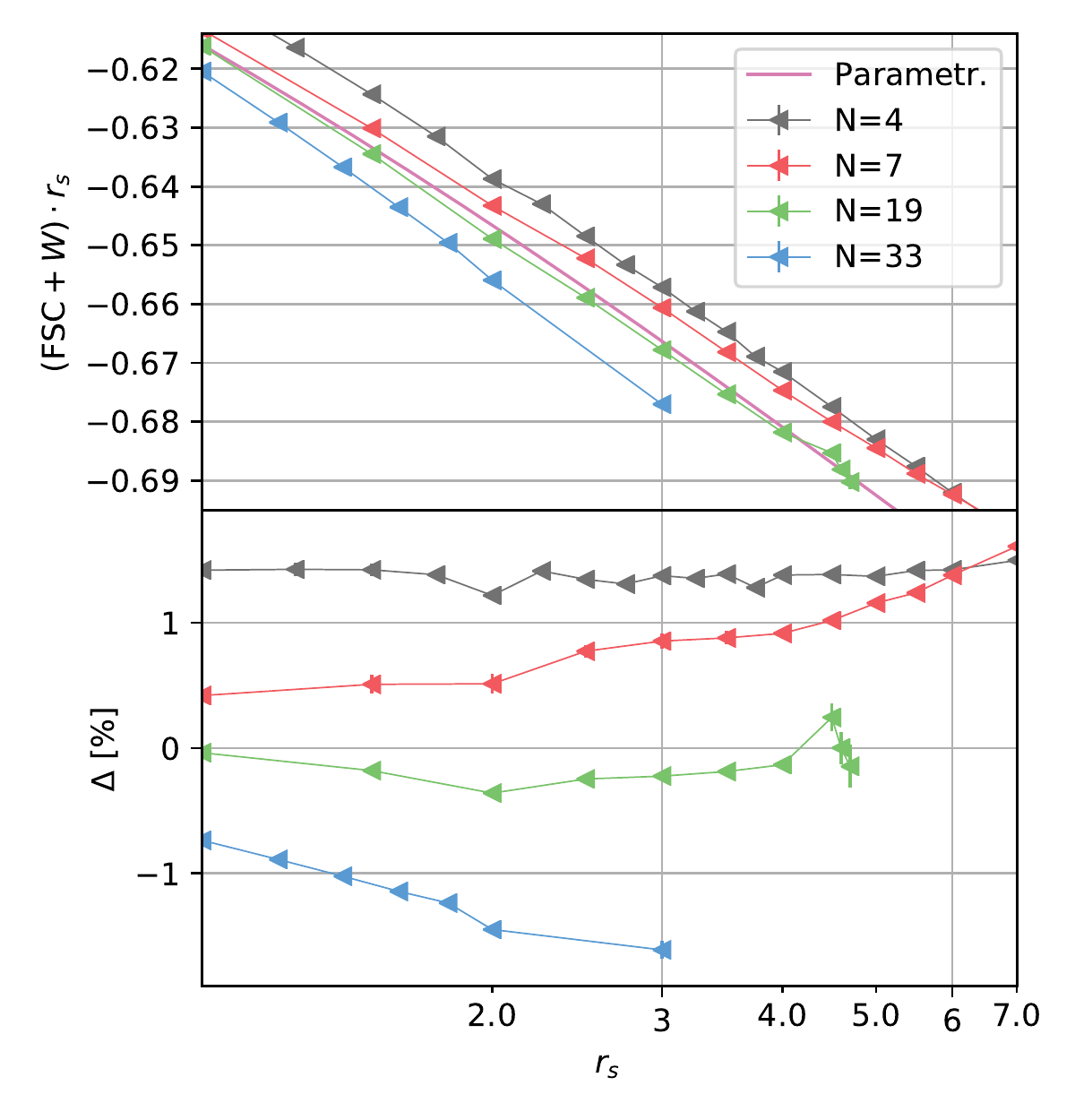}
\caption{Top: Potential energy of the macroscopic UEG for $\Theta=0.2$. On top of RCPIMC+ data for various particle numbers the finite size correction of Ref.~\cite{dornheim_prl16} is added and the result is compared to the analytical parametrization of Ref.~\cite{groth_prl17}. Bottom: relative deviation of the finite size corrected RCPIMC+ data from the parametrization. 
}
\label{fig:TDM_param_theta02}
\end{figure}

In Ref.~\cite{dornheim_prl16} 
a novel finite-size correction (FSC) for the potential energy was derived that accounts for the dominant discretization error in finite $N$ QMC simulations of the warm dense UEG. This FSC was found to have an error not exceeding $0.3\%$ over the relevant $r_s-\Theta$ range. 
It is, therefore, interesting to add this finite size correction to our novel RCPIMC+ data for different values of $N$.

This procedure is illustrated in Fig.~\ref{fig:TDM_param_theta02} for the   RCPIMC+ data at an intermediate temperature ($\Theta=0.2$) and several choices for the particle number, $4\le N\le 33$. As one can see this produces isotherms that all show a similar trend with $r_s$.
With increasing particle number the resulting prediction of the potential energy of the macroscopic system is systematically lowered.
This ambiguity can be overcome by comparing to reference data for the potential energy, i.e. the analytical parametrization (GDSMFB-parametrization) that was derived in Ref.~\cite{groth_prl17}. This result is shown in Fig.~\ref{fig:TDM_param_theta02} as the pink line without symbols. It is interesting to note that the parametrization result falls right into the middle of the finite size corrected curves. The results for $N=4$ and $7$ lie above the parametrization whereas the curve for $N=33$ is significantly below the reference. On the other hand, the result for $N=19$ is very close to the parametrization. This behavior is seen more clearly in the bottom panel where relative deviation of the different finite $N$ results from the parametrization are plotted. Again, we confirm that the $N=19$ curve exhibits the smallest deviations which are almost independent of $r_s$.

The deviations of the RPIMC and RPIMC+ results from the parametrization is the systematic error in the data for finite particle numbers, in the cases $N=19$ and $33$. There is also a small residual error left by the finite size correction that could be removed in Ref.~\cite{dornheim_prl16} but this error is small and of minor importance here. In contrast, for $N=4$ and $N=7$, the simulation results are very accurate, and the deviations are mainly caused by the finite size correction. After this comparison of different particle numbers for a single temperature, we now analyze more extensive data, separately for $N=4$ and $N=33$, in Secs.~\ref{ss:accuracy-rcpimc}, \ref{ss:accuracy+}, and \ref{ss:accuracy-tdl-n19}. 

\subsubsection{Accuracy of the RCPIMC+ results for the \\macroscopic UEG}\label{ss:accuracy+}
\begin{figure}[h]
\includegraphics[width=0.5\textwidth]{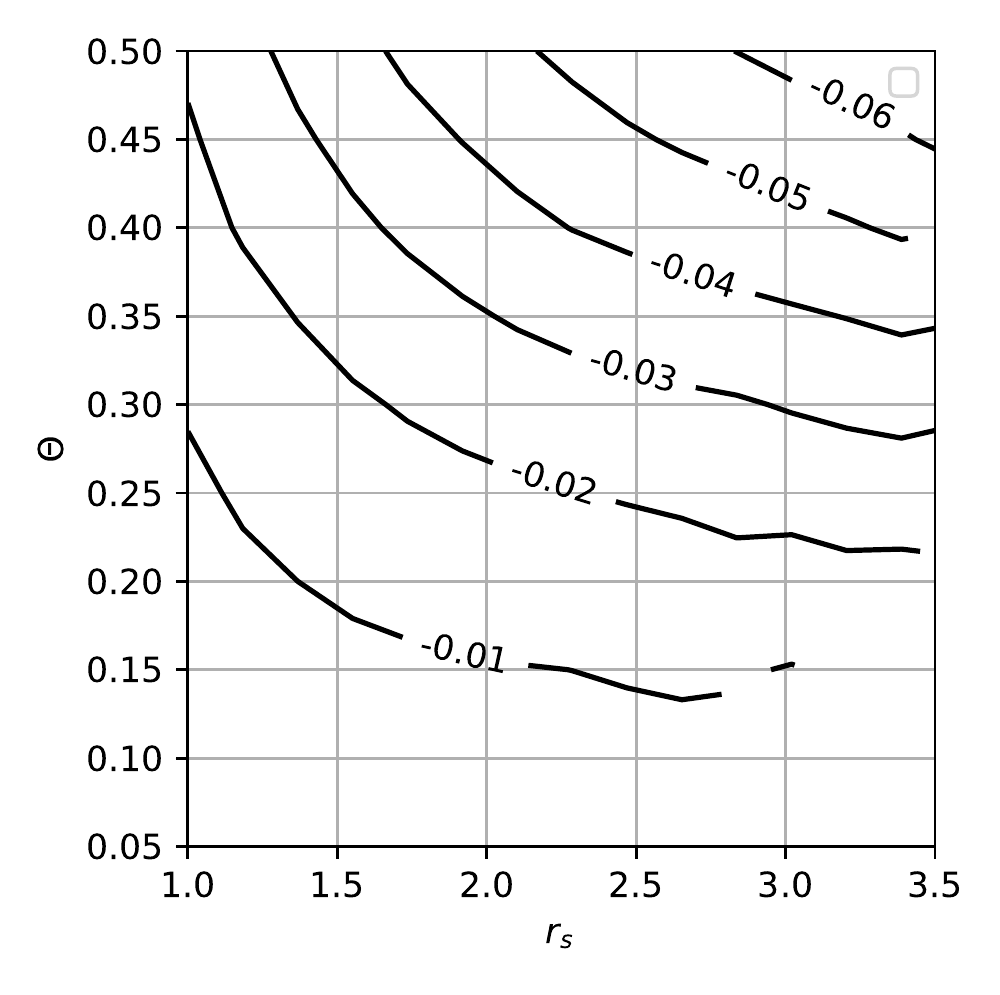}
\caption{Relative deviation of RCPIMC+ data for the potential energy of a ferromagnetic UEG from the analytical parametrization of Ref.~\cite{groth_prl17}. The RCPIMC+ simulations were performed for $N = 33$ particles and  extended to the thermodynamic limit using the the finite size correction of Ref. \cite{dornheim_prl16}. The left border of this plot corresponds to the lowest density ($r_s = 1$) which is accessible with \textit{ab initio} CPIMC employing a kink potential \cite{schoof_prl15}. The top end of this plot shows the lowest temperature ($\Theta = 0.5$) for which reliable PB-PIMC results are feasible.
Only data points with a statistical error below $0.2\%$ were used, the right end of the lines correspond to the points where this condition cannot be satisfied any more.
}
\label{fig:RCPIMC+Space}
\end{figure}
Even though the optimal starting point for RCPIMC+ results for the macroscopic UEG appear to be simulation results for $N=19$, it is interesting to investigate also larger particle numbers to map out their expected errors for different combinations of density and temperature. To this end, we first apply the finite size correction to RCPIMC+ simulations for $N=33$ particles. The results for the relevant parameter range are presented in Fig.~\ref{fig:RCPIMC+Space}. Note that we only consider $r_s$ values exceeding unity and temperatures below $\Theta=0.5$ because here no \textit{ab initio} results are available. 
First, we note that the systematic error increases with temperature.  This means, for the range of low temperatures that we are interested the most, RCPIMC+ is particularly well suited. The statistical error though increases with decreasing temperature. The open end of the lines show, up to which $r_s$ values data can be obtained with a statistical error below $0.2\%$. At $\Theta = 0.05$ we can obtain data up to $r_s = 2.2$.
Further, the error increases with density. Each line of a constant error level eventually terminates at some maximum density which is determined by the fermion sign problem. Based on this figure, we conclude that RCPIMC+ allows for an impressive extension of accurate simulations of the warm dense UEG to lower temperature and stronger coupling. This conclusion will be confirmed by our analysis of the $N=19$ case, in Sec.~\ref{ss:accuracy-tdl-n19}.

\subsubsection{Accuracy of the RCPIMC results for the macroscopic UEG}\label{ss:accuracy-rcpimc}
\begin{figure}[h]
\includegraphics[width=0.5\textwidth]{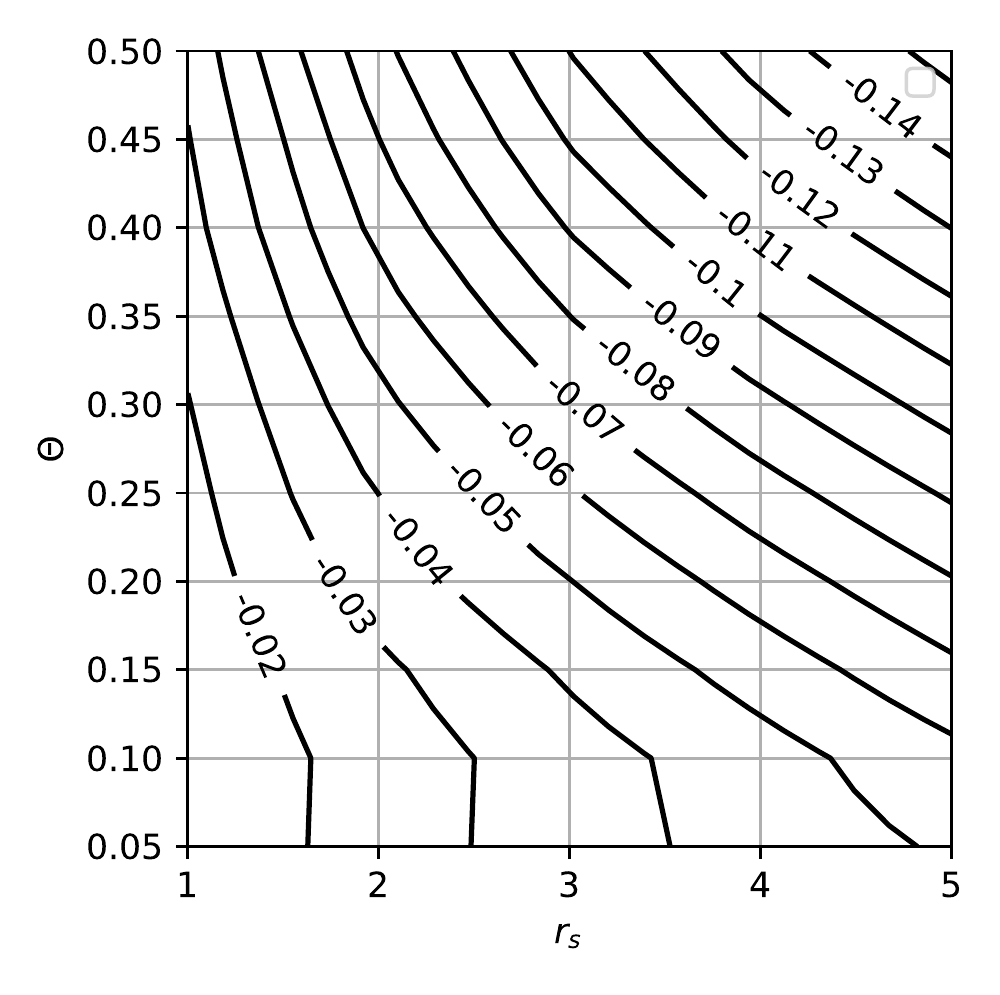}
\caption{Relative deviation of RCPIMC data for the potential energy of a ferromagnetic UEG from the analytical parametrization of Ref.~\cite{groth_prl17}. The RPIMC simulations were performed for $N = 33$ particles and  extended to the thermodynamic limit using the finite size correction of Ref. \cite{dornheim_prl16}. The left border of this plot corresponds to the lowest density ($r_s = 1$) which is accessible with \textit{ab initio} CPIMC employing a kink potential \cite{schoof_prl15}. The top end of this plot shows the lowest temperature ($\Theta = 0.5$) for which reliable PB-PIMC results are feasible.
}
\label{fig:RCPIMCSpace}
\end{figure}
Let us now do the same analysis for RCPIMC simulations of the macroscopic UEG. Again we use a finite simulation size of $N=33$ as the starting point. The results are collected in Fig.~\ref{fig:RCPIMCSpace}. The general trends are the same as observed in Fig.~\ref{fig:RCPIMC+Space}: the accuracy increases when the temperature is lowered, and lines of constant relative error have a similar shape. The main difference is that magnitude of the relative error is a factor $2\dots 3$ larger in case of RCPIMC. On the other hand, since RCPIMC has no sign problem, simulations are possible, in principle, for all parameter combinations (note that the x-axis extends more than twice as far, compared to Fig.~\ref{fig:RCPIMC+Space}) and are only limited by a threshold for the allowed error.

\subsubsection{Thermodynamic RCPIMC and RCPIMC+ results based on data for $N=19$}\label{ss:accuracy-tdl-n19}
\begin{figure}[h]
\includegraphics[width=0.5\textwidth]{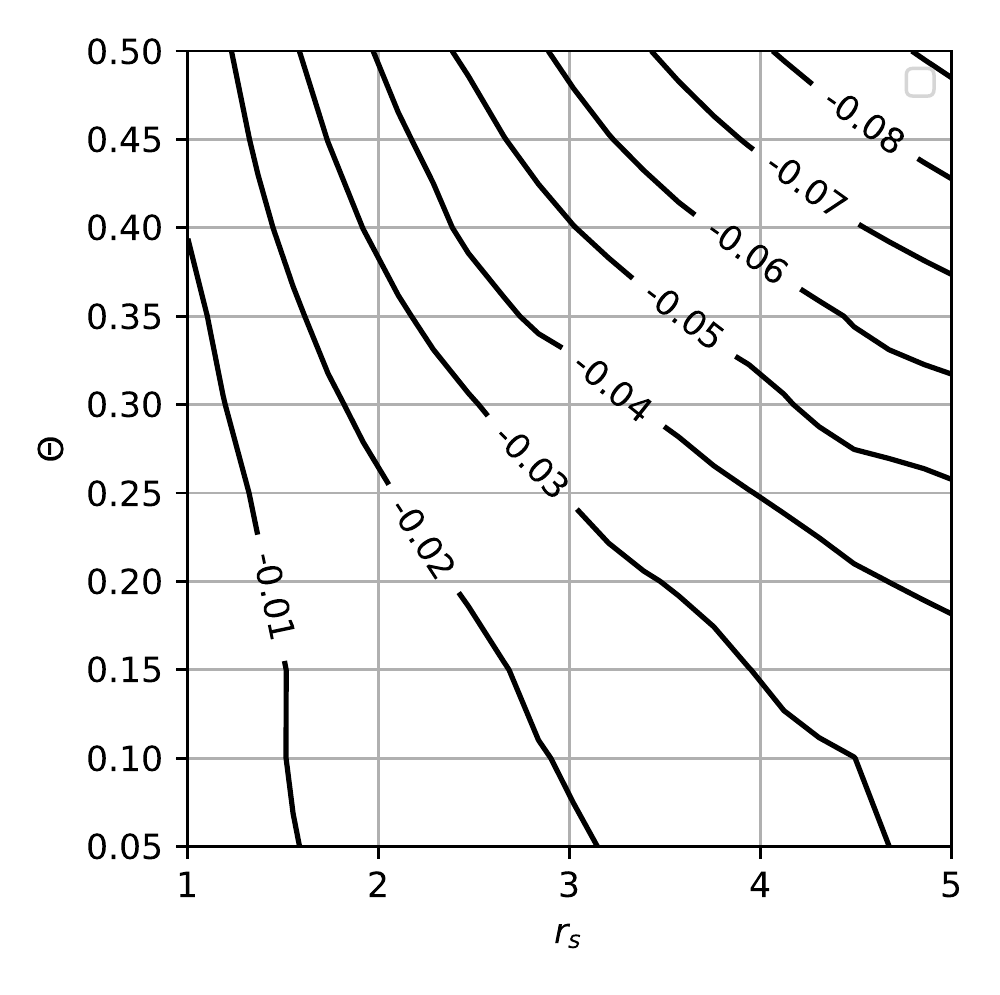}
\caption{Same as Fig.~\ref{fig:RCPIMCSpace}, but using RCPIMC simulations with $N=19$ particles as the starting point.
}
\label{fig:RCPIMCSpace19}
\end{figure}
\begin{figure}[t]
\includegraphics[width=0.5\textwidth]{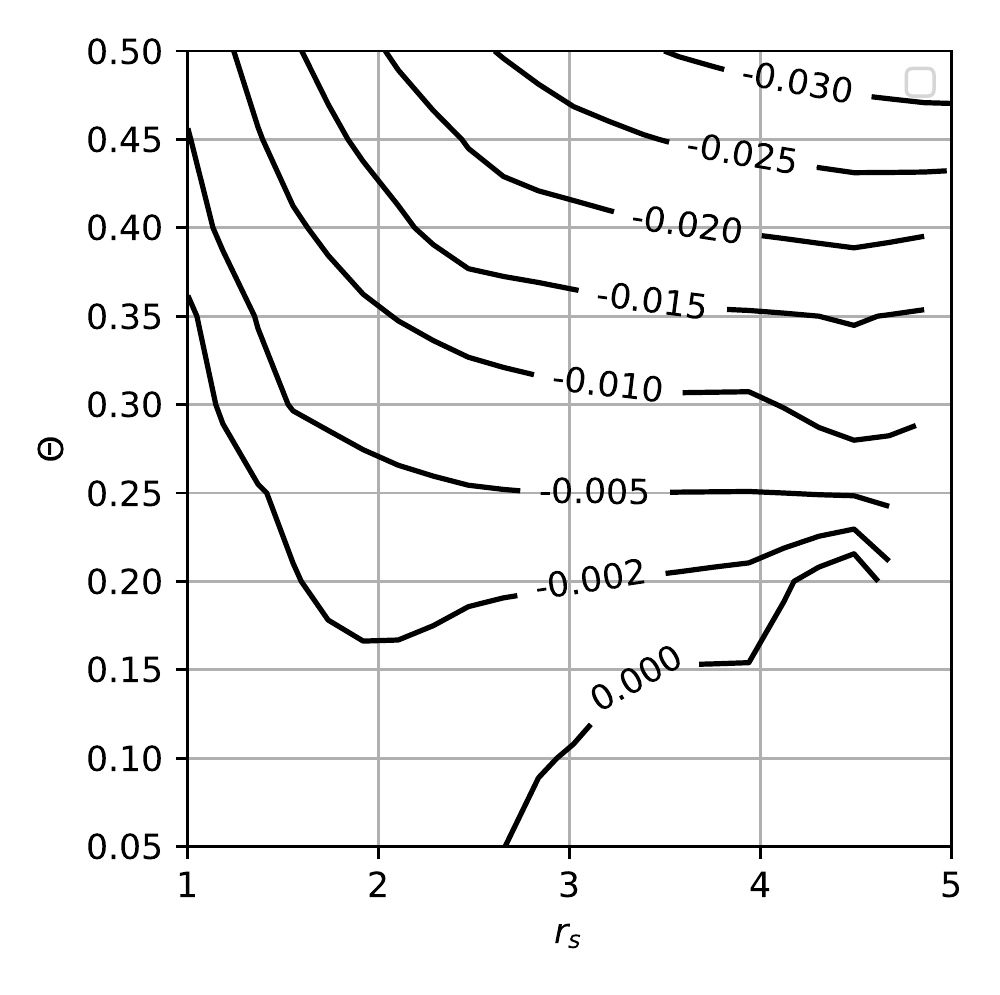}
\caption{Same as Fig.~\ref{fig:RCPIMCSpace19} but using RCPIMC+ data for $N=19$ as the starting point. Only Simulations with a statistical error below $0.02\%$ where used for this Plot. The open end of the lines correspond to the points, where this criteria is no longer met. At $\Theta = 0.05$ simulations are possible up to $r_s = 3$.
}
\label{fig:RCPIMC+Space19}
\end{figure}
As we saw in Fig.~\ref{fig:TDM_param_theta02} the optimum particle number to obtain observables in thermodynamic limit is around $N = 19$. For larger particle numbers the errors of the approximations will increase, while for smaller particle numbers the present finite size correction is most likely not adequate. 
The reason for the surprisingly good performance of the system with $N=19$ is that it constitutes a closed shell cluster for which finite size effects are known to be reduced compared to adjacent particle numbers.

While Fig.~\ref{fig:TDM_param_theta02} compared the different particle numbers only for the temperature $\Theta=0.2$ we can now extend the comparison to the entire interval $0.05 \le \Theta \le 0.5$. Comparing Figs.~\ref{fig:RCPIMCSpace} and \ref{fig:RCPIMCSpace19}, we see that the finite size corrected RCPIMC result for $N=19$ are indeed approximately a factor two more accurate
than those for $N=33$. Also, considering the surprisingly low magnitude of the relative errors, we conclude that approximately one half of the parameter square can be reasonably accurate predicted by RCPIMC simulations, taking e.g. $3\%$ accuracy as the threshold.

Let us now make the same comparison for RCPIMC+. The corresponding results for $N=19$ are presented in Fig.~\ref{fig:RCPIMC+Space19}. Again we observe a dramatic improvement compared to the case of $N=33$ in Fig.~\ref{fig:RCPIMC+Space}. Not only is the accuracy significantly higher, also the range of accessible $r_s$-values is much larger. The reason is that the number of kinks increases with the particle number, and the sign problem gets more severe for the same density and temperature.

\section{Conclusions and outlook} \label{s:conclusion}
In this article we have presented two novel approximations to configuration path integral Monte Carlo that were put forward in Ref.~\onlinecite{groth_ueg_2018} -- two variants of restricted CPIMC (RCPIMC). The motivation was to extend the range where QMC simulations in Fock space are possible because CPIMC is afflicted by a heavy sign problem for coupling parameters $r_s\gtrsim 1$.
The first of the new approximations -- RCPIMC -- neglects three of five classes of Monte Carlo updates and is free of the fermion sign problem. The second approximation takes into account four of five Monte Carlo updates, neglecting only Update E. It has a sign problem which is reduced as compared to full CPIMC. At the same time, RCPIMC+ is systematically more accurate than RCPIMC.

In order to test the accuracy of the two approximations we performed extensive simulations for the ferromagnetic warm dense UEG using various particle numbers, from $N=4$ to $N=33$.
In addition we also tested the predictions of RCPIMC and RCPIMC+ for the thermodynamic limit by applying the accurate finite size correction of Dornheim \textit{et al.} \cite{dornheim_prl16}.
Based on this analysis we conclude that the new approximations are indeed capable to present reliable data for the UEG in a parameter range that, until now was  not accessible to \textit{ab initio} simulations: temperatures below $\Theta=0.5$ and densities corresponding to $r_s\gtrsim 1$, as was indicated in Fig.~\ref{fig:rs-theta-overview}. 
We found that the optimal choice for the computation of macroscopic results is to start with simulations for the closed shell system $N=19$ and then apply the finite size correction. The results indicate that both approximations are particularly valuable at low temperatures, in the range of $0.05\le \Theta \le 0.5$, with the accuracy increasing towards lower temperatures. For the case $N=33$ we observe that the accuracy decreases again below $\Theta=0.1$. Thus, for low temperatures, both approximations provide useful data, up to $r_s \approx 5$, with RCPIMC+ achieving a higher accuracy than RCPIMC with approximately $2\dots 3$ times lower errors in the potential energy, additional results are presented in Ref.~\onlinecite{yilmaz_rcpimc_2020}. 

We found that RCPIMC+ alleviates the sign problem and, whenever simulations are possible, the results are reliable, deviating not more than $1\dots 2\%$ from the exact result for the potential energy. It can be expected that this applies also to interaction contributions of  other thermodynamic functions. Since the kinetic contributions to thermodynamic function are calculated accurately, 
the total quantities such as total internal energy and total pressure are delivered highly accurate by RCPIMC+ within its range of feasibility. We expect that this range can be further extended by applying a kink potential, as was done successfully for CPIMC before \cite{schoof_prl15,groth_prb16}.

RCPIMC, on the other hand, has significantly higher errors where the deviations of the potential energy, compared to RCPIMC+, are always negative. Since RCPIMC has no sign problem it can, in principle, be applied for arbitrary parameter combinations. The only limiting factor is the size $N_b$ of the single-particle basis which increases both with temperature and density (and the particle number). However, this increase is much slower than for CI\footnote{A peculiarity of both approximations is that the accuracy decreases with $N$. Apparantly the relative contribution of the kinks that are neglected increases faster than the particle number.}. The loss of accuracy towards larger coupling or/and temperature is exactly complementary to the behavior of configuration space PIMC
and also restricted PIMC. Therefore, a promising strategy is to combine the two methods to close (at least parts of the) existing gaps at low temperature and intermediate coupling, $1\lesssim r_s\lesssim 5$. 
Straightforward extensions of the present analysis with RCPIMC and RCPIMC+ include the properties of the paramagnetic UEG and the case of arbitrary spin polarization. Here comparison with existing ab \textit{initio data} and with parametrizations \cite{dornheim_physrep_18} can be made and new data points be produced. This can serve as valuable benchmark for other methods, e.g. for finite systems.

Another benefit from RCPIMC is that it gives direct access to other thermodynamic quantities that are related to derivatives of the potential or free energy where the existing parametrizations are not accurate enough to yield smooth results. This includes the heat capacity, the compressibility and equation of state that can be computed directly within RCPIMC. Similarly, it will be interesting to compute pair distributions, the static structure factor, or the momentum distribution of the UEG in the particularly interesting  region of moderate correlations \cite{hunger_20}. The same applies to dynamic quantities such as the dynamic structure factor, and the dynamic local field correction for which recently \textit{ab initio } PIMC results were reported by Dornheim \textit{et al.} \cite{dornheim_prl_18,bonitz_pop_20}. So far, these results were obtained with standard PIMC and, therefore, restricted to $\Theta \gtrsim 1$. The present approximations will allow one to extend the calculations to lower temperatures.

Finally, it will be interesting to extend RCPIMC and RCPIMC+ beyond the uniform electron gas. A particularly promising application would be many-electron atoms where the coupling parameter is right in the range where both approximations were found to perform very well. 
The crucial question to explore is how both approximations perform when spatial homogeneity is broken and one-particle excitations (kinks involving 2 orbitals) have to be included into the set of Monte Carlo updates. This has previously been explored for CPIMC with an external oscillator potential \cite{schoof_cpp11,cpimc_springer_14} as well as for a harmonic perturbation \cite{groth_jcp17}. Solving this problem would allow for an accurate study of finite temperature effects, including ionization potential depression and partial ionization. 

  \section*{Acknowledgments}
   We acknowledge stimulating discussions with Tim Schoof.
   This work has been supported by the Deutsche Forschungsgemeinschaft via grant BO1366/15 and by the Center of Advanced Systems Understanding (CASUS) which is financed by the German Federal Ministry of Education and Research (BMBF) and by the Saxon Ministry for Science, Art, and Tourism (SMWK) with tax funds on the basis of the budget approved by the Saxon State Parliament. We gratefully acknowledge computation time on the HLRN via grants shp00023 and shp00015, and on the Bull Cluster at the Center for Information Services and High Performace Computing (ZIH) at Technische Universit\"at Dresden.


\begin{thebibliography}{83}%
	\makeatletter
	\providecommand \@ifxundefined [1]{%
		\@ifx{#1\undefined}
	}%
	\providecommand \@ifnum [1]{%
		\ifnum #1\expandafter \@firstoftwo
		\else \expandafter \@secondoftwo
		\fi
	}%
	\providecommand \@ifx [1]{%
		\ifx #1\expandafter \@firstoftwo
		\else \expandafter \@secondoftwo
		\fi
	}%
	\providecommand \natexlab [1]{#1}%
	\providecommand \enquote  [1]{``#1''}%
	\providecommand \bibnamefont  [1]{#1}%
	\providecommand \bibfnamefont [1]{#1}%
	\providecommand \citenamefont [1]{#1}%
	\providecommand \href@noop [0]{\@secondoftwo}%
	\providecommand \href [0]{\begingroup \@sanitize@url \@href}%
	\providecommand \@href[1]{\@@startlink{#1}\@@href}%
	\providecommand \@@href[1]{\endgroup#1\@@endlink}%
	\providecommand \@sanitize@url [0]{\catcode `\\12\catcode `\$12\catcode
		`\&12\catcode `\#12\catcode `\^12\catcode `\_12\catcode `\%12\relax}%
	\providecommand \@@startlink[1]{}%
	\providecommand \@@endlink[0]{}%
	\providecommand \url  [0]{\begingroup\@sanitize@url \@url }%
	\providecommand \@url [1]{\endgroup\@href {#1}{\urlprefix }}%
	\providecommand \urlprefix  [0]{URL }%
	\providecommand \Eprint [0]{\href }%
	\providecommand \doibase [0]{http://dx.doi.org/}%
	\providecommand \selectlanguage [0]{\@gobble}%
	\providecommand \bibinfo  [0]{\@secondoftwo}%
	\providecommand \bibfield  [0]{\@secondoftwo}%
	\providecommand \translation [1]{[#1]}%
	\providecommand \BibitemOpen [0]{}%
	\providecommand \bibitemStop [0]{}%
	\providecommand \bibitemNoStop [0]{.\EOS\space}%
	\providecommand \EOS [0]{\spacefactor3000\relax}%
	\providecommand \BibitemShut  [1]{\csname bibitem#1\endcsname}%
	\let\auto@bib@innerbib\@empty
	\bibitem [{\citenamefont {Graziani}\ \emph {et~al.}(2014)\citenamefont
		{Graziani}, \citenamefont {Desjarlais}, \citenamefont {Redmer},\ and\
		\citenamefont {Trickey}}]{graziani-book}%
	\BibitemOpen
	\bibfield  {author} {\bibinfo {author} {\bibfnamefont {F.}~\bibnamefont
			{Graziani}}, \bibinfo {author} {\bibfnamefont {M.~P.}\ \bibnamefont
			{Desjarlais}}, \bibinfo {author} {\bibfnamefont {R.}~\bibnamefont {Redmer}},
		\ and\ \bibinfo {author} {\bibfnamefont {S.~B.}\ \bibnamefont {Trickey}},\
	}\href@noop {} {\emph {\bibinfo {title} {Frontiers and Challenges in Warm
				Dense Matter}}}\ (\bibinfo  {publisher} {Springer},\ \bibinfo {year}
	{2014})\BibitemShut {NoStop}%
	\bibitem [{\citenamefont {Fortov}(2016)}]{Fortov2016}%
	\BibitemOpen
	\bibfield  {author} {\bibinfo {author} {\bibfnamefont {V.~E.}\ \bibnamefont
			{Fortov}},\ }\href@noop {} {\emph {\bibinfo {title} {Extreme States of Matter
				(High Energy Density Physics, Second Edition)}}}\ (\bibinfo  {publisher}
	{Springer, Heidelberg},\ \bibinfo {year} {2016})\BibitemShut {NoStop}%
	\bibitem [{\citenamefont {Moldabekov}\ \emph {et~al.}(2018)\citenamefont
		{Moldabekov}, \citenamefont {Groth}, \citenamefont {Dornheim}, \citenamefont
		{K\"ahlert}, \citenamefont {Bonitz},\ and\ \citenamefont
		{Ramazanov}}]{moldabekov_pre_18}%
	\BibitemOpen
	\bibfield  {author} {\bibinfo {author} {\bibfnamefont {Z.~A.}\ \bibnamefont
			{Moldabekov}}, \bibinfo {author} {\bibfnamefont {S.}~\bibnamefont {Groth}},
		\bibinfo {author} {\bibfnamefont {T.}~\bibnamefont {Dornheim}}, \bibinfo
		{author} {\bibfnamefont {H.}~\bibnamefont {K\"ahlert}}, \bibinfo {author}
		{\bibfnamefont {M.}~\bibnamefont {Bonitz}}, \ and\ \bibinfo {author}
		{\bibfnamefont {T.~S.}\ \bibnamefont {Ramazanov}},\ }\bibfield  {title}
	{\enquote {\bibinfo {title} {Structural characteristics of strongly coupled
				ions in a dense quantum plasma},}\ }\href {\doibase
		10.1103/PhysRevE.98.023207} {\bibfield  {journal} {\bibinfo  {journal} {Phys.
				Rev. E}\ }\textbf {\bibinfo {volume} {98}},\ \bibinfo {pages} {023207}
		(\bibinfo {year} {2018})}\BibitemShut {NoStop}%
	\bibitem [{\citenamefont {Dornheim}, \citenamefont {Groth},\ and\ \citenamefont
		{Bonitz}(2018)}]{dornheim_physrep_18}%
	\BibitemOpen
	\bibfield  {author} {\bibinfo {author} {\bibfnamefont {T.}~\bibnamefont
			{Dornheim}}, \bibinfo {author} {\bibfnamefont {S.}~\bibnamefont {Groth}}, \
		and\ \bibinfo {author} {\bibfnamefont {M.}~\bibnamefont {Bonitz}},\
	}\bibfield  {title} {\enquote {\bibinfo {title} {The uniform electron gas at
				warm dense matter conditions},}\ }\href {\doibase
		10.1016/j.physrep.2018.04.001} {\bibfield  {journal} {\bibinfo  {journal}
			{Phys. Rep.}\ }\textbf {\bibinfo {volume} {744}},\ \bibinfo {pages} {1 -- 86}
		(\bibinfo {year} {2018})}\BibitemShut {NoStop}%
	\bibitem [{\citenamefont {{Saumon}}\ \emph {et~al.}(1992)\citenamefont
		{{Saumon}}, \citenamefont {{Hubbard}}, \citenamefont {{Chabrier}},\ and\
		\citenamefont {{van Horn}}}]{saumon_the_role_1992}%
	\BibitemOpen
	\bibfield  {author} {\bibinfo {author} {\bibfnamefont {D.}~\bibnamefont
			{{Saumon}}}, \bibinfo {author} {\bibfnamefont {W.~B.}\ \bibnamefont
			{{Hubbard}}}, \bibinfo {author} {\bibfnamefont {G.}~\bibnamefont
			{{Chabrier}}}, \ and\ \bibinfo {author} {\bibfnamefont {H.~M.}\ \bibnamefont
			{{van Horn}}},\ }\bibfield  {title} {\enquote {\bibinfo {title} {{The role of
					the molecular-metallic transition of hydrogen in the evolution of Jupiter,
					Saturn, and brown dwarfs}},}\ }\href {\doibase 10.1086/171391} {\bibfield
		{journal} {\bibinfo  {journal} {Astrophys.~J.}\ }\textbf {\bibinfo {volume}
			{391}},\ \bibinfo {pages} {827--831} (\bibinfo {year} {1992})}\BibitemShut
	{NoStop}%
	\bibitem [{\citenamefont {Chabrier}(1993)}]{chabrier_quantum_1993}%
	\BibitemOpen
	\bibfield  {author} {\bibinfo {author} {\bibfnamefont {G.}~\bibnamefont
			{Chabrier}},\ }\bibfield  {title} {\enquote {\bibinfo {title} {Quantum
				effects in dense {Coulumbic} matter - {Application} to the cooling of white
				dwarfs},}\ }\href {\doibase 10.1086/173115} {\bibfield  {journal} {\bibinfo
			{journal} {Astrophys.~J.}\ }\textbf {\bibinfo {volume} {414}},\ \bibinfo
		{pages} {695} (\bibinfo {year} {1993})}\BibitemShut {NoStop}%
	\bibitem [{\citenamefont {Chabrier}\ \emph {et~al.}(2000)\citenamefont
		{Chabrier}, \citenamefont {Brassard}, \citenamefont {Fontaine},\ and\
		\citenamefont {Saumon}}]{chabrier_cooling_2000}%
	\BibitemOpen
	\bibfield  {author} {\bibinfo {author} {\bibfnamefont {G.}~\bibnamefont
			{Chabrier}}, \bibinfo {author} {\bibfnamefont {P.}~\bibnamefont {Brassard}},
		\bibinfo {author} {\bibfnamefont {G.}~\bibnamefont {Fontaine}}, \ and\
		\bibinfo {author} {\bibfnamefont {D.}~\bibnamefont {Saumon}},\ }\bibfield
	{title} {\enquote {\bibinfo {title} {Cooling {Sequences} and
				{Color}-{Magnitude} {Diagrams} for {Cool} {White} {Dwarfs} with {Hydrogen}
				{Atmospheres}},}\ }\href {\doibase 10.1086/317092} {\bibfield  {journal}
		{\bibinfo  {journal} {Astrophys.~J.}\ }\textbf {\bibinfo {volume} {543}},\
		\bibinfo {pages} {216} (\bibinfo {year} {2000})}\BibitemShut {NoStop}%
	\bibitem [{\citenamefont {Schlanges}, \citenamefont {Bonitz},\ and\
		\citenamefont {Tschttschjan}(1995)}]{schlanges_cpp_95}%
	\BibitemOpen
	\bibfield  {author} {\bibinfo {author} {\bibfnamefont {M.}~\bibnamefont
			{Schlanges}}, \bibinfo {author} {\bibfnamefont {M.}~\bibnamefont {Bonitz}}, \
		and\ \bibinfo {author} {\bibfnamefont {A.}~\bibnamefont {Tschttschjan}},\
	}\bibfield  {title} {\enquote {\bibinfo {title} {Plasma phase transition in
				fluid hydrogen-helium mixtures},}\ }\href@noop {} {\bibfield  {journal}
		{\bibinfo  {journal} {Contributions to Plasma Physics}\ }\textbf {\bibinfo
			{volume} {35}},\ \bibinfo {pages} {109--125} (\bibinfo {year}
		{1995})}\BibitemShut {NoStop}%
	\bibitem [{\citenamefont {Bezkrovniy}\ \emph {et~al.}(2004)\citenamefont
		{Bezkrovniy}, \citenamefont {Filinov}, \citenamefont {Kremp}, \citenamefont
		{Bonitz}, \citenamefont {Schlanges}, \citenamefont {Kraeft}, \citenamefont
		{Levashov},\ and\ \citenamefont {Fortov}}]{bezkrovny_pre_4}%
	\BibitemOpen
	\bibfield  {author} {\bibinfo {author} {\bibfnamefont {V.}~\bibnamefont
			{Bezkrovniy}}, \bibinfo {author} {\bibfnamefont {V.~S.}\ \bibnamefont
			{Filinov}}, \bibinfo {author} {\bibfnamefont {D.}~\bibnamefont {Kremp}},
		\bibinfo {author} {\bibfnamefont {M.}~\bibnamefont {Bonitz}}, \bibinfo
		{author} {\bibfnamefont {M.}~\bibnamefont {Schlanges}}, \bibinfo {author}
		{\bibfnamefont {W.~D.}\ \bibnamefont {Kraeft}}, \bibinfo {author}
		{\bibfnamefont {P.~R.}\ \bibnamefont {Levashov}}, \ and\ \bibinfo {author}
		{\bibfnamefont {V.~E.}\ \bibnamefont {Fortov}},\ }\bibfield  {title}
	{\enquote {\bibinfo {title} {{Monte Carlo results for the hydrogen
					Hugoniot}},}\ }\href {\doibase 10.1103/PhysRevE.70.057401} {\bibfield
		{journal} {\bibinfo  {journal} {Phys. Rev. E}\ }\textbf {\bibinfo {volume}
			{70}},\ \bibinfo {pages} {057401} (\bibinfo {year} {2004})}\BibitemShut
	{NoStop}%
	\bibitem [{\citenamefont {Vorberger}\ \emph {et~al.}(2007)\citenamefont
		{Vorberger}, \citenamefont {Tamblyn}, \citenamefont {Militzer},\ and\
		\citenamefont {Bonev}}]{vorberger_hydrogen-helium_2007}%
	\BibitemOpen
	\bibfield  {author} {\bibinfo {author} {\bibfnamefont {J.}~\bibnamefont
			{Vorberger}}, \bibinfo {author} {\bibfnamefont {I.}~\bibnamefont {Tamblyn}},
		\bibinfo {author} {\bibfnamefont {B.}~\bibnamefont {Militzer}}, \ and\
		\bibinfo {author} {\bibfnamefont {S.~A.}\ \bibnamefont {Bonev}},\ }\bibfield
	{title} {\enquote {\bibinfo {title} {Hydrogen-helium mixtures in the
				interiors of giant planets},}\ }\href {\doibase 10.1103/PhysRevB.75.024206}
	{\bibfield  {journal} {\bibinfo  {journal} {Phys.~Rev.~B}\ }\textbf {\bibinfo
			{volume} {75}},\ \bibinfo {pages} {024206} (\bibinfo {year}
		{2007})}\BibitemShut {NoStop}%
	\bibitem [{\citenamefont {Militzer}\ \emph {et~al.}(2008)\citenamefont
		{Militzer}, \citenamefont {Hubbard}, \citenamefont {Vorberger}, \citenamefont
		{Tamblyn},\ and\ \citenamefont {Bonev}}]{militzer_massive_2008}%
	\BibitemOpen
	\bibfield  {author} {\bibinfo {author} {\bibfnamefont {B.}~\bibnamefont
			{Militzer}}, \bibinfo {author} {\bibfnamefont {W.~B.}\ \bibnamefont
			{Hubbard}}, \bibinfo {author} {\bibfnamefont {J.}~\bibnamefont {Vorberger}},
		\bibinfo {author} {\bibfnamefont {I.}~\bibnamefont {Tamblyn}}, \ and\
		\bibinfo {author} {\bibfnamefont {S.~A.}\ \bibnamefont {Bonev}},\ }\bibfield
	{title} {\enquote {\bibinfo {title} {A {Massive} {Core} in {Jupiter}
				{Predicted} from {First}-{Principles} {Simulations}},}\ }\href {\doibase
		10.1086/594364} {\bibfield  {journal} {\bibinfo  {journal}
			{Astrophys.~J.~Lett.}\ }\textbf {\bibinfo {volume} {688}},\ \bibinfo {pages}
		{L45} (\bibinfo {year} {2008})}\BibitemShut {NoStop}%
	\bibitem [{\citenamefont {Redmer}\ \emph {et~al.}(2011)\citenamefont {Redmer},
		\citenamefont {Mattsson}, \citenamefont {Nettelmann},\ and\ \citenamefont
		{French}}]{redmer_icarus_11}%
	\BibitemOpen
	\bibfield  {author} {\bibinfo {author} {\bibfnamefont {R.}~\bibnamefont
			{Redmer}}, \bibinfo {author} {\bibfnamefont {T.~R.}\ \bibnamefont
			{Mattsson}}, \bibinfo {author} {\bibfnamefont {N.}~\bibnamefont
			{Nettelmann}}, \ and\ \bibinfo {author} {\bibfnamefont {M.}~\bibnamefont
			{French}},\ }\bibfield  {title} {\enquote {\bibinfo {title} {The phase
				diagram of water and the magnetic fields of uranus and neptune},}\ }\href
	{\doibase https://doi.org/10.1016/j.icarus.2010.08.008} {\bibfield  {journal}
		{\bibinfo  {journal} {Icarus}\ }\textbf {\bibinfo {volume} {211}},\ \bibinfo
		{pages} {798 -- 803} (\bibinfo {year} {2011})}\BibitemShut {NoStop}%
	\bibitem [{\citenamefont {Nettelmann}, \citenamefont {P\"ustow},\ and\
		\citenamefont {Redmer}(2013)}]{nettelmann_saturn_2013}%
	\BibitemOpen
	\bibfield  {author} {\bibinfo {author} {\bibfnamefont {N.}~\bibnamefont
			{Nettelmann}}, \bibinfo {author} {\bibfnamefont {R.}~\bibnamefont
			{P\"ustow}}, \ and\ \bibinfo {author} {\bibfnamefont {R.}~\bibnamefont
			{Redmer}},\ }\bibfield  {title} {\enquote {\bibinfo {title} {Saturn layered
				structure and homogeneous evolution models with different {EOSs}},}\ }\href
	{\doibase 10.1016/j.icarus.2013.04.018} {\bibfield  {journal} {\bibinfo
			{journal} {Icarus}\ }\textbf {\bibinfo {volume} {225}},\ \bibinfo {pages}
		{548--557} (\bibinfo {year} {2013})}\BibitemShut {NoStop}%
	\bibitem [{\citenamefont {Haensel}, \citenamefont {Potekhin},\ and\
		\citenamefont {Yakovlev}(2006)}]{Haensel}%
	\BibitemOpen
	\bibfield  {author} {\bibinfo {author} {\bibfnamefont {P.}~\bibnamefont
			{Haensel}}, \bibinfo {author} {\bibfnamefont {A.~Y.}\ \bibnamefont
			{Potekhin}}, \ and\ \bibinfo {author} {\bibfnamefont {D.}~\bibnamefont
			{Yakovlev}},\ }\href@noop {} {\emph {\bibinfo {title} {Neutron Stars 1:
				Equation of State and Structure}}}\ (\bibinfo  {publisher} {New York:
		Springer},\ \bibinfo {year} {2006})\BibitemShut {NoStop}%
	\bibitem [{\citenamefont {Daligault}\ and\ \citenamefont
		{Gupta}(2009)}]{daligault_electronion_2009}%
	\BibitemOpen
	\bibfield  {author} {\bibinfo {author} {\bibfnamefont {J.}~\bibnamefont
			{Daligault}}\ and\ \bibinfo {author} {\bibfnamefont {S.}~\bibnamefont
			{Gupta}},\ }\bibfield  {title} {\enquote {\bibinfo {title} {Electron-{Ion}
				{Scattering} in {Dense} {Multi}-{Component} {Plasmas}: {Application} to the
				{Outer} {Crust} of an {Accreting} {Neutron} {Star}},}\ }\href {\doibase
		10.1088/0004-637X/703/1/994} {\bibfield  {journal} {\bibinfo  {journal}
			{Astrophys.~J.}\ }\textbf {\bibinfo {volume} {703}},\ \bibinfo {pages} {994}
		(\bibinfo {year} {2009})}\BibitemShut {NoStop}%
	\bibitem [{\citenamefont {Falk}(2018)}]{falk_2018}%
	\BibitemOpen
	\bibfield  {author} {\bibinfo {author} {\bibfnamefont {K.}~\bibnamefont
			{Falk}},\ }\bibfield  {title} {\enquote {\bibinfo {title} {Experimental
				methods for warm dense matter research},}\ }\href {\doibase
		10.1017/hpl.2018.53} {\bibfield  {journal} {\bibinfo  {journal} {High Power
				Laser Science and Engineering}\ }\textbf {\bibinfo {volume} {6}},\ \bibinfo
		{pages} {e59} (\bibinfo {year} {2018})}\BibitemShut {NoStop}%
	\bibitem [{\citenamefont {Moses}\ \emph {et~al.}(2009)\citenamefont {Moses},
		\citenamefont {Boyd}, \citenamefont {Remington}, \citenamefont {Keane},\ and\
		\citenamefont {Al-Ayat}}]{moses_national_2009}%
	\BibitemOpen
	\bibfield  {author} {\bibinfo {author} {\bibfnamefont {E.~I.}\ \bibnamefont
			{Moses}}, \bibinfo {author} {\bibfnamefont {R.~N.}\ \bibnamefont {Boyd}},
		\bibinfo {author} {\bibfnamefont {B.~A.}\ \bibnamefont {Remington}}, \bibinfo
		{author} {\bibfnamefont {C.~J.}\ \bibnamefont {Keane}}, \ and\ \bibinfo
		{author} {\bibfnamefont {R.}~\bibnamefont {Al-Ayat}},\ }\bibfield  {title}
	{\enquote {\bibinfo {title} {The {National} {Ignition} {Facility}: {Ushering}
				in a new age for high energy density science},}\ }\href {\doibase
		10.1063/1.3116505} {\bibfield  {journal} {\bibinfo  {journal}
			{Phys.~Plasmas}\ }\textbf {\bibinfo {volume} {16}},\ \bibinfo {pages}
		{041006} (\bibinfo {year} {2009})}\BibitemShut {NoStop}%
	\bibitem [{\citenamefont {Hurricane}\ \emph {et~al.}(2016)\citenamefont
		{Hurricane}, \citenamefont {Callahan}, \citenamefont {Casey}, \citenamefont
		{Dewald}, \citenamefont {Dittrich}, \citenamefont {D\"oppner}, \citenamefont
		{Haan}, \citenamefont {Hinkel}, \citenamefont {Berzak~Hopkins}, \citenamefont
		{Jones}, \citenamefont {Kritcher}, \citenamefont {Le~Pape}, \citenamefont
		{Ma}, \citenamefont {MacPhee}, \citenamefont {Milovich}, \citenamefont
		{Moody}, \citenamefont {Pak}, \citenamefont {Park}, \citenamefont {Patel},
		\citenamefont {Ralph}, \citenamefont {Robey}, \citenamefont {Ross},
		\citenamefont {Salmonson}, \citenamefont {Spears}, \citenamefont {Springer},
		\citenamefont {Tommasini}, \citenamefont {Albert}, \citenamefont {Benedetti},
		\citenamefont {Bionta}, \citenamefont {Bond}, \citenamefont {Bradley},
		\citenamefont {Caggiano}, \citenamefont {Celliers}, \citenamefont {Cerjan},
		\citenamefont {Church}, \citenamefont {Dylla-Spears}, \citenamefont {Edgell},
		\citenamefont {Edwards}, \citenamefont {Fittinghoff}, \citenamefont
		{Barrios~Garcia}, \citenamefont {Hamza}, \citenamefont {Hatarik},
		\citenamefont {Herrmann}, \citenamefont {Hohenberger}, \citenamefont
		{Hoover}, \citenamefont {Kline}, \citenamefont {Kyrala}, \citenamefont
		{Kozioziemski}, \citenamefont {Grim}, \citenamefont {Field}, \citenamefont
		{Frenje}, \citenamefont {Izumi}, \citenamefont {Gatu~Johnson}, \citenamefont
		{Khan}, \citenamefont {Knauer}, \citenamefont {Kohut}, \citenamefont
		{Landen}, \citenamefont {Merrill}, \citenamefont {Michel}, \citenamefont
		{Moore}, \citenamefont {Nagel}, \citenamefont {Nikroo}, \citenamefont
		{Parham}, \citenamefont {Rygg}, \citenamefont {Sayre}, \citenamefont
		{Schneider}, \citenamefont {Shaughnessy}, \citenamefont {Strozzi},
		\citenamefont {Town}, \citenamefont {Turnbull}, \citenamefont {Volegov},
		\citenamefont {Wan}, \citenamefont {Widmann}, \citenamefont {Wilde},\ and\
		\citenamefont {Yeamans}}]{hurricane_inertially_2016}%
	\BibitemOpen
	\bibfield  {author} {\bibinfo {author} {\bibfnamefont {O.~A.}\ \bibnamefont
			{Hurricane}}, \bibinfo {author} {\bibfnamefont {D.~A.}\ \bibnamefont
			{Callahan}}, \bibinfo {author} {\bibfnamefont {D.~T.}\ \bibnamefont {Casey}},
		\bibinfo {author} {\bibfnamefont {E.~L.}\ \bibnamefont {Dewald}}, \bibinfo
		{author} {\bibfnamefont {T.~R.}\ \bibnamefont {Dittrich}}, \bibinfo {author}
		{\bibfnamefont {T.}~\bibnamefont {D\"oppner}}, \bibinfo {author}
		{\bibfnamefont {S.}~\bibnamefont {Haan}}, \bibinfo {author} {\bibfnamefont
			{D.~E.}\ \bibnamefont {Hinkel}}, \bibinfo {author} {\bibfnamefont {L.~F.}\
			\bibnamefont {Berzak~Hopkins}}, \bibinfo {author} {\bibfnamefont
			{O.}~\bibnamefont {Jones}}, \bibinfo {author} {\bibfnamefont {A.~L.}\
			\bibnamefont {Kritcher}}, \bibinfo {author} {\bibfnamefont {S.}~\bibnamefont
			{Le~Pape}}, \bibinfo {author} {\bibfnamefont {T.}~\bibnamefont {Ma}},
		\bibinfo {author} {\bibfnamefont {A.~G.}\ \bibnamefont {MacPhee}}, \bibinfo
		{author} {\bibfnamefont {J.~L.}\ \bibnamefont {Milovich}}, \bibinfo {author}
		{\bibfnamefont {J.}~\bibnamefont {Moody}}, \bibinfo {author} {\bibfnamefont
			{A.}~\bibnamefont {Pak}}, \bibinfo {author} {\bibfnamefont {H.-S.}\
			\bibnamefont {Park}}, \bibinfo {author} {\bibfnamefont {P.~K.}\ \bibnamefont
			{Patel}}, \bibinfo {author} {\bibfnamefont {J.~E.}\ \bibnamefont {Ralph}},
		\bibinfo {author} {\bibfnamefont {H.~F.}\ \bibnamefont {Robey}}, \bibinfo
		{author} {\bibfnamefont {J.~S.}\ \bibnamefont {Ross}}, \bibinfo {author}
		{\bibfnamefont {J.~D.}\ \bibnamefont {Salmonson}}, \bibinfo {author}
		{\bibfnamefont {B.~K.}\ \bibnamefont {Spears}}, \bibinfo {author}
		{\bibfnamefont {P.~T.}\ \bibnamefont {Springer}}, \bibinfo {author}
		{\bibfnamefont {R.}~\bibnamefont {Tommasini}}, \bibinfo {author}
		{\bibfnamefont {F.}~\bibnamefont {Albert}}, \bibinfo {author} {\bibfnamefont
			{L.~R.}\ \bibnamefont {Benedetti}}, \bibinfo {author} {\bibfnamefont
			{R.}~\bibnamefont {Bionta}}, \bibinfo {author} {\bibfnamefont
			{E.}~\bibnamefont {Bond}}, \bibinfo {author} {\bibfnamefont {D.~K.}\
			\bibnamefont {Bradley}}, \bibinfo {author} {\bibfnamefont {J.}~\bibnamefont
			{Caggiano}}, \bibinfo {author} {\bibfnamefont {P.~M.}\ \bibnamefont
			{Celliers}}, \bibinfo {author} {\bibfnamefont {C.}~\bibnamefont {Cerjan}},
		\bibinfo {author} {\bibfnamefont {J.~A.}\ \bibnamefont {Church}}, \bibinfo
		{author} {\bibfnamefont {R.}~\bibnamefont {Dylla-Spears}}, \bibinfo {author}
		{\bibfnamefont {D.}~\bibnamefont {Edgell}}, \bibinfo {author} {\bibfnamefont
			{M.~J.}\ \bibnamefont {Edwards}}, \bibinfo {author} {\bibfnamefont
			{D.}~\bibnamefont {Fittinghoff}}, \bibinfo {author} {\bibfnamefont {M.~A.}\
			\bibnamefont {Barrios~Garcia}}, \bibinfo {author} {\bibfnamefont
			{A.}~\bibnamefont {Hamza}}, \bibinfo {author} {\bibfnamefont
			{R.}~\bibnamefont {Hatarik}}, \bibinfo {author} {\bibfnamefont
			{H.}~\bibnamefont {Herrmann}}, \bibinfo {author} {\bibfnamefont
			{M.}~\bibnamefont {Hohenberger}}, \bibinfo {author} {\bibfnamefont
			{D.}~\bibnamefont {Hoover}}, \bibinfo {author} {\bibfnamefont {J.~L.}\
			\bibnamefont {Kline}}, \bibinfo {author} {\bibfnamefont {G.}~\bibnamefont
			{Kyrala}}, \bibinfo {author} {\bibfnamefont {B.}~\bibnamefont
			{Kozioziemski}}, \bibinfo {author} {\bibfnamefont {G.}~\bibnamefont {Grim}},
		\bibinfo {author} {\bibfnamefont {J.~E.}\ \bibnamefont {Field}}, \bibinfo
		{author} {\bibfnamefont {J.}~\bibnamefont {Frenje}}, \bibinfo {author}
		{\bibfnamefont {N.}~\bibnamefont {Izumi}}, \bibinfo {author} {\bibfnamefont
			{M.}~\bibnamefont {Gatu~Johnson}}, \bibinfo {author} {\bibfnamefont {S.~F.}\
			\bibnamefont {Khan}}, \bibinfo {author} {\bibfnamefont {J.}~\bibnamefont
			{Knauer}}, \bibinfo {author} {\bibfnamefont {T.}~\bibnamefont {Kohut}},
		\bibinfo {author} {\bibfnamefont {O.}~\bibnamefont {Landen}}, \bibinfo
		{author} {\bibfnamefont {F.}~\bibnamefont {Merrill}}, \bibinfo {author}
		{\bibfnamefont {P.}~\bibnamefont {Michel}}, \bibinfo {author} {\bibfnamefont
			{A.}~\bibnamefont {Moore}}, \bibinfo {author} {\bibfnamefont {S.~R.}\
			\bibnamefont {Nagel}}, \bibinfo {author} {\bibfnamefont {A.}~\bibnamefont
			{Nikroo}}, \bibinfo {author} {\bibfnamefont {T.}~\bibnamefont {Parham}},
		\bibinfo {author} {\bibfnamefont {R.~R.}\ \bibnamefont {Rygg}}, \bibinfo
		{author} {\bibfnamefont {D.}~\bibnamefont {Sayre}}, \bibinfo {author}
		{\bibfnamefont {M.}~\bibnamefont {Schneider}}, \bibinfo {author}
		{\bibfnamefont {D.}~\bibnamefont {Shaughnessy}}, \bibinfo {author}
		{\bibfnamefont {D.}~\bibnamefont {Strozzi}}, \bibinfo {author} {\bibfnamefont
			{R.~P.~J.}\ \bibnamefont {Town}}, \bibinfo {author} {\bibfnamefont
			{D.}~\bibnamefont {Turnbull}}, \bibinfo {author} {\bibfnamefont
			{P.}~\bibnamefont {Volegov}}, \bibinfo {author} {\bibfnamefont
			{A.}~\bibnamefont {Wan}}, \bibinfo {author} {\bibfnamefont {K.}~\bibnamefont
			{Widmann}}, \bibinfo {author} {\bibfnamefont {C.}~\bibnamefont {Wilde}}, \
		and\ \bibinfo {author} {\bibfnamefont {C.}~\bibnamefont {Yeamans}},\
	}\bibfield  {title} {\enquote {\bibinfo {title} {Inertially confined fusion
				plasmas dominated by alpha-particle self-heating},}\ }\href {\doibase
		10.1038/nphys3720} {\bibfield  {journal} {\bibinfo  {journal} {Nat.~Phys.}\
		}\textbf {\bibinfo {volume} {12}},\ \bibinfo {pages} {800--806} (\bibinfo
		{year} {2016})}\BibitemShut {NoStop}%
	\bibitem [{\citenamefont {Matzen}\ \emph {et~al.}(2005)\citenamefont {Matzen},
		\citenamefont {Sweeney}, \citenamefont {Adams}, \citenamefont {Asay},
		\citenamefont {Bailey}, \citenamefont {Bennett}, \citenamefont {Bliss},
		\citenamefont {Bloomquist}, \citenamefont {Brunner}, \citenamefont
		{Campbell}, \citenamefont {Chandler}, \citenamefont {Coverdale},
		\citenamefont {Cuneo}, \citenamefont {Davis}, \citenamefont {Deeney},
		\citenamefont {Desjarlais}, \citenamefont {Donovan}, \citenamefont {Garasi},
		\citenamefont {Haill}, \citenamefont {Hall}, \citenamefont {Hanson},
		\citenamefont {Hurst}, \citenamefont {Jones}, \citenamefont {Knudson},
		\citenamefont {Leeper}, \citenamefont {Lemke}, \citenamefont {Mazarakis},
		\citenamefont {McDaniel}, \citenamefont {Mehlhorn}, \citenamefont {Nash},
		\citenamefont {Olson}, \citenamefont {Porter}, \citenamefont {Rambo},
		\citenamefont {Rosenthal}, \citenamefont {Rochau}, \citenamefont {Ruggles},
		\citenamefont {Ruiz}, \citenamefont {Sanford}, \citenamefont {Seamen},
		\citenamefont {Sinars}, \citenamefont {Slutz}, \citenamefont {Smith},
		\citenamefont {Struve}, \citenamefont {Stygar}, \citenamefont {Vesey},
		\citenamefont {Weinbrecht}, \citenamefont {Wenger},\ and\ \citenamefont
		{Yu}}]{matzen_pulsed-power-driven_2005}%
	\BibitemOpen
	\bibfield  {author} {\bibinfo {author} {\bibfnamefont {M.~K.}\ \bibnamefont
			{Matzen}}, \bibinfo {author} {\bibfnamefont {M.~A.}\ \bibnamefont {Sweeney}},
		\bibinfo {author} {\bibfnamefont {R.~G.}\ \bibnamefont {Adams}}, \bibinfo
		{author} {\bibfnamefont {J.~R.}\ \bibnamefont {Asay}}, \bibinfo {author}
		{\bibfnamefont {J.~E.}\ \bibnamefont {Bailey}}, \bibinfo {author}
		{\bibfnamefont {G.~R.}\ \bibnamefont {Bennett}}, \bibinfo {author}
		{\bibfnamefont {D.~E.}\ \bibnamefont {Bliss}}, \bibinfo {author}
		{\bibfnamefont {D.~D.}\ \bibnamefont {Bloomquist}}, \bibinfo {author}
		{\bibfnamefont {T.~A.}\ \bibnamefont {Brunner}}, \bibinfo {author}
		{\bibfnamefont {R.~B.}\ \bibnamefont {Campbell}}, \bibinfo {author}
		{\bibfnamefont {G.~A.}\ \bibnamefont {Chandler}}, \bibinfo {author}
		{\bibfnamefont {C.~A.}\ \bibnamefont {Coverdale}}, \bibinfo {author}
		{\bibfnamefont {M.~E.}\ \bibnamefont {Cuneo}}, \bibinfo {author}
		{\bibfnamefont {J.-P.}\ \bibnamefont {Davis}}, \bibinfo {author}
		{\bibfnamefont {C.}~\bibnamefont {Deeney}}, \bibinfo {author} {\bibfnamefont
			{M.~P.}\ \bibnamefont {Desjarlais}}, \bibinfo {author} {\bibfnamefont
			{G.~L.}\ \bibnamefont {Donovan}}, \bibinfo {author} {\bibfnamefont {C.~J.}\
			\bibnamefont {Garasi}}, \bibinfo {author} {\bibfnamefont {T.~A.}\
			\bibnamefont {Haill}}, \bibinfo {author} {\bibfnamefont {C.~A.}\ \bibnamefont
			{Hall}}, \bibinfo {author} {\bibfnamefont {D.~L.}\ \bibnamefont {Hanson}},
		\bibinfo {author} {\bibfnamefont {M.~J.}\ \bibnamefont {Hurst}}, \bibinfo
		{author} {\bibfnamefont {B.}~\bibnamefont {Jones}}, \bibinfo {author}
		{\bibfnamefont {M.~D.}\ \bibnamefont {Knudson}}, \bibinfo {author}
		{\bibfnamefont {R.~J.}\ \bibnamefont {Leeper}}, \bibinfo {author}
		{\bibfnamefont {R.~W.}\ \bibnamefont {Lemke}}, \bibinfo {author}
		{\bibfnamefont {M.~G.}\ \bibnamefont {Mazarakis}}, \bibinfo {author}
		{\bibfnamefont {D.~H.}\ \bibnamefont {McDaniel}}, \bibinfo {author}
		{\bibfnamefont {T.~A.}\ \bibnamefont {Mehlhorn}}, \bibinfo {author}
		{\bibfnamefont {T.~J.}\ \bibnamefont {Nash}}, \bibinfo {author}
		{\bibfnamefont {C.~L.}\ \bibnamefont {Olson}}, \bibinfo {author}
		{\bibfnamefont {J.~L.}\ \bibnamefont {Porter}}, \bibinfo {author}
		{\bibfnamefont {P.~K.}\ \bibnamefont {Rambo}}, \bibinfo {author}
		{\bibfnamefont {S.~E.}\ \bibnamefont {Rosenthal}}, \bibinfo {author}
		{\bibfnamefont {G.~A.}\ \bibnamefont {Rochau}}, \bibinfo {author}
		{\bibfnamefont {L.~E.}\ \bibnamefont {Ruggles}}, \bibinfo {author}
		{\bibfnamefont {C.~L.}\ \bibnamefont {Ruiz}}, \bibinfo {author}
		{\bibfnamefont {T.~W.~L.}\ \bibnamefont {Sanford}}, \bibinfo {author}
		{\bibfnamefont {J.~F.}\ \bibnamefont {Seamen}}, \bibinfo {author}
		{\bibfnamefont {D.~B.}\ \bibnamefont {Sinars}}, \bibinfo {author}
		{\bibfnamefont {S.~A.}\ \bibnamefont {Slutz}}, \bibinfo {author}
		{\bibfnamefont {I.~C.}\ \bibnamefont {Smith}}, \bibinfo {author}
		{\bibfnamefont {K.~W.}\ \bibnamefont {Struve}}, \bibinfo {author}
		{\bibfnamefont {W.~A.}\ \bibnamefont {Stygar}}, \bibinfo {author}
		{\bibfnamefont {R.~A.}\ \bibnamefont {Vesey}}, \bibinfo {author}
		{\bibfnamefont {E.~A.}\ \bibnamefont {Weinbrecht}}, \bibinfo {author}
		{\bibfnamefont {D.~F.}\ \bibnamefont {Wenger}}, \ and\ \bibinfo {author}
		{\bibfnamefont {E.~P.}\ \bibnamefont {Yu}},\ }\bibfield  {title} {\enquote
		{\bibinfo {title} {Pulsed-power-driven high energy density phys. and inertial
				confinement fusion research},}\ }\href {\doibase 10.1063/1.1891746}
	{\bibfield  {journal} {\bibinfo  {journal} {Phys.~Plasmas}\ }\textbf
		{\bibinfo {volume} {12}},\ \bibinfo {pages} {055503} (\bibinfo {year}
		{2005})}\BibitemShut {NoStop}%
	\bibitem [{\citenamefont {Knudson}\ \emph {et~al.}(2015)\citenamefont
		{Knudson}, \citenamefont {Desjarlais}, \citenamefont {Becker}, \citenamefont
		{Lemke}, \citenamefont {Cochrane}, \citenamefont {Savage}, \citenamefont
		{Bliss}, \citenamefont {Mattsson},\ and\ \citenamefont
		{Redmer}}]{knudson_direct_2015}%
	\BibitemOpen
	\bibfield  {author} {\bibinfo {author} {\bibfnamefont {M.~D.}\ \bibnamefont
			{Knudson}}, \bibinfo {author} {\bibfnamefont {M.~P.}\ \bibnamefont
			{Desjarlais}}, \bibinfo {author} {\bibfnamefont {A.}~\bibnamefont {Becker}},
		\bibinfo {author} {\bibfnamefont {R.~W.}\ \bibnamefont {Lemke}}, \bibinfo
		{author} {\bibfnamefont {K.~R.}\ \bibnamefont {Cochrane}}, \bibinfo {author}
		{\bibfnamefont {M.~E.}\ \bibnamefont {Savage}}, \bibinfo {author}
		{\bibfnamefont {D.~E.}\ \bibnamefont {Bliss}}, \bibinfo {author}
		{\bibfnamefont {T.~R.}\ \bibnamefont {Mattsson}}, \ and\ \bibinfo {author}
		{\bibfnamefont {R.}~\bibnamefont {Redmer}},\ }\bibfield  {title} {\enquote
		{\bibinfo {title} {Direct observation of an abrupt insulator-to-metal
				transition in dense liquid deuterium},}\ }\href {\doibase
		10.1126/science.aaa7471} {\bibfield  {journal} {\bibinfo  {journal}
			{Science}\ }\textbf {\bibinfo {volume} {348}},\ \bibinfo {pages} {1455--1460}
		(\bibinfo {year} {2015})}\BibitemShut {NoStop}%
	\bibitem [{\citenamefont {Nora}\ \emph {et~al.}(2015)\citenamefont {Nora},
		\citenamefont {Theobald}, \citenamefont {Betti}, \citenamefont {Marshall},
		\citenamefont {Michel}, \citenamefont {Seka}, \citenamefont {Yaakobi},
		\citenamefont {Lafon}, \citenamefont {Stoeckl}, \citenamefont {Delettrez},
		\citenamefont {Solodov}, \citenamefont {Casner}, \citenamefont {Reverdin},
		\citenamefont {Ribeyre}, \citenamefont {Vallet}, \citenamefont {Peebles},
		\citenamefont {Beg},\ and\ \citenamefont {Wei}}]{nora_gigabar_2015}%
	\BibitemOpen
	\bibfield  {author} {\bibinfo {author} {\bibfnamefont {R.}~\bibnamefont
			{Nora}}, \bibinfo {author} {\bibfnamefont {W.}~\bibnamefont {Theobald}},
		\bibinfo {author} {\bibfnamefont {R.}~\bibnamefont {Betti}}, \bibinfo
		{author} {\bibfnamefont {F.}~\bibnamefont {Marshall}}, \bibinfo {author}
		{\bibfnamefont {D.}~\bibnamefont {Michel}}, \bibinfo {author} {\bibfnamefont
			{W.}~\bibnamefont {Seka}}, \bibinfo {author} {\bibfnamefont {B.}~\bibnamefont
			{Yaakobi}}, \bibinfo {author} {\bibfnamefont {M.}~\bibnamefont {Lafon}},
		\bibinfo {author} {\bibfnamefont {C.}~\bibnamefont {Stoeckl}}, \bibinfo
		{author} {\bibfnamefont {J.}~\bibnamefont {Delettrez}}, \bibinfo {author}
		{\bibfnamefont {A.}~\bibnamefont {Solodov}}, \bibinfo {author} {\bibfnamefont
			{A.}~\bibnamefont {Casner}}, \bibinfo {author} {\bibfnamefont
			{C.}~\bibnamefont {Reverdin}}, \bibinfo {author} {\bibfnamefont
			{X.}~\bibnamefont {Ribeyre}}, \bibinfo {author} {\bibfnamefont
			{A.}~\bibnamefont {Vallet}}, \bibinfo {author} {\bibfnamefont
			{J.}~\bibnamefont {Peebles}}, \bibinfo {author} {\bibfnamefont
			{F.}~\bibnamefont {Beg}}, \ and\ \bibinfo {author} {\bibfnamefont
			{M.}~\bibnamefont {Wei}},\ }\bibfield  {title} {\enquote {\bibinfo {title}
			{Gigabar {Spherical} {Shock} {Generation} on the {OMEGA} {Laser}},}\ }\href
	{\doibase 10.1103/PhysRevLett.114.045001} {\bibfield  {journal} {\bibinfo
			{journal} {Phys.~Rev.~Lett.}\ }\textbf {\bibinfo {volume} {114}},\ \bibinfo
		{pages} {045001} (\bibinfo {year} {2015})}\BibitemShut {NoStop}%
	\bibitem [{\citenamefont {Sperling}\ \emph {et~al.}(2015)\citenamefont
		{Sperling}, \citenamefont {Gamboa}, \citenamefont {Lee}, \citenamefont
		{Chung}, \citenamefont {Galtier}, \citenamefont {Omarbakiyeva}, \citenamefont
		{Reinholz}, \citenamefont {R\"opke}, \citenamefont {Zastrau}, \citenamefont
		{Hastings}, \citenamefont {Fletcher},\ and\ \citenamefont
		{Glenzer}}]{sperling_free-electron_2015}%
	\BibitemOpen
	\bibfield  {author} {\bibinfo {author} {\bibfnamefont {P.}~\bibnamefont
			{Sperling}}, \bibinfo {author} {\bibfnamefont {E.}~\bibnamefont {Gamboa}},
		\bibinfo {author} {\bibfnamefont {H.}~\bibnamefont {Lee}}, \bibinfo {author}
		{\bibfnamefont {H.}~\bibnamefont {Chung}}, \bibinfo {author} {\bibfnamefont
			{E.}~\bibnamefont {Galtier}}, \bibinfo {author} {\bibfnamefont
			{Y.}~\bibnamefont {Omarbakiyeva}}, \bibinfo {author} {\bibfnamefont
			{H.}~\bibnamefont {Reinholz}}, \bibinfo {author} {\bibfnamefont
			{G.}~\bibnamefont {R\"opke}}, \bibinfo {author} {\bibfnamefont
			{U.}~\bibnamefont {Zastrau}}, \bibinfo {author} {\bibfnamefont
			{J.}~\bibnamefont {Hastings}}, \bibinfo {author} {\bibfnamefont
			{L.}~\bibnamefont {Fletcher}}, \ and\ \bibinfo {author} {\bibfnamefont
			{S.}~\bibnamefont {Glenzer}},\ }\bibfield  {title} {\enquote {\bibinfo
			{title} {Free-{Electron} {X}-{Ray} {Laser} {Measurements} of
				{Collisional}-{Damped} {Plasmons} in {Isochorically} {Heated} {Warm} {Dense}
				{Matter}},}\ }\href {\doibase 10.1103/PhysRevLett.115.115001} {\bibfield
		{journal} {\bibinfo  {journal} {Phys.~Rev.~Lett.}\ }\textbf {\bibinfo
			{volume} {115}},\ \bibinfo {pages} {115001} (\bibinfo {year}
		{2015})}\BibitemShut {NoStop}%
	\bibitem [{\citenamefont {Glenzer}\ \emph {et~al.}(2016)\citenamefont
		{Glenzer}, \citenamefont {Fletcher}, \citenamefont {Galtier}, \citenamefont
		{Nagler}, \citenamefont {Alonso-Mori}, \citenamefont {Barbrel}, \citenamefont
		{Brown}, \citenamefont {Chapman}, \citenamefont {Chen}, \citenamefont
		{Curry}, \citenamefont {Fiuza}, \citenamefont {Gamboa}, \citenamefont
		{Gauthier}, \citenamefont {Gericke}, \citenamefont {Gleason}, \citenamefont
		{Goede}, \citenamefont {Granados}, \citenamefont {Heimann}, \citenamefont
		{Kim}, \citenamefont {Kraus}, \citenamefont {MacDonald}, \citenamefont
		{Mackinnon}, \citenamefont {Mishra}, \citenamefont {Ravasio}, \citenamefont
		{Roedel}, \citenamefont {Sperling}, \citenamefont {Schumaker}, \citenamefont
		{Tsui}, \citenamefont {Vorberger}, \citenamefont {{U Zastrau}}, \citenamefont
		{Fry}, \citenamefont {White}, \citenamefont {Hasting},\ and\ \citenamefont
		{Lee}}]{glenzer_matter_2016}%
	\BibitemOpen
	\bibfield  {author} {\bibinfo {author} {\bibfnamefont {S.~H.}\ \bibnamefont
			{Glenzer}}, \bibinfo {author} {\bibfnamefont {L.~B.}\ \bibnamefont
			{Fletcher}}, \bibinfo {author} {\bibfnamefont {E.}~\bibnamefont {Galtier}},
		\bibinfo {author} {\bibfnamefont {B.}~\bibnamefont {Nagler}}, \bibinfo
		{author} {\bibfnamefont {R.}~\bibnamefont {Alonso-Mori}}, \bibinfo {author}
		{\bibfnamefont {B.}~\bibnamefont {Barbrel}}, \bibinfo {author} {\bibfnamefont
			{S.~B.}\ \bibnamefont {Brown}}, \bibinfo {author} {\bibfnamefont {D.~A.}\
			\bibnamefont {Chapman}}, \bibinfo {author} {\bibfnamefont {Z.}~\bibnamefont
			{Chen}}, \bibinfo {author} {\bibfnamefont {C.~B.}\ \bibnamefont {Curry}},
		\bibinfo {author} {\bibfnamefont {F.}~\bibnamefont {Fiuza}}, \bibinfo
		{author} {\bibfnamefont {E.}~\bibnamefont {Gamboa}}, \bibinfo {author}
		{\bibfnamefont {M.}~\bibnamefont {Gauthier}}, \bibinfo {author}
		{\bibfnamefont {D.~O.}\ \bibnamefont {Gericke}}, \bibinfo {author}
		{\bibfnamefont {A.}~\bibnamefont {Gleason}}, \bibinfo {author} {\bibfnamefont
			{S.}~\bibnamefont {Goede}}, \bibinfo {author} {\bibfnamefont
			{E.}~\bibnamefont {Granados}}, \bibinfo {author} {\bibfnamefont
			{P.}~\bibnamefont {Heimann}}, \bibinfo {author} {\bibfnamefont
			{J.}~\bibnamefont {Kim}}, \bibinfo {author} {\bibfnamefont {D.}~\bibnamefont
			{Kraus}}, \bibinfo {author} {\bibfnamefont {M.~J.}\ \bibnamefont
			{MacDonald}}, \bibinfo {author} {\bibfnamefont {A.~J.}\ \bibnamefont
			{Mackinnon}}, \bibinfo {author} {\bibfnamefont {R.}~\bibnamefont {Mishra}},
		\bibinfo {author} {\bibfnamefont {A.}~\bibnamefont {Ravasio}}, \bibinfo
		{author} {\bibfnamefont {C.}~\bibnamefont {Roedel}}, \bibinfo {author}
		{\bibfnamefont {P.}~\bibnamefont {Sperling}}, \bibinfo {author}
		{\bibfnamefont {W.}~\bibnamefont {Schumaker}}, \bibinfo {author}
		{\bibfnamefont {Y.~Y.}\ \bibnamefont {Tsui}}, \bibinfo {author}
		{\bibfnamefont {J.}~\bibnamefont {Vorberger}}, \bibinfo {author}
		{\bibnamefont {{U Zastrau}}}, \bibinfo {author} {\bibfnamefont
			{A.}~\bibnamefont {Fry}}, \bibinfo {author} {\bibfnamefont {W.~E.}\
			\bibnamefont {White}}, \bibinfo {author} {\bibfnamefont {J.~B.}\ \bibnamefont
			{Hasting}}, \ and\ \bibinfo {author} {\bibfnamefont {H.~J.}\ \bibnamefont
			{Lee}},\ }\bibfield  {title} {\enquote {\bibinfo {title} {Matter under
				extreme conditions experiments at the {Linac} {Coherent} {Light} {Source}},}\
	}\href {\doibase 10.1088/0953-4075/49/9/092001} {\bibfield  {journal}
		{\bibinfo  {journal} {J.~Phys.~B}\ }\textbf {\bibinfo {volume} {49}},\
		\bibinfo {pages} {092001} (\bibinfo {year} {2016})}\BibitemShut {NoStop}%
	\bibitem [{\citenamefont {Zastrau}\ \emph {et~al.}(2014)\citenamefont
		{Zastrau}, \citenamefont {Sperling}, \citenamefont {Harmand}, \citenamefont
		{Becker}, \citenamefont {Bornath}, \citenamefont {Bredow}, \citenamefont
		{Dziarzhytski}, \citenamefont {Fennel}, \citenamefont {Fletcher},
		\citenamefont {F\"orster}, \citenamefont {G\"ode}, \citenamefont {Gregori},
		\citenamefont {Hilbert}, \citenamefont {Hochhaus}, \citenamefont {Holst},
		\citenamefont {Laarmann}, \citenamefont {Lee}, \citenamefont {Ma},
		\citenamefont {Mithen}, \citenamefont {Mitzner}, \citenamefont {Murphy},
		\citenamefont {Nakatsutsumi}, \citenamefont {Neumayer}, \citenamefont
		{Przystawik}, \citenamefont {Roling}, \citenamefont {Schulz}, \citenamefont
		{Siemer}, \citenamefont {Skruszewicz}, \citenamefont {Tiggesb\"aumker},
		\citenamefont {Toleikis}, \citenamefont {Tschentscher}, \citenamefont
		{White}, \citenamefont {W\"ostmann}, \citenamefont {Zacharias}, \citenamefont
		{D\"oppner}, \citenamefont {Glenzer},\ and\ \citenamefont
		{Redmer}}]{zastrau_resolving_2014}%
	\BibitemOpen
	\bibfield  {author} {\bibinfo {author} {\bibfnamefont {U.}~\bibnamefont
			{Zastrau}}, \bibinfo {author} {\bibfnamefont {P.}~\bibnamefont {Sperling}},
		\bibinfo {author} {\bibfnamefont {M.}~\bibnamefont {Harmand}}, \bibinfo
		{author} {\bibfnamefont {A.}~\bibnamefont {Becker}}, \bibinfo {author}
		{\bibfnamefont {T.}~\bibnamefont {Bornath}}, \bibinfo {author} {\bibfnamefont
			{R.}~\bibnamefont {Bredow}}, \bibinfo {author} {\bibfnamefont
			{S.}~\bibnamefont {Dziarzhytski}}, \bibinfo {author} {\bibfnamefont
			{T.}~\bibnamefont {Fennel}}, \bibinfo {author} {\bibfnamefont
			{L.}~\bibnamefont {Fletcher}}, \bibinfo {author} {\bibfnamefont
			{E.}~\bibnamefont {F\"orster}}, \bibinfo {author} {\bibfnamefont
			{S.}~\bibnamefont {G\"ode}}, \bibinfo {author} {\bibfnamefont
			{G.}~\bibnamefont {Gregori}}, \bibinfo {author} {\bibfnamefont
			{V.}~\bibnamefont {Hilbert}}, \bibinfo {author} {\bibfnamefont
			{D.}~\bibnamefont {Hochhaus}}, \bibinfo {author} {\bibfnamefont
			{B.}~\bibnamefont {Holst}}, \bibinfo {author} {\bibfnamefont
			{T.}~\bibnamefont {Laarmann}}, \bibinfo {author} {\bibfnamefont
			{H.}~\bibnamefont {Lee}}, \bibinfo {author} {\bibfnamefont {T.}~\bibnamefont
			{Ma}}, \bibinfo {author} {\bibfnamefont {J.}~\bibnamefont {Mithen}}, \bibinfo
		{author} {\bibfnamefont {R.}~\bibnamefont {Mitzner}}, \bibinfo {author}
		{\bibfnamefont {C.}~\bibnamefont {Murphy}}, \bibinfo {author} {\bibfnamefont
			{M.}~\bibnamefont {Nakatsutsumi}}, \bibinfo {author} {\bibfnamefont
			{P.}~\bibnamefont {Neumayer}}, \bibinfo {author} {\bibfnamefont
			{A.}~\bibnamefont {Przystawik}}, \bibinfo {author} {\bibfnamefont
			{S.}~\bibnamefont {Roling}}, \bibinfo {author} {\bibfnamefont
			{M.}~\bibnamefont {Schulz}}, \bibinfo {author} {\bibfnamefont
			{B.}~\bibnamefont {Siemer}}, \bibinfo {author} {\bibfnamefont
			{S.}~\bibnamefont {Skruszewicz}}, \bibinfo {author} {\bibfnamefont
			{J.}~\bibnamefont {Tiggesb\"aumker}}, \bibinfo {author} {\bibfnamefont
			{S.}~\bibnamefont {Toleikis}}, \bibinfo {author} {\bibfnamefont
			{T.}~\bibnamefont {Tschentscher}}, \bibinfo {author} {\bibfnamefont
			{T.}~\bibnamefont {White}}, \bibinfo {author} {\bibfnamefont
			{M.}~\bibnamefont {W\"ostmann}}, \bibinfo {author} {\bibfnamefont
			{H.}~\bibnamefont {Zacharias}}, \bibinfo {author} {\bibfnamefont
			{T.}~\bibnamefont {D\"oppner}}, \bibinfo {author} {\bibfnamefont
			{S.}~\bibnamefont {Glenzer}}, \ and\ \bibinfo {author} {\bibfnamefont
			{R.}~\bibnamefont {Redmer}},\ }\bibfield  {title} {\enquote {\bibinfo {title}
			{Resolving {Ultrafast} {Heating} of {Dense} {Cryogenic} {Hydrogen}},}\ }\href
	{\doibase 10.1103/PhysRevLett.112.105002} {\bibfield  {journal} {\bibinfo
			{journal} {Phys.~Rev.~Lett.}\ }\textbf {\bibinfo {volume} {112}},\ \bibinfo
		{pages} {105002} (\bibinfo {year} {2014})}\BibitemShut {NoStop}%
	\bibitem [{\citenamefont {Tschentscher}\ \emph {et~al.}(2017)\citenamefont
		{Tschentscher}, \citenamefont {Bressler}, \citenamefont {Gr\"unert},
		\citenamefont {Madsen}, \citenamefont {Mancuso}, \citenamefont {Meyer},
		\citenamefont {Scherz}, \citenamefont {Sinn},\ and\ \citenamefont
		{Zastrau}}]{tschentscher_photon_2017}%
	\BibitemOpen
	\bibfield  {author} {\bibinfo {author} {\bibfnamefont {T.}~\bibnamefont
			{Tschentscher}}, \bibinfo {author} {\bibfnamefont {C.}~\bibnamefont
			{Bressler}}, \bibinfo {author} {\bibfnamefont {J.}~\bibnamefont {Gr\"unert}},
		\bibinfo {author} {\bibfnamefont {A.}~\bibnamefont {Madsen}}, \bibinfo
		{author} {\bibfnamefont {A.~P.}\ \bibnamefont {Mancuso}}, \bibinfo {author}
		{\bibfnamefont {M.}~\bibnamefont {Meyer}}, \bibinfo {author} {\bibfnamefont
			{A.}~\bibnamefont {Scherz}}, \bibinfo {author} {\bibfnamefont
			{H.}~\bibnamefont {Sinn}}, \ and\ \bibinfo {author} {\bibfnamefont
			{U.}~\bibnamefont {Zastrau}},\ }\bibfield  {title} {\enquote {\bibinfo
			{title} {Photon {Beam} {Transport} and {Scientific} {Instruments} at the
				{European} {XFEL}},}\ }\href {\doibase 10.3390/app7060592} {\bibfield
		{journal} {\bibinfo  {journal} {Appl.~Sci.}\ }\textbf {\bibinfo {volume}
			{7}},\ \bibinfo {pages} {592} (\bibinfo {year} {2017})}\BibitemShut {NoStop}%
	\bibitem [{\citenamefont {Ren}\ \emph {et~al.}(2018)\citenamefont {Ren},
		\citenamefont {Maurer}, \citenamefont {Katrik}, \citenamefont {Lang},
		\citenamefont {Golubev}, \citenamefont {Mintsev}, \citenamefont {Zhao},\ and\
		\citenamefont {Hoffmann}}]{hoffmann_cpp_18}%
	\BibitemOpen
	\bibfield  {author} {\bibinfo {author} {\bibfnamefont {J.}~\bibnamefont
			{Ren}}, \bibinfo {author} {\bibfnamefont {C.}~\bibnamefont {Maurer}},
		\bibinfo {author} {\bibfnamefont {P.}~\bibnamefont {Katrik}}, \bibinfo
		{author} {\bibfnamefont {P.}~\bibnamefont {Lang}}, \bibinfo {author}
		{\bibfnamefont {A.}~\bibnamefont {Golubev}}, \bibinfo {author} {\bibfnamefont
			{V.}~\bibnamefont {Mintsev}}, \bibinfo {author} {\bibfnamefont
			{Y.}~\bibnamefont {Zhao}}, \ and\ \bibinfo {author} {\bibfnamefont
			{D.}~\bibnamefont {Hoffmann}},\ }\bibfield  {title} {\enquote {\bibinfo
			{title} {Accelerator-driven high-energy-density physics: Status and
				chances},}\ }\href {\doibase 10.1002/ctpp.201700110} {\bibfield  {journal}
		{\bibinfo  {journal} {Contrib. Plasma Phys.}\ }\textbf {\bibinfo {volume}
			{58}},\ \bibinfo {pages} {82--92} (\bibinfo {year} {2018})}\BibitemShut
	{NoStop}%
	\bibitem [{\citenamefont {Tahir}\ \emph {et~al.}(2019)\citenamefont {Tahir},
		\citenamefont {Neumayer}, \citenamefont {Shutov}, \citenamefont {Piriz},
		\citenamefont {Lomonosov}, \citenamefont {Bagnoud}, \citenamefont {Piriz},\
		and\ \citenamefont {Deutsch}}]{tahir_cpp19}%
	\BibitemOpen
	\bibfield  {author} {\bibinfo {author} {\bibfnamefont {N.~A.}\ \bibnamefont
			{Tahir}}, \bibinfo {author} {\bibfnamefont {P.}~\bibnamefont {Neumayer}},
		\bibinfo {author} {\bibfnamefont {A.}~\bibnamefont {Shutov}}, \bibinfo
		{author} {\bibfnamefont {A.~R.}\ \bibnamefont {Piriz}}, \bibinfo {author}
		{\bibfnamefont {I.~V.}\ \bibnamefont {Lomonosov}}, \bibinfo {author}
		{\bibfnamefont {V.}~\bibnamefont {Bagnoud}}, \bibinfo {author} {\bibfnamefont
			{S.~A.}\ \bibnamefont {Piriz}}, \ and\ \bibinfo {author} {\bibfnamefont
			{C.}~\bibnamefont {Deutsch}},\ }\bibfield  {title} {\enquote {\bibinfo
			{title} {Equation-of-state studies of high-energy-density matter using
				intense ion beams at the facility for antiprotons and ion research},}\ }\href
	{\doibase 10.1002/ctpp.201800143} {\bibfield  {journal} {\bibinfo  {journal}
			{Contrib. Plasma Phys.}\ }\textbf {\bibinfo {volume} {59}},\ \bibinfo {pages}
		{e201800143} (\bibinfo {year} {2019})}\BibitemShut {NoStop}%
	\bibitem [{\citenamefont {Ernstorfer}\ \emph {et~al.}(2009)\citenamefont
		{Ernstorfer}, \citenamefont {Harb}, \citenamefont {Hebeisen}, \citenamefont
		{Sciaini}, \citenamefont {Dartigalongue},\ and\ \citenamefont
		{Miller}}]{Ernstorfer1033}%
	\BibitemOpen
	\bibfield  {author} {\bibinfo {author} {\bibfnamefont {R.}~\bibnamefont
			{Ernstorfer}}, \bibinfo {author} {\bibfnamefont {M.}~\bibnamefont {Harb}},
		\bibinfo {author} {\bibfnamefont {C.~T.}\ \bibnamefont {Hebeisen}}, \bibinfo
		{author} {\bibfnamefont {G.}~\bibnamefont {Sciaini}}, \bibinfo {author}
		{\bibfnamefont {T.}~\bibnamefont {Dartigalongue}}, \ and\ \bibinfo {author}
		{\bibfnamefont {R.~J.~D.}\ \bibnamefont {Miller}},\ }\bibfield  {title}
	{\enquote {\bibinfo {title} {The formation of warm dense matter: Experimental
				evidence for electronic bond hardening in gold},}\ }\href {\doibase
		10.1126/science.1162697} {\bibfield  {journal} {\bibinfo  {journal}
			{Science}\ }\textbf {\bibinfo {volume} {323}},\ \bibinfo {pages} {1033--1037}
		(\bibinfo {year} {2009})}\BibitemShut {NoStop}%
	\bibitem [{\citenamefont {Waldecker}\ \emph {et~al.}(2016)\citenamefont
		{Waldecker}, \citenamefont {Bertoni}, \citenamefont {Ernstorfer},\ and\
		\citenamefont {Vorberger}}]{PhysRevX.6.021003}%
	\BibitemOpen
	\bibfield  {author} {\bibinfo {author} {\bibfnamefont {L.}~\bibnamefont
			{Waldecker}}, \bibinfo {author} {\bibfnamefont {R.}~\bibnamefont {Bertoni}},
		\bibinfo {author} {\bibfnamefont {R.}~\bibnamefont {Ernstorfer}}, \ and\
		\bibinfo {author} {\bibfnamefont {J.}~\bibnamefont {Vorberger}},\ }\bibfield
	{title} {\enquote {\bibinfo {title} {Electron-phonon coupling and energy flow
				in a simple metal beyond the two-temperature approximation},}\ }\href
	{\doibase 10.1103/PhysRevX.6.021003} {\bibfield  {journal} {\bibinfo
			{journal} {Phys. Rev. X}\ }\textbf {\bibinfo {volume} {6}},\ \bibinfo {pages}
		{021003} (\bibinfo {year} {2016})}\BibitemShut {NoStop}%
	\bibitem [{\citenamefont {Bonitz}\ \emph {et~al.}(2020)\citenamefont {Bonitz},
		\citenamefont {Dornheim}, \citenamefont {Moldabekov}, \citenamefont {Zhang},
		\citenamefont {Hamann}, \citenamefont {Kählert}, \citenamefont {Filinov},
		\citenamefont {Ramakrishna},\ and\ \citenamefont
		{Vorberger}}]{bonitz_pop_20}%
	\BibitemOpen
	\bibfield  {author} {\bibinfo {author} {\bibfnamefont {M.}~\bibnamefont
			{Bonitz}}, \bibinfo {author} {\bibfnamefont {T.}~\bibnamefont {Dornheim}},
		\bibinfo {author} {\bibfnamefont {Z.~A.}\ \bibnamefont {Moldabekov}},
		\bibinfo {author} {\bibfnamefont {S.}~\bibnamefont {Zhang}}, \bibinfo
		{author} {\bibfnamefont {P.}~\bibnamefont {Hamann}}, \bibinfo {author}
		{\bibfnamefont {H.}~\bibnamefont {Kählert}}, \bibinfo {author}
		{\bibfnamefont {A.}~\bibnamefont {Filinov}}, \bibinfo {author} {\bibfnamefont
			{K.}~\bibnamefont {Ramakrishna}}, \ and\ \bibinfo {author} {\bibfnamefont
			{J.}~\bibnamefont {Vorberger}},\ }\bibfield  {title} {\enquote {\bibinfo
			{title} {Ab initio simulation of warm dense matter},}\ }\href {\doibase
		10.1063/1.5143225} {\bibfield  {journal} {\bibinfo  {journal} {Physics of
				Plasmas}\ }\textbf {\bibinfo {volume} {27}},\ \bibinfo {pages} {042710}
		(\bibinfo {year} {2020})},\ \Eprint
	{http://arxiv.org/abs/https://doi.org/10.1063/1.5143225}
	{https://doi.org/10.1063/1.5143225} \BibitemShut {NoStop}%
	\bibitem [{\citenamefont {Kadanoff}\ and\ \citenamefont
		{Baym}(1989)}]{kadanoff-baym}%
	\BibitemOpen
	\bibfield  {author} {\bibinfo {author} {\bibfnamefont {L.}~\bibnamefont
			{Kadanoff}}\ and\ \bibinfo {author} {\bibfnamefont {G.}~\bibnamefont
			{Baym}},\ }\href@noop {} {\emph {\bibinfo {title} {Quantum Statistical
				Mechanics}}},\ \bibinfo {edition} {2nd}\ ed.\ (\bibinfo  {publisher}
	{Addison-Wesley Publ. Co. Inc.},\ \bibinfo {year} {1989})\BibitemShut
	{NoStop}%
	\bibitem [{\citenamefont {Keldysh}(1964)}]{keldysh65}%
	\BibitemOpen
	\bibfield  {author} {\bibinfo {author} {\bibfnamefont {L.}~\bibnamefont
			{Keldysh}},\ }\bibfield  {title} {\enquote {\bibinfo {title} {Diagram
				technique for nonequilibrium processes},}\ }\href@noop {} {\bibfield
		{journal} {\bibinfo  {journal} {ZhETF}\ }\textbf {\bibinfo {volume} {{\bf
					47}}},\ \bibinfo {pages} {1515} (\bibinfo {year} {1964})},\ \bibinfo {note}
	{[Soviet Phys. JETP, {\bf 20}, 1018 (1965)]}\BibitemShut {NoStop}%
	\bibitem [{\citenamefont {Kraeft}\ \emph {et~al.}(1986)\citenamefont {Kraeft},
		\citenamefont {Kremp}, \citenamefont {Ebeling},\ and\ \citenamefont
		{R\"opke}}]{green-book}%
	\BibitemOpen
	\bibfield  {author} {\bibinfo {author} {\bibfnamefont {W.-D.}\ \bibnamefont
			{Kraeft}}, \bibinfo {author} {\bibfnamefont {D.}~\bibnamefont {Kremp}},
		\bibinfo {author} {\bibfnamefont {W.}~\bibnamefont {Ebeling}}, \ and\
		\bibinfo {author} {\bibfnamefont {G.}~\bibnamefont {R\"opke}},\ }\href@noop
	{} {\emph {\bibinfo {title} {Quantum Statistics of Charged Particle
				Systems}}}\ (\bibinfo  {publisher} {Akademie-Verlag},\ \bibinfo {address}
	{Berlin},\ \bibinfo {year} {1986})\BibitemShut {NoStop}%
	\bibitem [{\citenamefont {Bonitz}(2012)}]{bonitz_aip_12}%
	\BibitemOpen
	\bibfield  {author} {\bibinfo {author} {\bibfnamefont {M.}~\bibnamefont
			{Bonitz}},\ }\bibfield  {title} {\enquote {\bibinfo {title} {Kinetic theory
				for quantum plasmas},}\ }\href {\doibase http://dx.doi.org/10.1063/1.3679593}
	{\bibfield  {journal} {\bibinfo  {journal} {AIP Conference Proceedings}\
		}\textbf {\bibinfo {volume} {1421}},\ \bibinfo {pages} {135--155} (\bibinfo
		{year} {2012})}\BibitemShut {NoStop}%
	\bibitem [{\citenamefont {Balzer}, \citenamefont {Bauch},\ and\ \citenamefont
		{Bonitz}(2010{\natexlab{a}})}]{balzer_pra_10}%
	\BibitemOpen
	\bibfield  {author} {\bibinfo {author} {\bibfnamefont {K.}~\bibnamefont
			{Balzer}}, \bibinfo {author} {\bibfnamefont {S.}~\bibnamefont {Bauch}}, \
		and\ \bibinfo {author} {\bibfnamefont {M.}~\bibnamefont {Bonitz}},\
	}\bibfield  {title} {\enquote {\bibinfo {title} {Efficient grid-based method
				in nonequilibrium {Green's function calculations: Application to model atoms
					and molecules}},}\ }\href {\doibase 10.1103/PhysRevA.81.022510} {\bibfield
		{journal} {\bibinfo  {journal} {Phys. Rev. A}\ }\textbf {\bibinfo {volume}
			{81}},\ \bibinfo {pages} {022510} (\bibinfo {year}
		{2010}{\natexlab{a}})}\BibitemShut {NoStop}%
	\bibitem [{\citenamefont {Balzer}, \citenamefont {Bauch},\ and\ \citenamefont
		{Bonitz}(2010{\natexlab{b}})}]{balzer_pra_10_2}%
	\BibitemOpen
	\bibfield  {author} {\bibinfo {author} {\bibfnamefont {K.}~\bibnamefont
			{Balzer}}, \bibinfo {author} {\bibfnamefont {S.}~\bibnamefont {Bauch}}, \
		and\ \bibinfo {author} {\bibfnamefont {M.}~\bibnamefont {Bonitz}},\
	}\bibfield  {title} {\enquote {\bibinfo {title} {Time-dependent second-order
				{Born} calculations for model atoms and molecules in strong laser fields},}\
	}\href {\doibase 10.1103/PhysRevA.82.033427} {\bibfield  {journal} {\bibinfo
			{journal} {Phys. Rev. A}\ }\textbf {\bibinfo {volume} {82}},\ \bibinfo
		{pages} {033427} (\bibinfo {year} {2010}{\natexlab{b}})}\BibitemShut
	{NoStop}%
	\bibitem [{\citenamefont {Schlünzen}, \citenamefont {Joost},\ and\
		\citenamefont {Bonitz}(2020)}]{schluenzen_19_prl}%
	\BibitemOpen
	\bibfield  {author} {\bibinfo {author} {\bibfnamefont {N.}~\bibnamefont
			{Schlünzen}}, \bibinfo {author} {\bibfnamefont {J.-P.}\ \bibnamefont
			{Joost}}, \ and\ \bibinfo {author} {\bibfnamefont {M.}~\bibnamefont
			{Bonitz}},\ }\bibfield  {title} {\enquote {\bibinfo {title} {{Achieving the
					Scaling Limit for Nonequilibrium Green Functions Simulations}},}\ }\href
	{\doibase 10.1103/PhysRevLett.124.076601} {\bibfield  {journal} {\bibinfo
			{journal} {Phys. Rev. Lett.}\ }\textbf {\bibinfo {volume} {124}},\ \bibinfo
		{pages} {076601} (\bibinfo {year} {2020})}\BibitemShut {NoStop}%
	\bibitem [{\citenamefont {Schl{\"u}nzen}\ \emph {et~al.}(2019)\citenamefont
		{Schl{\"u}nzen}, \citenamefont {Balzer}, \citenamefont {Bonitz},
		\citenamefont {Deuchler},\ and\ \citenamefont {Pehlke}}]{schluenzen_cpp_18}%
	\BibitemOpen
	\bibfield  {author} {\bibinfo {author} {\bibfnamefont {N.}~\bibnamefont
			{Schl{\"u}nzen}}, \bibinfo {author} {\bibfnamefont {K.}~\bibnamefont
			{Balzer}}, \bibinfo {author} {\bibfnamefont {M.}~\bibnamefont {Bonitz}},
		\bibinfo {author} {\bibfnamefont {L.}~\bibnamefont {Deuchler}}, \ and\
		\bibinfo {author} {\bibfnamefont {E.}~\bibnamefont {Pehlke}},\ }\bibfield
	{title} {\enquote {\bibinfo {title} {Time-dependent simulation of ion
				stopping: charge transfer and electronic excitations},}\ }\href {\doibase
		10.1002/ctpp.201800184} {\bibfield  {journal} {\bibinfo  {journal} {Contrib.
				Plasma Phys.}\ }\textbf {\bibinfo {volume} {59}},\ \bibinfo {pages}
		{e201800184} (\bibinfo {year} {2019})}\BibitemShut {NoStop}%
	\bibitem [{\citenamefont {Bonitz}, \citenamefont {Pehlke},\ and\ \citenamefont
		{Schoof}(2013)}]{bonitz_pre_13}%
	\BibitemOpen
	\bibfield  {author} {\bibinfo {author} {\bibfnamefont {M.}~\bibnamefont
			{Bonitz}}, \bibinfo {author} {\bibfnamefont {E.}~\bibnamefont {Pehlke}}, \
		and\ \bibinfo {author} {\bibfnamefont {T.}~\bibnamefont {Schoof}},\
	}\bibfield  {title} {\enquote {\bibinfo {title} {Attractive forces between
				ions in quantum plasmas: Failure of linearized quantum hydrodynamics},}\
	}\href {\doibase 10.1103/PhysRevE.87.033105} {\bibfield  {journal} {\bibinfo
			{journal} {Phys. Rev. E}\ }\textbf {\bibinfo {volume} {87}},\ \bibinfo
		{pages} {033105} (\bibinfo {year} {2013})}\BibitemShut {NoStop}%
	\bibitem [{\citenamefont {Moldabekov}, \citenamefont {Bonitz},\ and\
		\citenamefont {Ramazanov}(2018)}]{zhandos_pop18}%
	\BibitemOpen
	\bibfield  {author} {\bibinfo {author} {\bibfnamefont {Z.~A.}\ \bibnamefont
			{Moldabekov}}, \bibinfo {author} {\bibfnamefont {M.}~\bibnamefont {Bonitz}},
		\ and\ \bibinfo {author} {\bibfnamefont {T.~S.}\ \bibnamefont {Ramazanov}},\
	}\bibfield  {title} {\enquote {\bibinfo {title} {Theoretical foundations of
				quantum hydrodynamics for plasmas},}\ }\href {\doibase 10.1063/1.5003910}
	{\bibfield  {journal} {\bibinfo  {journal} {Phys. Plasmas}\ }\textbf
		{\bibinfo {volume} {25}},\ \bibinfo {pages} {031903} (\bibinfo {year}
		{2018})}\BibitemShut {NoStop}%
	\bibitem [{\citenamefont {Moldabekov}\ \emph {et~al.}(2019)\citenamefont
		{Moldabekov}, \citenamefont {Amirov}, \citenamefont {Ludwig}, \citenamefont
		{Bonitz},\ and\ \citenamefont {Ramazanov}}]{zhandos_cpp_19}%
	\BibitemOpen
	\bibfield  {author} {\bibinfo {author} {\bibfnamefont {Z.}~\bibnamefont
			{Moldabekov}}, \bibinfo {author} {\bibfnamefont {S.}~\bibnamefont {Amirov}},
		\bibinfo {author} {\bibfnamefont {P.}~\bibnamefont {Ludwig}}, \bibinfo
		{author} {\bibfnamefont {M.}~\bibnamefont {Bonitz}}, \ and\ \bibinfo {author}
		{\bibfnamefont {T.}~\bibnamefont {Ramazanov}},\ }\bibfield  {title} {\enquote
		{\bibinfo {title} {Effect of the dynamical collision frequency on quantum
				wakefields},}\ }\href {\doibase 10.1002/ctpp.201800161} {\bibfield  {journal}
		{\bibinfo  {journal} {Contributions to Plasma Physics}\ }\textbf {\bibinfo
			{volume} {59}},\ \bibinfo {pages} {e201800161} (\bibinfo {year}
		{2019})}\BibitemShut {NoStop}%
	\bibitem [{\citenamefont {Collins}\ \emph {et~al.}(1995)\citenamefont
		{Collins}, \citenamefont {Kwon}, \citenamefont {Kress}, \citenamefont
		{Troullier},\ and\ \citenamefont {Lynch}}]{collins_pre_95}%
	\BibitemOpen
	\bibfield  {author} {\bibinfo {author} {\bibfnamefont {L.}~\bibnamefont
			{Collins}}, \bibinfo {author} {\bibfnamefont {I.}~\bibnamefont {Kwon}},
		\bibinfo {author} {\bibfnamefont {J.}~\bibnamefont {Kress}}, \bibinfo
		{author} {\bibfnamefont {N.}~\bibnamefont {Troullier}}, \ and\ \bibinfo
		{author} {\bibfnamefont {D.}~\bibnamefont {Lynch}},\ }\bibfield  {title}
	{\enquote {\bibinfo {title} {Quantum molecular dynamics simulations of hot,
				dense hydrogen},}\ }\href {\doibase 10.1103/PhysRevE.52.6202} {\bibfield
		{journal} {\bibinfo  {journal} {Phys. Rev. E}\ }\textbf {\bibinfo {volume}
			{52}},\ \bibinfo {pages} {6202--6219} (\bibinfo {year} {1995})}\BibitemShut
	{NoStop}%
	\bibitem [{\citenamefont {Plagemann}\ \emph {et~al.}(2012)\citenamefont
		{Plagemann}, \citenamefont {Sperling}, \citenamefont {Thiele}, \citenamefont
		{Desjarlais}, \citenamefont {Fortmann}, \citenamefont {D\"oppner},
		\citenamefont {Lee}, \citenamefont {Glenzer},\ and\ \citenamefont
		{Redmer}}]{plagemann_njp_2012}%
	\BibitemOpen
	\bibfield  {author} {\bibinfo {author} {\bibfnamefont {K.-U.}\ \bibnamefont
			{Plagemann}}, \bibinfo {author} {\bibfnamefont {P.}~\bibnamefont {Sperling}},
		\bibinfo {author} {\bibfnamefont {R.}~\bibnamefont {Thiele}}, \bibinfo
		{author} {\bibfnamefont {M.~P.}\ \bibnamefont {Desjarlais}}, \bibinfo
		{author} {\bibfnamefont {C.}~\bibnamefont {Fortmann}}, \bibinfo {author}
		{\bibfnamefont {T.}~\bibnamefont {D\"oppner}}, \bibinfo {author}
		{\bibfnamefont {H.~J.}\ \bibnamefont {Lee}}, \bibinfo {author} {\bibfnamefont
			{S.~H.}\ \bibnamefont {Glenzer}}, \ and\ \bibinfo {author} {\bibfnamefont
			{R.}~\bibnamefont {Redmer}},\ }\bibfield  {title} {\enquote {\bibinfo {title}
			{Dynamic structure factor in warm dense beryllium},}\ }\href {\doibase
		10.1088/1367-2630/14/5/055020} {\bibfield  {journal} {\bibinfo  {journal}
			{New Journal of Physics}\ }\textbf {\bibinfo {volume} {14}},\ \bibinfo
		{pages} {055020} (\bibinfo {year} {2012})}\BibitemShut {NoStop}%
	\bibitem [{\citenamefont {Witte}\ \emph {et~al.}(2017)\citenamefont {Witte},
		\citenamefont {Fletcher}, \citenamefont {Galtier}, \citenamefont {Gamboa},
		\citenamefont {Lee}, \citenamefont {Zastrau}, \citenamefont {Redmer},
		\citenamefont {Glenzer},\ and\ \citenamefont {Sperling}}]{witte_prl_17}%
	\BibitemOpen
	\bibfield  {author} {\bibinfo {author} {\bibfnamefont {B.~B.~L.}\
			\bibnamefont {Witte}}, \bibinfo {author} {\bibfnamefont {L.~B.}\ \bibnamefont
			{Fletcher}}, \bibinfo {author} {\bibfnamefont {E.}~\bibnamefont {Galtier}},
		\bibinfo {author} {\bibfnamefont {E.}~\bibnamefont {Gamboa}}, \bibinfo
		{author} {\bibfnamefont {H.~J.}\ \bibnamefont {Lee}}, \bibinfo {author}
		{\bibfnamefont {U.}~\bibnamefont {Zastrau}}, \bibinfo {author} {\bibfnamefont
			{R.}~\bibnamefont {Redmer}}, \bibinfo {author} {\bibfnamefont {S.~H.}\
			\bibnamefont {Glenzer}}, \ and\ \bibinfo {author} {\bibfnamefont
			{P.}~\bibnamefont {Sperling}},\ }\bibfield  {title} {\enquote {\bibinfo
			{title} {Warm dense matter demonstrating non-drude conductivity from
				observations of nonlinear plasmon damping},}\ }\href {\doibase
		10.1103/PhysRevLett.118.225001} {\bibfield  {journal} {\bibinfo  {journal}
			{Phys. Rev. Lett.}\ }\textbf {\bibinfo {volume} {118}},\ \bibinfo {pages}
		{225001} (\bibinfo {year} {2017})}\BibitemShut {NoStop}%
	\bibitem [{\citenamefont {Ceperley}(1996)}]{sign_cite}%
	\BibitemOpen
	\bibfield  {author} {\bibinfo {author} {\bibfnamefont {D.}~\bibnamefont
			{Ceperley}},\ }\bibfield  {title} {\enquote {\bibinfo {title} {{Path integral
					Monte Carlo methods for fermions}},}\ }in\ \href@noop {} {\emph {\bibinfo
			{booktitle} {Monte Carlo and Molecular Dynamics of Condensed Matter
				Systems}}},\ \bibinfo {editor} {edited by\ \bibinfo {editor} {\bibfnamefont
			{K.}~\bibnamefont {Binder}}\ and\ \bibinfo {editor} {\bibfnamefont
			{G.}~\bibnamefont {Ciccotti}}}\ (\bibinfo  {publisher} {Italian Physical
		Society},\ \bibinfo {address} {Bologna},\ \bibinfo {year} {1996})\BibitemShut
	{NoStop}%
	\bibitem [{\citenamefont {Militzer}\ and\ \citenamefont
		{Ceperley}(2000)}]{militzer_path_2000}%
	\BibitemOpen
	\bibfield  {author} {\bibinfo {author} {\bibfnamefont {B.}~\bibnamefont
			{Militzer}}\ and\ \bibinfo {author} {\bibfnamefont {D.~M.}\ \bibnamefont
			{Ceperley}},\ }\bibfield  {title} {\enquote {\bibinfo {title} {Path
				{Integral} {Monte} {Carlo} {Calculation} of the {Deuterium} {Hugoniot}},}\
	}\href {\doibase 10.1103/PhysRevLett.85.1890} {\bibfield  {journal} {\bibinfo
			{journal} {Phys.~Rev.~Lett.}\ }\textbf {\bibinfo {volume} {85}},\ \bibinfo
		{pages} {1890--1893} (\bibinfo {year} {2000})}\BibitemShut {NoStop}%
	\bibitem [{\citenamefont {Filinov}\ \emph
		{et~al.}(2001{\natexlab{a}})\citenamefont {Filinov}, \citenamefont {Bonitz},
		\citenamefont {Ebeling},\ and\ \citenamefont {Fortov}}]{filinov_ppcf_01}%
	\BibitemOpen
	\bibfield  {author} {\bibinfo {author} {\bibfnamefont {V.~S.}\ \bibnamefont
			{Filinov}}, \bibinfo {author} {\bibfnamefont {M.}~\bibnamefont {Bonitz}},
		\bibinfo {author} {\bibfnamefont {W.}~\bibnamefont {Ebeling}}, \ and\
		\bibinfo {author} {\bibfnamefont {V.~E.}\ \bibnamefont {Fortov}},\ }\bibfield
	{title} {\enquote {\bibinfo {title} {Thermodynamics of hot dense
				{H}-plasmas: path integral {M}onte {C}arlo simulations and analytical
				approximations},}\ }\href {http://stacks.iop.org/0741-3335/43/i=6/a=301}
	{\bibfield  {journal} {\bibinfo  {journal} {Plasma Phys. Control. Fusion}\
		}\textbf {\bibinfo {volume} {{\bf 43}}},\ \bibinfo {pages} {743} (\bibinfo
		{year} {2001}{\natexlab{a}})}\BibitemShut {NoStop}%
	\bibitem [{\citenamefont {Filinov}, \citenamefont {Lozovik},\ and\
		\citenamefont {Bonitz}(2000)}]{filinov_pss_00}%
	\BibitemOpen
	\bibfield  {author} {\bibinfo {author} {\bibfnamefont {A.}~\bibnamefont
			{Filinov}}, \bibinfo {author} {\bibfnamefont {Y.}~\bibnamefont {Lozovik}}, \
		and\ \bibinfo {author} {\bibfnamefont {M.}~\bibnamefont {Bonitz}},\
	}\bibfield  {title} {\enquote {\bibinfo {title} {Path integral simulations of
				crystallization of quantum confined electrons},}\ }\href {\doibase
		10.1002/1521-3951(200009)221:1<231::AID-PSSB231>3.0.CO;2-D} {\bibfield
		{journal} {\bibinfo  {journal} {Phys. Status Solidi B}\ }\textbf {\bibinfo
			{volume} {221}},\ \bibinfo {pages} {231--234} (\bibinfo {year}
		{2000})}\BibitemShut {NoStop}%
	\bibitem [{\citenamefont {Schoof}\ \emph {et~al.}(2011)\citenamefont {Schoof},
		\citenamefont {Bonitz}, \citenamefont {Filinov}, \citenamefont {Hochstuhl},\
		and\ \citenamefont {Dufty}}]{schoof_cpp11}%
	\BibitemOpen
	\bibfield  {author} {\bibinfo {author} {\bibfnamefont {T.}~\bibnamefont
			{Schoof}}, \bibinfo {author} {\bibfnamefont {M.}~\bibnamefont {Bonitz}},
		\bibinfo {author} {\bibfnamefont {A.}~\bibnamefont {Filinov}}, \bibinfo
		{author} {\bibfnamefont {D.}~\bibnamefont {Hochstuhl}}, \ and\ \bibinfo
		{author} {\bibfnamefont {J.}~\bibnamefont {Dufty}},\ }\bibfield  {title}
	{\enquote {\bibinfo {title} {Configuration path integral {M}onte {C}arlo},}\
	}\href@noop {} {\bibfield  {journal} {\bibinfo  {journal} {Contrib. Plasma
				Phys.}\ }\textbf {\bibinfo {volume} {84}},\ \bibinfo {pages} {687--697}
		(\bibinfo {year} {2011})}\BibitemShut {NoStop}%
	\bibitem [{\citenamefont {Filinov}\ \emph {et~al.}(2015)\citenamefont
		{Filinov}, \citenamefont {Fortov}, \citenamefont {Bonitz},\ and\
		\citenamefont {Moldabekov}}]{filinov_pre15}%
	\BibitemOpen
	\bibfield  {author} {\bibinfo {author} {\bibfnamefont {V.~S.}\ \bibnamefont
			{Filinov}}, \bibinfo {author} {\bibfnamefont {V.~E.}\ \bibnamefont {Fortov}},
		\bibinfo {author} {\bibfnamefont {M.}~\bibnamefont {Bonitz}}, \ and\ \bibinfo
		{author} {\bibfnamefont {Z.}~\bibnamefont {Moldabekov}},\ }\bibfield  {title}
	{\enquote {\bibinfo {title} {{Fermionic path-integral Monte Carlo results for
					the uniform electron gas at finite temperature}},}\ }\href {\doibase
		10.1103/PhysRevE.91.033108} {\bibfield  {journal} {\bibinfo  {journal} {Phys.
				Rev. E}\ }\textbf {\bibinfo {volume} {91}},\ \bibinfo {pages} {033108}
		(\bibinfo {year} {2015})}\BibitemShut {NoStop}%
	\bibitem [{\citenamefont {Schoof}\ \emph {et~al.}(2015)\citenamefont {Schoof},
		\citenamefont {Groth}, \citenamefont {Vorberger},\ and\ \citenamefont
		{Bonitz}}]{schoof_prl15}%
	\BibitemOpen
	\bibfield  {author} {\bibinfo {author} {\bibfnamefont {T.}~\bibnamefont
			{Schoof}}, \bibinfo {author} {\bibfnamefont {S.}~\bibnamefont {Groth}},
		\bibinfo {author} {\bibfnamefont {J.}~\bibnamefont {Vorberger}}, \ and\
		\bibinfo {author} {\bibfnamefont {M.}~\bibnamefont {Bonitz}},\ }\bibfield
	{title} {\enquote {\bibinfo {title} {\textit{Ab Initio} thermodynamic results
				for the degenerate electron gas at finite temperature},}\ }\href {\doibase
		10.1103/PhysRevLett.115.130402} {\bibfield  {journal} {\bibinfo  {journal}
			{Phys. Rev. Lett.}\ }\textbf {\bibinfo {volume} {115}},\ \bibinfo {pages}
		{130402} (\bibinfo {year} {2015})}\BibitemShut {NoStop}%
	\bibitem [{\citenamefont {Dornheim}\ \emph
		{et~al.}(2015{\natexlab{a}})\citenamefont {Dornheim}, \citenamefont {Groth},
		\citenamefont {Filinov},\ and\ \citenamefont {Bonitz}}]{dornheim_njp15}%
	\BibitemOpen
	\bibfield  {author} {\bibinfo {author} {\bibfnamefont {T.}~\bibnamefont
			{Dornheim}}, \bibinfo {author} {\bibfnamefont {S.}~\bibnamefont {Groth}},
		\bibinfo {author} {\bibfnamefont {A.}~\bibnamefont {Filinov}}, \ and\
		\bibinfo {author} {\bibfnamefont {M.}~\bibnamefont {Bonitz}},\ }\bibfield
	{title} {\enquote {\bibinfo {title} {{Permutation blocking path integral
					Monte Carlo: a highly efficient approach to the simulation of strongly
					degenerate non-ideal fermions}},}\ }\href
	{http://stacks.iop.org/1367-2630/17/i=7/a=073017} {\bibfield  {journal}
		{\bibinfo  {journal} {New J. Phys.}\ }\textbf {\bibinfo {volume} {17}},\
		\bibinfo {pages} {073017} (\bibinfo {year} {2015}{\natexlab{a}})}\BibitemShut
	{NoStop}%
	\bibitem [{\citenamefont {Gorelov}, \citenamefont {Pierleoni},\ and\
		\citenamefont {Ceperley}(2019)}]{pierleoni_cpp19}%
	\BibitemOpen
	\bibfield  {author} {\bibinfo {author} {\bibfnamefont {V.}~\bibnamefont
			{Gorelov}}, \bibinfo {author} {\bibfnamefont {C.}~\bibnamefont {Pierleoni}},
		\ and\ \bibinfo {author} {\bibfnamefont {D.~M.}\ \bibnamefont {Ceperley}},\
	}\bibfield  {title} {\enquote {\bibinfo {title} {{Benchmarking vdW-DF
					first-principles predictions against Coupled Electron–Ion Monte Carlo for
					high-pressure liquid hydrogen}},}\ }\href {\doibase 10.1002/ctpp.201800185}
	{\bibfield  {journal} {\bibinfo  {journal} {Contributions to Plasma Physics}\
		}\textbf {\bibinfo {volume} {59}},\ \bibinfo {pages} {e201800185} (\bibinfo
		{year} {2019})}\BibitemShut {NoStop}%
	\bibitem [{\citenamefont {Ceperley}(1991)}]{Ceperley1991}%
	\BibitemOpen
	\bibfield  {author} {\bibinfo {author} {\bibfnamefont {D.~M.}\ \bibnamefont
			{Ceperley}},\ }\bibfield  {title} {\enquote {\bibinfo {title} {Fermion
				nodes},}\ }\href {\doibase 10.1007/BF01030009} {\bibfield  {journal}
		{\bibinfo  {journal} {Journal of Statistical Physics}\ }\textbf {\bibinfo
			{volume} {63}},\ \bibinfo {pages} {1237--1267} (\bibinfo {year}
		{1991})}\BibitemShut {NoStop}%
	\bibitem [{\citenamefont {Filinov}\ \emph
		{et~al.}(2001{\natexlab{b}})\citenamefont {Filinov}, \citenamefont {Fortov},
		\citenamefont {Bonitz},\ and\ \citenamefont {Levashov}}]{filinov_jetpl_01}%
	\BibitemOpen
	\bibfield  {author} {\bibinfo {author} {\bibfnamefont {V.}~\bibnamefont
			{Filinov}}, \bibinfo {author} {\bibfnamefont {V.}~\bibnamefont {Fortov}},
		\bibinfo {author} {\bibfnamefont {M.}~\bibnamefont {Bonitz}}, \ and\ \bibinfo
		{author} {\bibfnamefont {P.}~\bibnamefont {Levashov}},\ }\bibfield  {title}
	{\enquote {\bibinfo {title} {Phase transition in strongly degenerate hydrogen
				plasma},}\ }\href@noop {} {\bibfield  {journal} {\bibinfo  {journal} {JETP
				Lett.}\ }\textbf {\bibinfo {volume} {74}},\ \bibinfo {pages} {384} (\bibinfo
		{year} {2001}{\natexlab{b}})}\BibitemShut {NoStop}%
	\bibitem [{\citenamefont {Dornheim}(2019)}]{dornheim_pre_2019}%
	\BibitemOpen
	\bibfield  {author} {\bibinfo {author} {\bibfnamefont {T.}~\bibnamefont
			{Dornheim}},\ }\bibfield  {title} {\enquote {\bibinfo {title} {Fermion sign
				problem in path integral monte carlo simulations: Quantum dots, ultracold
				atoms, and warm dense matter},}\ }\href {\doibase
		10.1103/PhysRevE.100.023307} {\bibfield  {journal} {\bibinfo  {journal}
			{Phys. Rev. E}\ }\textbf {\bibinfo {volume} {100}},\ \bibinfo {pages}
		{023307} (\bibinfo {year} {2019})}\BibitemShut {NoStop}%
	\bibitem [{\citenamefont {Bonitz}\ \emph {et~al.}(2005)\citenamefont {Bonitz},
		\citenamefont {Filinov}, \citenamefont {Fortov}, \citenamefont {Levashov},\
		and\ \citenamefont {Fehske}}]{bonitz_prl_5}%
	\BibitemOpen
	\bibfield  {author} {\bibinfo {author} {\bibfnamefont {M.}~\bibnamefont
			{Bonitz}}, \bibinfo {author} {\bibfnamefont {V.~S.}\ \bibnamefont {Filinov}},
		\bibinfo {author} {\bibfnamefont {V.~E.}\ \bibnamefont {Fortov}}, \bibinfo
		{author} {\bibfnamefont {P.~R.}\ \bibnamefont {Levashov}}, \ and\ \bibinfo
		{author} {\bibfnamefont {H.}~\bibnamefont {Fehske}},\ }\bibfield  {title}
	{\enquote {\bibinfo {title} {{Crystallization in Two-Component Coulomb
					Systems}},}\ }\href {\doibase 10.1103/PhysRevLett.95.235006} {\bibfield
		{journal} {\bibinfo  {journal} {Phys. Rev. Lett.}\ }\textbf {\bibinfo
			{volume} {95}},\ \bibinfo {pages} {235006} (\bibinfo {year}
		{2005})}\BibitemShut {NoStop}%
	\bibitem [{\citenamefont {Brown}\ \emph
		{et~al.}(2013{\natexlab{a}})\citenamefont {Brown}, \citenamefont {Clark},
		\citenamefont {DuBois},\ and\ \citenamefont {Ceperley}}]{Brown_2014}%
	\BibitemOpen
	\bibfield  {author} {\bibinfo {author} {\bibfnamefont {E.~W.}\ \bibnamefont
			{Brown}}, \bibinfo {author} {\bibfnamefont {B.~K.}\ \bibnamefont {Clark}},
		\bibinfo {author} {\bibfnamefont {J.~L.}\ \bibnamefont {DuBois}}, \ and\
		\bibinfo {author} {\bibfnamefont {D.~M.}\ \bibnamefont {Ceperley}},\
	}\bibfield  {title} {\enquote {\bibinfo {title} {{Path-Integral Monte Carlo
					Simulation of the Warm Dense Homogeneous Electron Gas}},}\ }\href {\doibase
		10.1103/PhysRevLett.110.146405} {\bibfield  {journal} {\bibinfo  {journal}
			{Phys. Rev. Lett.}\ }\textbf {\bibinfo {volume} {110}},\ \bibinfo {pages}
		{146405} (\bibinfo {year} {2013}{\natexlab{a}})}\BibitemShut {NoStop}%
	\bibitem [{\citenamefont {Brown}\ \emph
		{et~al.}(2013{\natexlab{b}})\citenamefont {Brown}, \citenamefont {DuBois},
		\citenamefont {Holzmann},\ and\ \citenamefont {Ceperley}}]{Brown_PRB_2013}%
	\BibitemOpen
	\bibfield  {author} {\bibinfo {author} {\bibfnamefont {E.~W.}\ \bibnamefont
			{Brown}}, \bibinfo {author} {\bibfnamefont {J.~L.}\ \bibnamefont {DuBois}},
		\bibinfo {author} {\bibfnamefont {M.}~\bibnamefont {Holzmann}}, \ and\
		\bibinfo {author} {\bibfnamefont {D.~M.}\ \bibnamefont {Ceperley}},\
	}\bibfield  {title} {\enquote {\bibinfo {title} {Exchange-correlation energy
				for the three-dimensional homogeneous electron gas at arbitrary
				temperature},}\ }\href {\doibase 10.1103/PhysRevB.88.081102} {\bibfield
		{journal} {\bibinfo  {journal} {Phys. Rev. B}\ }\textbf {\bibinfo {volume}
			{88}},\ \bibinfo {pages} {081102} (\bibinfo {year}
		{2013}{\natexlab{b}})}\BibitemShut {NoStop}%
	\bibitem [{\citenamefont {Dornheim}\ \emph
		{et~al.}(2015{\natexlab{b}})\citenamefont {Dornheim}, \citenamefont {Schoof},
		\citenamefont {Groth}, \citenamefont {Filinov},\ and\ \citenamefont
		{Bonitz}}]{dornheim_jcp15}%
	\BibitemOpen
	\bibfield  {author} {\bibinfo {author} {\bibfnamefont {T.}~\bibnamefont
			{Dornheim}}, \bibinfo {author} {\bibfnamefont {T.}~\bibnamefont {Schoof}},
		\bibinfo {author} {\bibfnamefont {S.}~\bibnamefont {Groth}}, \bibinfo
		{author} {\bibfnamefont {A.}~\bibnamefont {Filinov}}, \ and\ \bibinfo
		{author} {\bibfnamefont {M.}~\bibnamefont {Bonitz}},\ }\bibfield  {title}
	{\enquote {\bibinfo {title} {Permutation blocking path integral {M}onte
				{C}arlo approach to the uniform electron gas at finite temperature},}\ }\href
	{\doibase http://dx.doi.org/10.1063/1.4936145} {\bibfield  {journal}
		{\bibinfo  {journal} {J. Chem. Phys.}\ }\textbf {\bibinfo {volume} {143}},\
		\bibinfo {pages} {204101} (\bibinfo {year} {2015}{\natexlab{b}})}\BibitemShut
	{NoStop}%
	\bibitem [{\citenamefont {Dornheim}, \citenamefont {Groth},\ and\ \citenamefont
		{Bonitz}(2019)}]{dornheim_cpp_19}%
	\BibitemOpen
	\bibfield  {author} {\bibinfo {author} {\bibfnamefont {T.}~\bibnamefont
			{Dornheim}}, \bibinfo {author} {\bibfnamefont {S.}~\bibnamefont {Groth}}, \
		and\ \bibinfo {author} {\bibfnamefont {M.}~\bibnamefont {Bonitz}},\
	}\bibfield  {title} {\enquote {\bibinfo {title} {{Permutation blocking path
					integral Monte Carlo simulations of degenerate electrons at finite
					temperature}},}\ }\href {\doibase 10.1002/ctpp.201800157} {\bibfield
		{journal} {\bibinfo  {journal} {Contrib. Plasma Phys.}\ }\textbf {\bibinfo
			{volume} {59}},\ \bibinfo {pages} {e201800157} (\bibinfo {year}
		{2019})}\BibitemShut {NoStop}%
	\bibitem [{\citenamefont {Malone}\ \emph {et~al.}(2016)\citenamefont {Malone},
		\citenamefont {Blunt}, \citenamefont {Brown}, \citenamefont {Lee},
		\citenamefont {Spencer}, \citenamefont {Foulkes},\ and\ \citenamefont
		{Shepherd}}]{Malone_PRL_2016}%
	\BibitemOpen
	\bibfield  {author} {\bibinfo {author} {\bibfnamefont {F.~D.}\ \bibnamefont
			{Malone}}, \bibinfo {author} {\bibfnamefont {N.~S.}\ \bibnamefont {Blunt}},
		\bibinfo {author} {\bibfnamefont {E.~W.}\ \bibnamefont {Brown}}, \bibinfo
		{author} {\bibfnamefont {D.~K.~K.}\ \bibnamefont {Lee}}, \bibinfo {author}
		{\bibfnamefont {J.~S.}\ \bibnamefont {Spencer}}, \bibinfo {author}
		{\bibfnamefont {W.~M.~C.}\ \bibnamefont {Foulkes}}, \ and\ \bibinfo {author}
		{\bibfnamefont {J.~J.}\ \bibnamefont {Shepherd}},\ }\bibfield  {title}
	{\enquote {\bibinfo {title} {Accurate exchange-correlation energies for the
				warm dense electron gas},}\ }\href {\doibase 10.1103/PhysRevLett.117.115701}
	{\bibfield  {journal} {\bibinfo  {journal} {Phys. Rev. Lett.}\ }\textbf
		{\bibinfo {volume} {117}},\ \bibinfo {pages} {115701} (\bibinfo {year}
		{2016})}\BibitemShut {NoStop}%
	\bibitem [{\citenamefont {Dornheim}\ \emph
		{et~al.}(2016{\natexlab{a}})\citenamefont {Dornheim}, \citenamefont {Groth},
		\citenamefont {Sjostrom}, \citenamefont {Malone}, \citenamefont {Foulkes},\
		and\ \citenamefont {Bonitz}}]{dornheim_prl16}%
	\BibitemOpen
	\bibfield  {author} {\bibinfo {author} {\bibfnamefont {T.}~\bibnamefont
			{Dornheim}}, \bibinfo {author} {\bibfnamefont {S.}~\bibnamefont {Groth}},
		\bibinfo {author} {\bibfnamefont {T.}~\bibnamefont {Sjostrom}}, \bibinfo
		{author} {\bibfnamefont {F.~D.}\ \bibnamefont {Malone}}, \bibinfo {author}
		{\bibfnamefont {W.~M.~C.}\ \bibnamefont {Foulkes}}, \ and\ \bibinfo {author}
		{\bibfnamefont {M.}~\bibnamefont {Bonitz}},\ }\bibfield  {title} {\enquote
		{\bibinfo {title} {{Ab Initio Quantum Monte Carlo Simulation of the Warm
					Dense Electron Gas in the Thermodynamic Limit}},}\ }\href {\doibase
		10.1103/PhysRevLett.117.156403} {\bibfield  {journal} {\bibinfo  {journal}
			{Phys. Rev. Lett.}\ }\textbf {\bibinfo {volume} {117}},\ \bibinfo {pages}
		{156403} (\bibinfo {year} {2016}{\natexlab{a}})}\BibitemShut {NoStop}%
	\bibitem [{\citenamefont {Spink}, \citenamefont {Needs},\ and\ \citenamefont
		{Drummond}(2013)}]{Spink_PRB_2013}%
	\BibitemOpen
	\bibfield  {author} {\bibinfo {author} {\bibfnamefont {G.~G.}\ \bibnamefont
			{Spink}}, \bibinfo {author} {\bibfnamefont {R.~J.}\ \bibnamefont {Needs}}, \
		and\ \bibinfo {author} {\bibfnamefont {N.~D.}\ \bibnamefont {Drummond}},\
	}\bibfield  {title} {\enquote {\bibinfo {title} {Quantum monte carlo study of
				the three-dimensional spin-polarized homogeneous electron gas},}\ }\href
	{\doibase 10.1103/PhysRevB.88.085121} {\bibfield  {journal} {\bibinfo
			{journal} {Phys. Rev. B}\ }\textbf {\bibinfo {volume} {88}},\ \bibinfo
		{pages} {085121} (\bibinfo {year} {2013})}\BibitemShut {NoStop}%
	\bibitem [{\citenamefont {Groth}\ \emph {et~al.}(2017)\citenamefont {Groth},
		\citenamefont {Dornheim}, \citenamefont {Sjostrom}, \citenamefont {Malone},
		\citenamefont {Foulkes},\ and\ \citenamefont {Bonitz}}]{groth_prl17}%
	\BibitemOpen
	\bibfield  {author} {\bibinfo {author} {\bibfnamefont {S.}~\bibnamefont
			{Groth}}, \bibinfo {author} {\bibfnamefont {T.}~\bibnamefont {Dornheim}},
		\bibinfo {author} {\bibfnamefont {T.}~\bibnamefont {Sjostrom}}, \bibinfo
		{author} {\bibfnamefont {F.~D.}\ \bibnamefont {Malone}}, \bibinfo {author}
		{\bibfnamefont {W.~M.~C.}\ \bibnamefont {Foulkes}}, \ and\ \bibinfo {author}
		{\bibfnamefont {M.}~\bibnamefont {Bonitz}},\ }\bibfield  {title} {\enquote
		{\bibinfo {title} {{Ab initio Exchange-Correlation Free Energy of the Uniform
					Electron Gas at Warm Dense Matter Conditions}},}\ }\href {\doibase
		10.1103/PhysRevLett.119.135001} {\bibfield  {journal} {\bibinfo  {journal}
			{Phys. Rev. Lett.}\ }\textbf {\bibinfo {volume} {119}},\ \bibinfo {pages}
		{135001} (\bibinfo {year} {2017})}\BibitemShut {NoStop}%
	\bibitem [{\citenamefont {Karasiev}, \citenamefont {Trickey},\ and\
		\citenamefont {Dufty}(2019)}]{karasiev_PRB_2019}%
	\BibitemOpen
	\bibfield  {author} {\bibinfo {author} {\bibfnamefont {V.~V.}\ \bibnamefont
			{Karasiev}}, \bibinfo {author} {\bibfnamefont {S.~B.}\ \bibnamefont
			{Trickey}}, \ and\ \bibinfo {author} {\bibfnamefont {J.~W.}\ \bibnamefont
			{Dufty}},\ }\bibfield  {title} {\enquote {\bibinfo {title} {Status of
				free-energy representations for the homogeneous electron gas},}\ }\href
	{\doibase 10.1103/PhysRevB.99.195134} {\bibfield  {journal} {\bibinfo
			{journal} {Phys. Rev. B}\ }\textbf {\bibinfo {volume} {99}},\ \bibinfo
		{pages} {195134} (\bibinfo {year} {2019})}\BibitemShut {NoStop}%
	\bibitem [{\citenamefont {Dornheim}\ \emph {et~al.}(2018)\citenamefont
		{Dornheim}, \citenamefont {Groth}, \citenamefont {Vorberger},\ and\
		\citenamefont {Bonitz}}]{dornheim_prl_18}%
	\BibitemOpen
	\bibfield  {author} {\bibinfo {author} {\bibfnamefont {T.}~\bibnamefont
			{Dornheim}}, \bibinfo {author} {\bibfnamefont {S.}~\bibnamefont {Groth}},
		\bibinfo {author} {\bibfnamefont {J.}~\bibnamefont {Vorberger}}, \ and\
		\bibinfo {author} {\bibfnamefont {M.}~\bibnamefont {Bonitz}},\ }\bibfield
	{title} {\enquote {\bibinfo {title} {{Ab initio Path Integral Monte Carlo
					Results for the Dynamic Structure Factor of Correlated Electrons: From the
					Electron Liquid to Warm Dense Matter}},}\ }\href {\doibase
		10.1103/PhysRevLett.121.255001} {\bibfield  {journal} {\bibinfo  {journal}
			{Phys. Rev. Lett.}\ }\textbf {\bibinfo {volume} {121}},\ \bibinfo {pages}
		{255001} (\bibinfo {year} {2018})}\BibitemShut {NoStop}%
	\bibitem [{\citenamefont {Dornheim}\ \emph {et~al.}(2019)\citenamefont
		{Dornheim}, \citenamefont {Vorberger}, \citenamefont {Groth}, \citenamefont
		{Hoffmann}, \citenamefont {Moldabekov},\ and\ \citenamefont
		{Bonitz}}]{dornheim_jcp_19-nn}%
	\BibitemOpen
	\bibfield  {author} {\bibinfo {author} {\bibfnamefont {T.}~\bibnamefont
			{Dornheim}}, \bibinfo {author} {\bibfnamefont {J.}~\bibnamefont {Vorberger}},
		\bibinfo {author} {\bibfnamefont {S.}~\bibnamefont {Groth}}, \bibinfo
		{author} {\bibfnamefont {N.}~\bibnamefont {Hoffmann}}, \bibinfo {author}
		{\bibfnamefont {Z.}~\bibnamefont {Moldabekov}}, \ and\ \bibinfo {author}
		{\bibfnamefont {M.}~\bibnamefont {Bonitz}},\ }\bibfield  {title} {\enquote
		{\bibinfo {title} {{The Static Local Field Correction of the Warm Dense
					Electron Gas: An ab initio Path Integral Monte Carlo Study and Machine
					Learning Representation}},}\ }\href {\doibase 10.1063/1.5123013} {\bibfield
		{journal} {\bibinfo  {journal} {The Journal of Chemical Physics}\ }\textbf
		{\bibinfo {volume} {151}},\ \bibinfo {pages} {194104} (\bibinfo {year}
		{2019})}\BibitemShut {NoStop}%
	\bibitem [{\citenamefont {Groth}, \citenamefont {Dornheim},\ and\ \citenamefont
		{Vorberger}(2019)}]{groth_prb_19}%
	\BibitemOpen
	\bibfield  {author} {\bibinfo {author} {\bibfnamefont {S.}~\bibnamefont
			{Groth}}, \bibinfo {author} {\bibfnamefont {T.}~\bibnamefont {Dornheim}}, \
		and\ \bibinfo {author} {\bibfnamefont {J.}~\bibnamefont {Vorberger}},\
	}\bibfield  {title} {\enquote {\bibinfo {title} {Ab initio path integral
				monte carlo approach to the static and dynamic density response of the
				uniform electron gas},}\ }\href {\doibase 10.1103/PhysRevB.99.235122}
	{\bibfield  {journal} {\bibinfo  {journal} {Phys. Rev. B}\ }\textbf {\bibinfo
			{volume} {99}},\ \bibinfo {pages} {235122} (\bibinfo {year}
		{2019})}\BibitemShut {NoStop}%
	\bibitem [{\citenamefont {{Dornheim}}\ and\ \citenamefont
		{{Vorberger}}(2020)}]{dornheim_HEDP_2020}%
	\BibitemOpen
	\bibfield  {author} {\bibinfo {author} {\bibfnamefont {T.}~\bibnamefont
			{{Dornheim}}}\ and\ \bibinfo {author} {\bibfnamefont {J.}~\bibnamefont
			{{Vorberger}}},\ }\bibfield  {title} {\enquote {\bibinfo {title}
			{{Finite-size effects in the reconstruction of dynamic properties from ab
					initio path integral Monte-Carlo simulations}},}\ }\href@noop {} {\bibfield
		{journal} {\bibinfo  {journal} {arXiv e-prints}\ ,\ \bibinfo {eid}
			{arXiv:2004.13429}} (\bibinfo {year} {2020})},\ \Eprint
	{http://arxiv.org/abs/2004.13429} {arXiv:2004.13429 [cond-mat.str-el]}
	\BibitemShut {NoStop}%
	\bibitem [{\citenamefont {Dornheim}, \citenamefont {Vorberger},\ and\
		\citenamefont {Bonitz}(2020)}]{dornheim_prl_20}%
	\BibitemOpen
	\bibfield  {author} {\bibinfo {author} {\bibfnamefont {T.}~\bibnamefont
			{Dornheim}}, \bibinfo {author} {\bibfnamefont {J.}~\bibnamefont {Vorberger}},
		\ and\ \bibinfo {author} {\bibfnamefont {M.}~\bibnamefont {Bonitz}},\
	}\bibfield  {title} {\enquote {\bibinfo {title} {{Nonlinear Electronic
					Density Response in Warm Dense Matter}},}\ }\href@noop {} {\bibfield
		{journal} {\bibinfo  {journal} {Phys. Rev. Lett.}\ } (\bibinfo {year}
		{2020})}\BibitemShut {NoStop}%
	\bibitem [{\citenamefont {Dornheim}\ \emph {et~al.}(2017)\citenamefont
		{Dornheim}, \citenamefont {Groth}, \citenamefont {Malone}, \citenamefont
		{Schoof}, \citenamefont {Sjostrom}, \citenamefont {Foulkes},\ and\
		\citenamefont {Bonitz}}]{dornheim_pop17}%
	\BibitemOpen
	\bibfield  {author} {\bibinfo {author} {\bibfnamefont {T.}~\bibnamefont
			{Dornheim}}, \bibinfo {author} {\bibfnamefont {S.}~\bibnamefont {Groth}},
		\bibinfo {author} {\bibfnamefont {F.~D.}\ \bibnamefont {Malone}}, \bibinfo
		{author} {\bibfnamefont {T.}~\bibnamefont {Schoof}}, \bibinfo {author}
		{\bibfnamefont {T.}~\bibnamefont {Sjostrom}}, \bibinfo {author}
		{\bibfnamefont {W.~M.~C.}\ \bibnamefont {Foulkes}}, \ and\ \bibinfo {author}
		{\bibfnamefont {M.}~\bibnamefont {Bonitz}},\ }\bibfield  {title} {\enquote
		{\bibinfo {title} {{Ab initio quantum Monte Carlo simulation of the warm
					dense electron gas}},}\ }\href {\doibase 10.1063/1.4977920} {\bibfield
		{journal} {\bibinfo  {journal} {Phys. Plasmas}\ }\textbf {\bibinfo {volume}
			{24}},\ \bibinfo {pages} {056303} (\bibinfo {year} {2017})},\ \Eprint
	{http://arxiv.org/abs/https://doi.org/10.1063/1.4977920}
	{https://doi.org/10.1063/1.4977920} \BibitemShut {NoStop}%
	\bibitem [{\citenamefont {Schoof}, \citenamefont {Groth},\ and\ \citenamefont
		{Bonitz}(2015)}]{schoof_cpp15}%
	\BibitemOpen
	\bibfield  {author} {\bibinfo {author} {\bibfnamefont {T.}~\bibnamefont
			{Schoof}}, \bibinfo {author} {\bibfnamefont {S.}~\bibnamefont {Groth}}, \
		and\ \bibinfo {author} {\bibfnamefont {M.}~\bibnamefont {Bonitz}},\
	}\bibfield  {title} {\enquote {\bibinfo {title} {Towards ab initio
				thermodynamics of the electron gas at strong degeneracy},}\ }\href {\doibase
		10.1002/ctpp.201400072} {\bibfield  {journal} {\bibinfo  {journal} {Contrib.
				Plasma Phys.}\ }\textbf {\bibinfo {volume} {55}},\ \bibinfo {pages}
		{136--143} (\bibinfo {year} {2015})}\BibitemShut {NoStop}%
	\bibitem [{\citenamefont {Schoof}, \citenamefont {Groth},\ and\ \citenamefont
		{Bonitz}(2014)}]{cpimc_springer_14}%
	\BibitemOpen
	\bibfield  {author} {\bibinfo {author} {\bibfnamefont {T.}~\bibnamefont
			{Schoof}}, \bibinfo {author} {\bibfnamefont {S.}~\bibnamefont {Groth}}, \
		and\ \bibinfo {author} {\bibfnamefont {M.}~\bibnamefont {Bonitz}},\
	}\bibfield  {title} {\enquote {\bibinfo {title} {{Introduction to
					Configuration Path Integral Monte Carlo}},}\ }in\ \href {\doibase
		10.1007/978-3-319-05437-7_5} {\emph {\bibinfo {booktitle} {Complex
				Plasmas}}},\ \bibinfo {series} {Springer Ser. At., Opt., Plasma Phys.},
	Vol.~\bibinfo {volume} {82},\ \bibinfo {editor} {edited by\ \bibinfo {editor}
		{\bibfnamefont {M.}~\bibnamefont {Bonitz}}, \bibinfo {editor} {\bibfnamefont
			{J.}~\bibnamefont {Lopez}}, \bibinfo {editor} {\bibfnamefont
			{K.}~\bibnamefont {Becker}}, \ and\ \bibinfo {editor} {\bibfnamefont
			{H.}~\bibnamefont {Thomsen}}}\ (\bibinfo  {publisher} {Springer International
		Publishing},\ \bibinfo {year} {2014})\ pp.\ \bibinfo {pages}
	{153--194}\BibitemShut {NoStop}%
	\bibitem [{\citenamefont {Groth}, \citenamefont {Dornheim},\ and\ \citenamefont
		{Bonitz}(2017)}]{groth_jcp17}%
	\BibitemOpen
	\bibfield  {author} {\bibinfo {author} {\bibfnamefont {S.}~\bibnamefont
			{Groth}}, \bibinfo {author} {\bibfnamefont {T.}~\bibnamefont {Dornheim}}, \
		and\ \bibinfo {author} {\bibfnamefont {M.}~\bibnamefont {Bonitz}},\
	}\bibfield  {title} {\enquote {\bibinfo {title} {{Configuration path integral
					Monte Carlo approach to the static density response of the warm dense
					electron gas}},}\ }\href {\doibase 10.1063/1.4999907} {\bibfield  {journal}
		{\bibinfo  {journal} {J. Chem. Phys.}\ }\textbf {\bibinfo {volume} {147}},\
		\bibinfo {pages} {164108} (\bibinfo {year} {2017})}\BibitemShut {NoStop}%
	\bibitem [{\citenamefont {Groth}\ \emph {et~al.}(2016)\citenamefont {Groth},
		\citenamefont {Schoof}, \citenamefont {Dornheim},\ and\ \citenamefont
		{Bonitz}}]{groth_prb16}%
	\BibitemOpen
	\bibfield  {author} {\bibinfo {author} {\bibfnamefont {S.}~\bibnamefont
			{Groth}}, \bibinfo {author} {\bibfnamefont {T.}~\bibnamefont {Schoof}},
		\bibinfo {author} {\bibfnamefont {T.}~\bibnamefont {Dornheim}}, \ and\
		\bibinfo {author} {\bibfnamefont {M.}~\bibnamefont {Bonitz}},\ }\bibfield
	{title} {\enquote {\bibinfo {title} {{\textit{Ab initio} quantum Monte Carlo
					simulations of the uniform electron gas without fixed nodes}},}\ }\href
	{\doibase 10.1103/PhysRevB.93.085102} {\bibfield  {journal} {\bibinfo
			{journal} {Phys. Rev. B}\ }\textbf {\bibinfo {volume} {93}},\ \bibinfo
		{pages} {085102} (\bibinfo {year} {2016})}\BibitemShut {NoStop}%
	\bibitem [{\citenamefont {Dornheim}, \citenamefont {Filinov},\ and\
		\citenamefont {Bonitz}(2015)}]{dornheim_prb15}%
	\BibitemOpen
	\bibfield  {author} {\bibinfo {author} {\bibfnamefont {T.}~\bibnamefont
			{Dornheim}}, \bibinfo {author} {\bibfnamefont {A.}~\bibnamefont {Filinov}}, \
		and\ \bibinfo {author} {\bibfnamefont {M.}~\bibnamefont {Bonitz}},\
	}\bibfield  {title} {\enquote {\bibinfo {title} {Superfluidity of strongly
				correlated bosons in two- and three-dimensional traps},}\ }\href {\doibase
		10.1103/PhysRevB.91.054503} {\bibfield  {journal} {\bibinfo  {journal} {Phys.
				Rev. B}\ }\textbf {\bibinfo {volume} {91}},\ \bibinfo {pages} {054503}
		(\bibinfo {year} {2015})}\BibitemShut {NoStop}%
	\bibitem [{\citenamefont {Kas}\ and\ \citenamefont
		{Rehr}(2017)}]{Kas_PRL_2017}%
	\BibitemOpen
	\bibfield  {author} {\bibinfo {author} {\bibfnamefont {J.~J.}\ \bibnamefont
			{Kas}}\ and\ \bibinfo {author} {\bibfnamefont {J.~J.}\ \bibnamefont {Rehr}},\
	}\bibfield  {title} {\enquote {\bibinfo {title} {Finite temperature green's
				function approach for excited state and thermodynamic properties of cool to
				warm dense matter},}\ }\href {\doibase 10.1103/PhysRevLett.119.176403}
	{\bibfield  {journal} {\bibinfo  {journal} {Phys. Rev. Lett.}\ }\textbf
		{\bibinfo {volume} {119}},\ \bibinfo {pages} {176403} (\bibinfo {year}
		{2017})}\BibitemShut {NoStop}%
	\bibitem [{\citenamefont {Dornheim}\ \emph
		{et~al.}(2016{\natexlab{b}})\citenamefont {Dornheim}, \citenamefont {Groth},
		\citenamefont {Schoof}, \citenamefont {Hann},\ and\ \citenamefont
		{Bonitz}}]{dornheim_prb16}%
	\BibitemOpen
	\bibfield  {author} {\bibinfo {author} {\bibfnamefont {T.}~\bibnamefont
			{Dornheim}}, \bibinfo {author} {\bibfnamefont {S.}~\bibnamefont {Groth}},
		\bibinfo {author} {\bibfnamefont {T.}~\bibnamefont {Schoof}}, \bibinfo
		{author} {\bibfnamefont {C.}~\bibnamefont {Hann}}, \ and\ \bibinfo {author}
		{\bibfnamefont {M.}~\bibnamefont {Bonitz}},\ }\bibfield  {title} {\enquote
		{\bibinfo {title} {{Ab initio quantum Monte Carlo simulations of the uniform
					electron gas without fixed nodes: The unpolarized case}},}\ }\href {\doibase
		10.1103/PhysRevB.93.205134} {\bibfield  {journal} {\bibinfo  {journal} {Phys.
				Rev. B}\ }\textbf {\bibinfo {volume} {93}},\ \bibinfo {pages} {205134}
		(\bibinfo {year} {2016}{\natexlab{b}})}\BibitemShut {NoStop}%
	\bibitem [{\citenamefont {Groth}()}]{groth_ueg_2018}%
	\BibitemOpen
	\bibfield  {author} {\bibinfo {author} {\bibfnamefont {S.}~\bibnamefont
			{Groth}},\ }\href@noop {} {\enquote {\bibinfo {title} {{The uniform electron
					gas at the warm dense matter conditions: A Configuration Path Integral Monte
					Carlo perspective}},}\ }\BibitemShut {NoStop}%
	\bibitem [{\citenamefont {Yilmaz}()}]{yilmaz_rcpimc_2020}%
	\BibitemOpen
	\bibfield  {author} {\bibinfo {author} {\bibfnamefont {A.}~\bibnamefont
			{Yilmaz}},\ }\href@noop {} {\enquote {\bibinfo {title} {{Configuration
					Pfadintegral-Monte Carlo unter Einschränkung des Update-Sets}},}\
	}\BibitemShut {NoStop}%
	\bibitem [{Note1()}]{Note1}%
	\BibitemOpen
	\bibinfo {note} {A peculiarity of both approximations is that the accuracy
		decreases with $N$. Apparantly the relative contribution of the kinks that
		are neglected increases faster than the particle number.}\BibitemShut {Stop}%
	\bibitem [{\citenamefont {Hunger}\ \emph {et~al.}(2020)\citenamefont {Hunger},
		\citenamefont {Schoof}, \citenamefont {Dornheim}, \citenamefont {Groth},\
		and\ \citenamefont {Bonitz}}]{hunger_20}%
	\BibitemOpen
	\bibfield  {author} {\bibinfo {author} {\bibfnamefont {K.}~\bibnamefont
			{Hunger}}, \bibinfo {author} {\bibfnamefont {T.}~\bibnamefont {Schoof}},
		\bibinfo {author} {\bibfnamefont {T.}~\bibnamefont {Dornheim}}, \bibinfo
		{author} {\bibfnamefont {S.}~\bibnamefont {Groth}}, \ and\ \bibinfo {author}
		{\bibfnamefont {M.}~\bibnamefont {Bonitz}},\ }\bibfield  {title} {\enquote
		{\bibinfo {title} {{Short-range correlations and momentum distribution
					function of the warm dense electron gas -- \textit{ab initio} quantum Monte
					Carlo results}},}\ }\href@noop {} {\bibfield  {journal} {\bibinfo  {journal}
			{to be published}\ } (\bibinfo {year} {2020})}\BibitemShut {NoStop}%
\end{thebibliography}


%

\end{document}